\documentclass[3p]{elsarticle}

\usepackage{amssymb,bm}
\usepackage{mathrsfs,amsmath}
\usepackage{amsfonts}
\usepackage{subfigure}
\usepackage{color}
\usepackage{graphicx}
\usepackage{tabu}
\usepackage{undertilde}
\usepackage{lineno,hyperref}
\usepackage{multirow}
\newcommand{\captionabove}[2][]{%
    \vskip-\abovecaptionskip
    \vskip+\belowcaptionskip
    \ifx\@nnil#1\@nnil
        \caption{#2}%
        
    \else
        \caption[#1]{#2}%
    \fi
    \vskip+\abovecaptionskip
    \vskip-\belowcaptionskip
}

\usepackage{amsmath}

\numberwithin{equation}{section}

\makeatletter
\def\ps@pprintTitle{%
	\let\@oddhead\@empty
	\let\@evenhead\@empty
	\def\@oddfoot{}%
	\let\@evenfoot\@oddfoot}
\makeatother
\journal{}

\begin{document}

\begin{frontmatter}

\title{Exploring the 3D architectures of deep material network in data-driven multiscale mechanics}
\address[Zeliangaddress]{Livermore Software Technology Corporation (LSTC), Livermore, CA 94551, USA}
\cortext[mycorrespondingauthor]{Corresponding author.}
\author[Zeliangaddress]{Zeliang Liu\corref{mycorrespondingauthor}}
\ead{zlliu@lstc.com}
\author[Zeliangaddress]{C.T. Wu}
\begin{abstract}
This paper extends the deep material network (DMN) proposed by Liu et al. (2019) \cite{liu2019deep} to tackle general 3-dimensional (3D) problems with arbitrary material and geometric nonlinearities. It discovers a new way of describing multiscale heterogeneous materials by a multi-layer network structure and mechanistic building blocks. The data-driven framework of DMN is discussed in detail about the offline training and online extrapolation stages. Analytical solutions of the 3D building block with a two-layer structure in both small- and finite-strain formulations are derived based on interfacial equilibrium conditions and kinematic constraints. With linear elastic data generated by direct numerical simulations on a representative volume element (RVE), the network can be effectively trained in the offline stage using stochastic gradient descent and advanced model compression algorithms. Efficiency and accuracy of DMN on addressing the long-standing 3D RVE challenges with complex morphologies and material laws are validated through numerical experiments, including 1) hyperelastic particle-reinforced rubber composite with Mullins effect; 2) polycrystalline materials with rate-dependent crystal plasticity; 3) carbon fiber reinforced polymer (CFRP) composites with fiber anisotropic elasticity and matrix plasticity. In particular, we demonstrate a three-scale homogenization procedure of CFRP system by concatenating the microscale and mesoscale material networks. The complete learning and extrapolation procedures of DMN establish a reliable data-driven framework for multiscale material modeling and design.
\end{abstract}
\end{frontmatter}
\section{Introduction}
Modern material systems with properly designed microstructures offer new material engineering avenues for producing advantageous mechanical properties and functionalities in various applications. In aerospace and automotive industries, carbon fiber reinforced polymer (CFRP) composites have become attractive alternatives of metal materials due to their high strength/weight ratio induced by the interplay between carbon fibers and epoxy matrix \cite{daniel1994engineering,deng2015isogeometric,ren2018peridynamic}. Another typical example of heterogeneous materials is the particle-reinforced rubber composite where nanoparticles are added into the matrix to manipulate the overall mechanical properties, such as stiffness and viscoelastic behaviors \cite{balazs2006nanoparticle,ramanathan2008functionalized,xu2015machine,liu2016extended}. Moreover, in metal additive manufacturing, the performance of the final product is strongly affected by the microscopic polycrystalline microstructure, which could be controlled by the deposition and cooling processes \cite{yan2018data,lian2018parallelized}. Efficient and physical descriptions of such multiscale materials will not only help to predict the performance of large-scale structures, but also accelerate the pace of discovery and design of new material systems.

The multiscale nature of heterogeneous materials poses limitations on single-scale empirical models, which lose the sight of microstructural interactions. They tend to fail in attempting to capture nonlinear or anisotropic responses, and may require burdensome calibration to find the model parameters. As a result, homogenization based on the concept of representative volume element (RVE) \cite{hill1963elastic} has become an important approach to model multiscale materials \cite{geers2010multi}. Many analytical methods have been proposed, which adopt some micromechanics assumptions to simplify the full-field RVE problem, such as Hashin-Shtrikman bounds \cite{hashin1962variational,hashin1963variational}, the Mori-Tanaka method \cite{mori1973average} and self-consistent methods\cite{hill1965self}. Since most analytical methods are derived based on the solution for regular geometries and simple material models (e.g. Eshelby's solution for isotropic elastic materials \cite{eshelby1957determination}), it is usually difficult for them to consider complex microstructural morphologies, history-dependent materials and large deformations. Additionally, direct numerical simulation (DNS) tools, such as finite element \cite{feyel2000a}, meshfree \cite{wu2012three,wu2013numerical} and fast Fourier transform (FFT)-based micromechanics methods \cite{moulinec1998a,de2017finite}, are both flexible and accurate. However, a DNS model involves a detailed meshing of the RVE microstructures and requires tremendous computational cost, especially for 3D problems. Therefore, one of the fundamental issues in multiscale material modeling and design is how to find an accurate low-dimensional representation of the RVE for arbitrary morphologies and nonlinearities. 

In the past decade, a plethora of data-driven material modeling methods have been proposed based on existing machine learning techniques. Depending on the type of training data in the offline stage, we classify these methods mainly into macroscopic and microscopic approaches. In macroscopic approaches, the stress-strain relations, or strain energy density functions, are directly fitted by regression methods, like deep neural network (DNN) \cite{ghaboussi1991knowledge,unger2008coupling,le2015computational,bessa2017} and Kriging methods \cite{bessa2017,krige1951statistical}. Enabled by recent progresses in computer hardware systems, DNN becomes one of the most popular tools due to its large model generalities \cite{goodfellow2014generative,lecun2015deep}, and has also stimulated applications across different engineering disciplines \cite{decost2017exploring,gu2018novo,gu2018bioinspired,lu2018data}. However, the extrapolation capability of the macroscopic approaches to unknown material and loading spaces is usually limited by the lack of microscale physics. To overcome this problem, microscopic approaches take the full RVE stress or strain fields as the training data and use dimension-reduction techniques to find the reduced basis, such as non-uniform transformation field analysis\cite{michel2003a,roussette2009nonuniform}, proper orthogonal decomposition \cite{yvonnet2007a,oliver2017reduced}, manifold learning methods \cite{bhattacharjee2016nonlinear,ibanez2018manifold} and self-consistent clustering analysis (SCA)\cite{liu2016self,liu2018microstructural,tang2018virtual}. One potential limitation of the microscopic approaches is the difficulty of obtaining stress or strain fields from experiments, though advances in digital image correlation methods \cite{chu1985applications,buljac2017numerical} may provide a solution.

We recently proposed a novel data-driven material modeling method called ``deep material network" (DMN) \cite{liu2019deep} and demonstrated its effectiveness for various challenging 2-dimensional RVE problems, including matrix-inclusion composite, amorphous material and anisotropic material with penetrating phase. It discovers a reduced representation of RVE based on a binary-tree network of mechanistic two-layer building blocks. Comparing to other data-driven methods, DMN has the following intriguing features: 1) Avoiding an extensive offline sampling stage; 2) eliminating the need for extra calibration and micromechanics assumption; 3) efficient online prediction without the danger of extrapolation \cite{liu2019deep}. With the physics embedded in the DMN model, we are able to extract essential microstructural information (e.g. volume fractions) from pure macroscopic mechanical data, which can also be measured from experiments. In the online extrapolation stage, efficient and accurate predictions are achieved for unknown materials, such as linear elasticity with high contrast of phase properties, nonlinear history-dependent plasticity and finite-strain hyperelasticity under large deformations. Different from methods based on solving the continuum equilibrium equation of the RVE, either in partial differential from (e.g. FEM) or integral forms (e.g. FFT-based methods and SCA), DMN uses a hierarchical network structure to propagate all physical quantities, such as the stress and strain. As a result, its computational time is proportional to the number of degrees of freedom (DOFs), which is advantageous over most existing homogenization methods.

This paper adds an important piece to the DMN framework to treat general 3-dimensional (3D) problems with both material and geometric nonlinearities. The global framework of DMN, including the offline and online stages, is introduced in Section \ref{sec:framework}. Theories of the 3D mechanistic building block are derived in Section \ref{sec:theory}. We apply the DMN to three representative multiscale material systems in Section \ref{sec:application}, including the particle-reinforced rubber composite, polycrystalline materials and CFRP composites. Moreover, a three-scale homogenization procedure of CFRP material system is accomplished by simple concatenation of microscale and mesoscale material networks. We provide the computational cost of DMN in the offline and online stages in Section \ref{sec:time}, and further discuss its efficiency comparing to DNS and reference FE models. Concluding remarks are given in Section \ref{sec:conclusion}.

\section{The global framework of deep material network} \label{sec:framework}
\subsection{Preliminaries}\label{sec:pre}
\begin{figure}[!t]
	\centering
	\graphicspath{{Figures/}}
	\includegraphics[clip=true,trim = 4cm 3cm 4cm 3cm, width = 0.75\textwidth]{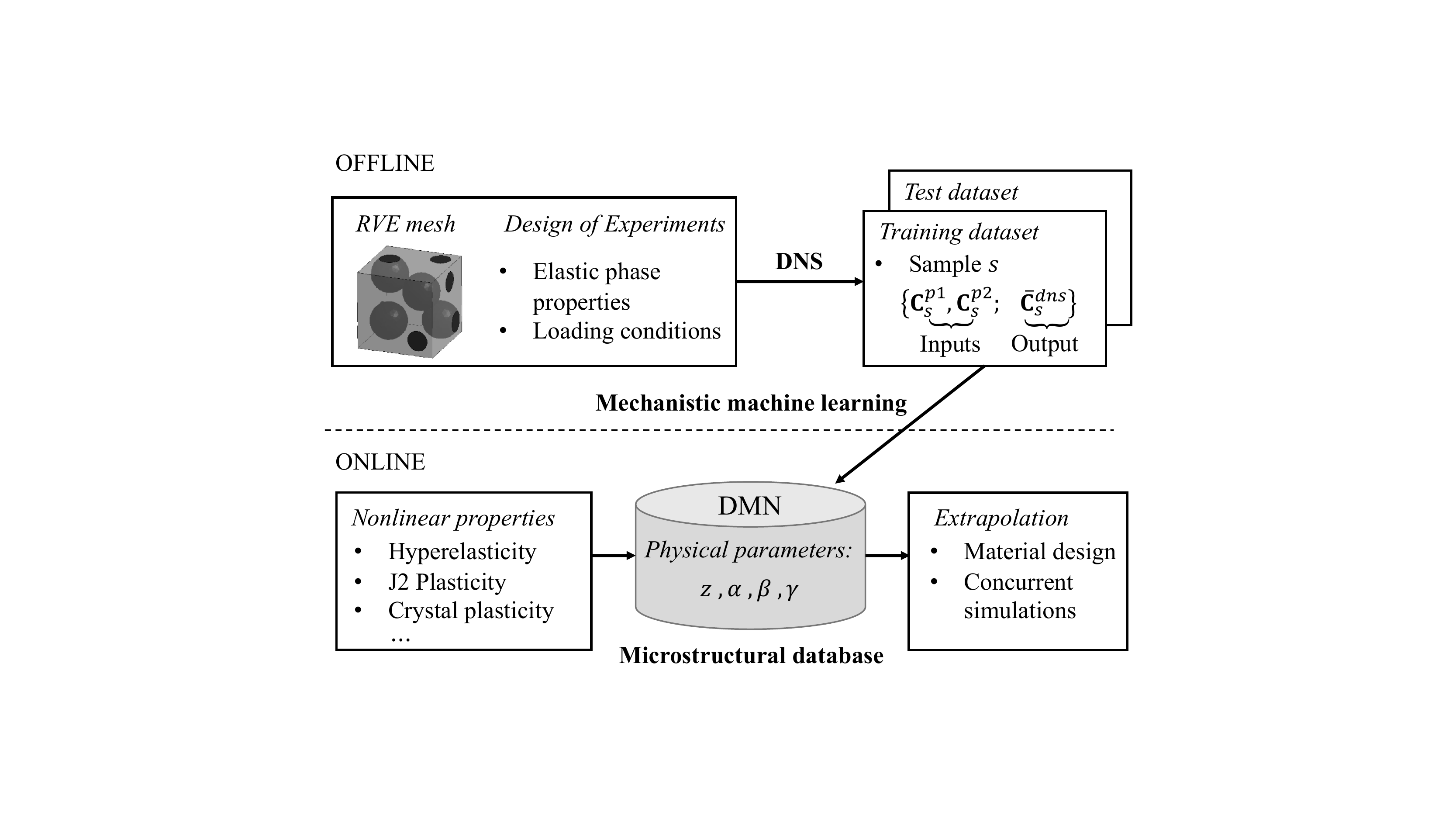}
	\caption{Global framework of deep material network illustrated for a 3D two-phase RVE. The stiffness matrices of the two microscale phases are $\textbf{C}_s^{p1}$ and $\textbf{C}_s^{p2}$, and $\bar{\textbf{C}}^{dns}_s$ is the overall stiffness matrices generated by DNS of RVE homogenization. Fitting parameters of DMN include activation $z$ and rotation angles ($\alpha,\beta,\gamma$).}
	\label{fig:framework}
\end{figure}

The global framework of deep material network is presented in Figure \ref{fig:framework}. As a common practice of data-driven methods, an offline stage is first performed to collect the training datasets and search for the optimum model that represents the input-output relation. We start from a given RVE mesh, and for each sample $s$, we assign a set of random elastic stiffness matrices to microscale material phases as the inputs.  For a two-phase RVE, these inputs are denoted by $\textbf{C}^{p1}_s$ and $\textbf{C}^{p2}_s$. Design of experiments (DoE) is performed to effectively explore the high-dimensional sampling space of input variables. The full-field RVE models can be then analyzed by various DNS tools, such as finite element method and FFT-based method. In general, six orthogonal loading conditions need to be simulated for each sample in order to output all the components in an elastic stiffness matrix $\bar{\textbf{C}}^{dns}_s$.  Other than the training dataset, it is always important to prepare the test dataset with extra points sampled in the same design space, which can be used later on to check the quality of a fitted model for generalization.

\textbf{Mechanistic machine learning.}
Once the training dataset is collected, it can be evaluated by machine learning algorithms to train the parameters in DMN. As shown in Figure \ref{fig:framework}, the fitting parameters in the model are activation $z$ and rotation angles $\alpha,\beta,\gamma$, all of which have physical meanings related to microstructural geometry and mechanics, as will be discussed in Section \ref{sec:mechgeoflow}. The training process is formulated as a supervised optimization problem, where the objective is to minimize the distance between the stiffness matrix predicted by the DMN and the one from DNS. Gradient-based methods are adopted to update the fitting parameters. Meanwhile, several network compression algorithms are also introduced to reduced the number of fitting parameters and improve the convergence speed during the training. Details on the data generation and machine learning of DMN can be found in Section \ref{sec:offline}.

\textbf{Microstructural database.}
The trained DMN discovers a reduced topological representation of the original full-field RVE with fewer DOFs. Based on the mechanistic building block with embedded physics, it can be extrapolated to unknown materials and loading paths outside the original linear elastic sampling space, without violating any physical law in the homogenization. By virtue of its unique capability of extrapolation, DMN creates a so-called ``microstructural database", which has the potential to drive high-fidelity concurrent multiscale simulations. Its intrinsically parameterized structure also provides a mechanistic understanding of structure-property relations and enables material design across length scales.  

\subsection{Multi-layer network structure}\label{sec:mechgeoflow}
\begin{figure}[!t]
	\centering
	\graphicspath{{Figures/}}
	\subfigure[Data flow of stiffness matrices for a two-phase material.]{\includegraphics[clip=true,trim = 9.5cm 2.5cm 7.5cm 2.0cm,width=0.49\textwidth]{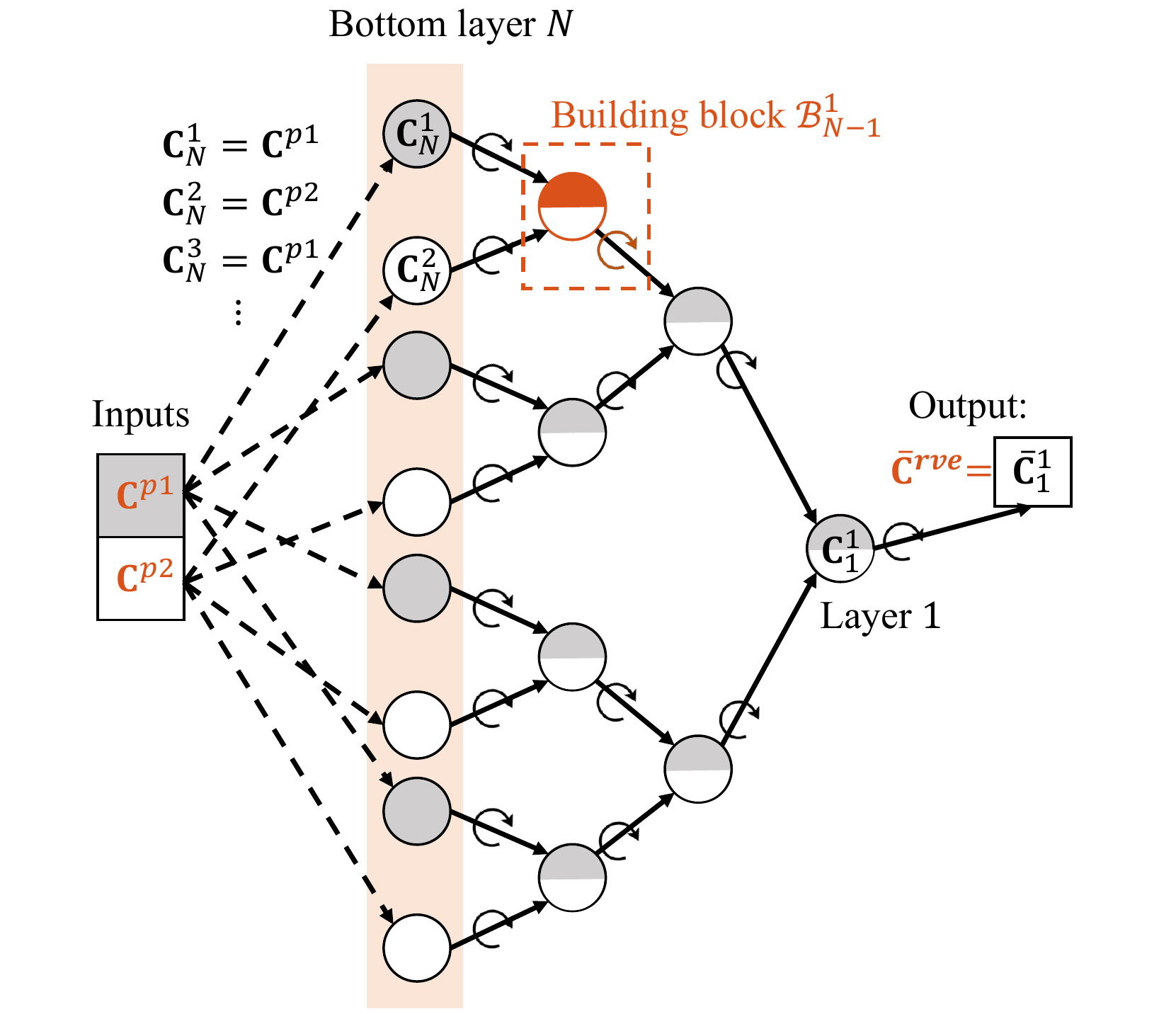}}
	\subfigure[Data flow of weights.]{\includegraphics[clip=true,trim = 9.5cm 2.5cm 7.5cm 2.0cm,width=0.49\textwidth]{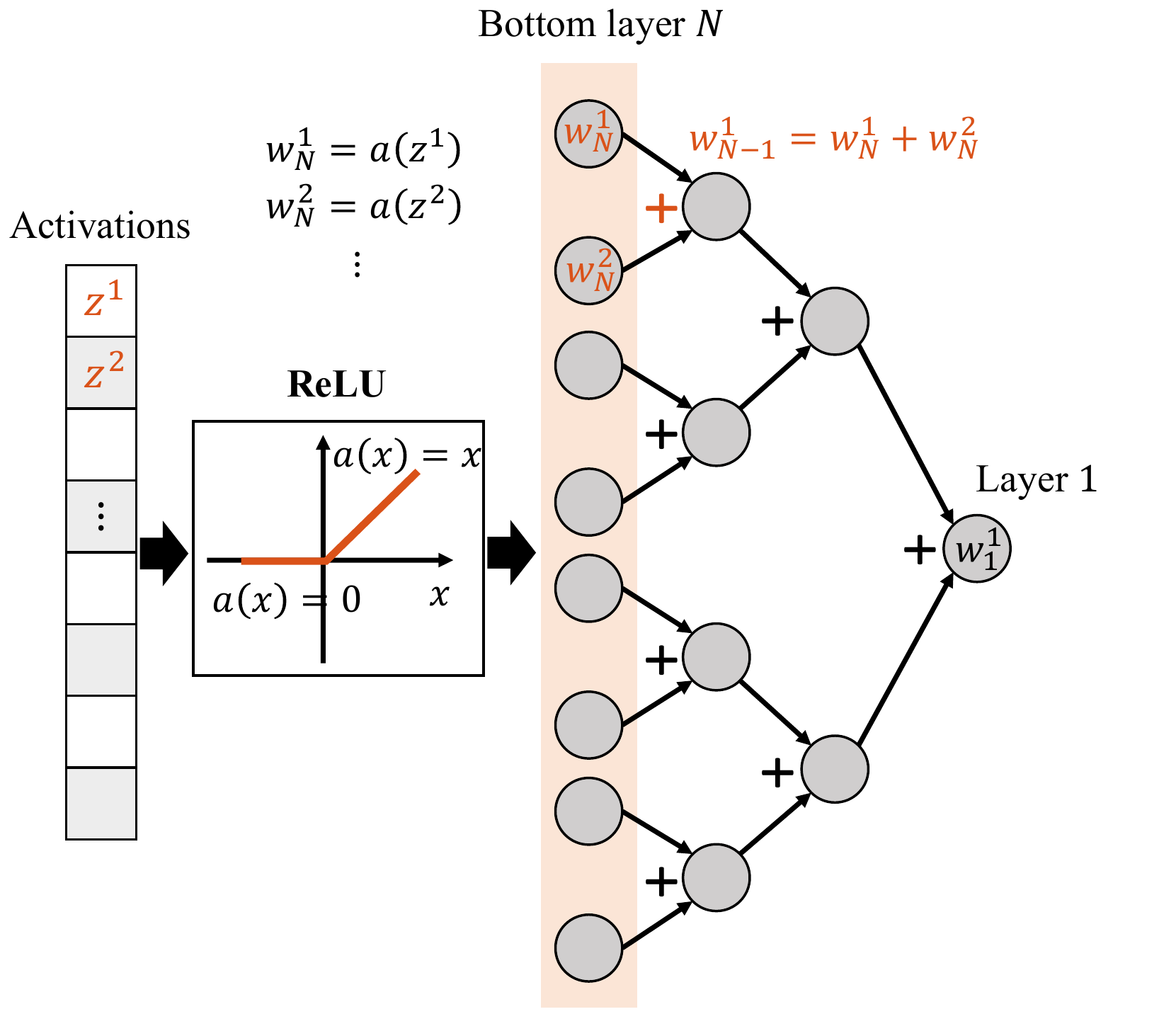}}
	\caption{Illustrations of the multi-layer network structure of DMN, as well as the (a) mechanical and  (b) geometric data flows within the network. The depth of the network is $N$, and the $k$-th building block at layer $i$ is denoted by $\mathfrak{B}_i^k$.}
	\label{fig:dataFlow}
\end{figure}

We propose to use a collection of connected mechanistic building blocks to represent the RVE model. As shown in Figure \ref{fig:dataFlow}, a binary-tree structure is chosen as the base of DMN, where each node has two child nodes. The top layer is denoted as Layer $1$, and the bottom layer is denoted as Layer $N$. Given a network with depth $N$, there are $2^{N-1}$ nodes in the bottom layer and in total  $(2^N-1)$ nodes within the whole network. Nodes can be deactivated during the training process, as a result, some nodes may only have one child node in the final network structure. 

\textbf{Physics of building blocks}
Other than the binary-tree network structure, another key ingredient of DMN is the use of mechanistic building blocks. The rule of a building block is similar to the one of ``neurons" in artificial neural network, however, the transformation functions in the building block are derived based on the analytical solutions of mechanistic structures with physical meanings. Generally speaking, each building block contains one homogenization operation and one rotation operation. As will be shown in Section \ref{sec:theory}, the building block is proposed as a 3D two-layer structure, which is probably the simplest form of a heterogeneous composite. The material phases interact through the interface between the two layers. The effect of phase fraction is captured by the homogenization operation, which is related to the fitting parameter $z$. To increase the flexibility of the DMN representation, the rotation operation. controlled by the fitting parameters ($\alpha,\beta,\gamma$), is introduced to vary the global orientation of the block. Our hypothesis is that the original 3D RVE can be represented by patterns of the two-layer structures with unknown phase fractions and block orientations. At layer $i$, the $k$-th building block is denoted as $\mathfrak{B}_i^k$. We use $\textbf{C}$ to represent a stiffness matrix before the rotation operation, and the one after rotation is denoted by $\bar{\textbf{C}}$.

\textbf{Mechanical data flow.}
Inputs and outputs of current DMN models are all mechanical data, such as stiffness matrix and residual stress. In particular, a material network in the offline stage takes the stiffness matrices of microscale phases as the inputs and output the overall RVE stiffness matrix $\bar{\textbf{C}}^{rve}$.  Illustrations of how the stiffness matrices propagate from the input layer to the output layer are depicted in Figure \ref{fig:dataFlow} (a) for a two-phase material. The inputs $\textbf{C}^{p1}$ and $\textbf{C}^{p2}$ are first assigned to the nodes in the bottom layer $N$ in a scheme designed as below,
\begin{equation}\label{eq:input}
\textbf{C}_N^{odd}=\textbf{C}^{p1}, \quad \textbf{C}_N^{even}=\textbf{C}^{p2}.
\end{equation}
At the end of forward propagation, the output of DMN is obtained as
\begin{equation}\label{eq:output}
\bar{\textbf{C}}^{rve} = \bar{\textbf{C}}^1_1.
\end{equation}
The framework can be generalized to single-phase or multi-phase materials. For a RVE with one single material phase, like the polycrystalline RVE that will be investigated in Section \ref{sec:poly}, the whole framework of DMN does not need to be altered, except that all the bottom-layer nodes in the network receive the same stiffness matrix $\textbf{C}^{p1}$. Extensions to RVE problems with more than two material phases are also possible, such as redesigning the network structure to let each node take more child nodes, or introducing advanced schemes of assigning multiple material phases to the bottom layer without changing binary-tree structure. Nevertheless, we will focus on single-phase and two-phase RVEs in this paper. In addition, to consider both material and geometric nonlinearities in the online stage, we also need to add the residual stress to the mechanical data flow, as will be discussed in Section \ref{sec:online}.

\textbf{Geometric data flow.}
As shown in Figure \ref{fig:dataFlow} (b), the weights $w$ is added into the data flow to keep track of the physical portion of each node in the network.  We use $w_i^k$ to denote the weight of the $k$-th node at Layer $i$. An important feature of DMN is that the weights in the bottom layer are activated through the rectified linear unit (ReLU) \cite{glorot2011deep}. For $j\in[1,2^{N-1}]$, we have
\begin{equation}\label{eq:relu}
w_N^{j}=a(z^j)=\max(z^j,0),
\end{equation}
where $a$ is the ReLU activation function and $z$ is called the ``activation". 
It should be noted that once a unit is deactivated ($z^j<0$), the unit will never be activated again during the training process due to its vanishing gradient ($a'(z^j)=0$). This feature is helpful for automatically simplifying the material network and increasing the training speed. Without losing the physical meaning of $w$,  weights of two child nodes are summed up and passed to their parent node.  From Layer $i+1$ to $i$, we have 
\begin{equation}
w_i^k=w_{i+1}^{2k-1}+w_{i+1}^{2k}.
\end{equation}
By performing the summation recursively, any weight $w_i^k$ can be expressed as a summation of weights of its descendant nodes at the bottom layer $N$,
\begin{equation}\label{eq:wsum}
w_i^k=\sum_{j=2^{N-i}(k-1)+1}^{2^{N-i}k}w_N^j = \sum_{j=2^{N-i}(k-1)+1}^{2^{N-i}k}a(z^j) .
\end{equation}
The activations $z^j$ determine all the weights in the network, hence, they are chosen as the independent fitting parameters in DMN. Additionally, the partial derivative of $w_i^k$ with respect to $z^j$ can be written as
\begin{equation}
\dfrac{\partial w_i^k }{\partial z^j}=
\begin{cases}
a'(z^j) \quad\quad\text{if } k=\lceil{j/2^{N-i}}\rceil\\
0\quad\quad \text{otherwise}
\end{cases},
\end{equation} 
where the operator $\lceil\;\rceil$ rounds a number to the next larger integer. 

\subsection{Offline training based on linear elasticity}\label{sec:offline}
\begin{figure}[!t]
	\centering
	\graphicspath{{Figures/}}
	\includegraphics[clip=true,trim = 4cm 6.5cm 4cm 6.5cm, width = 0.75\textwidth]{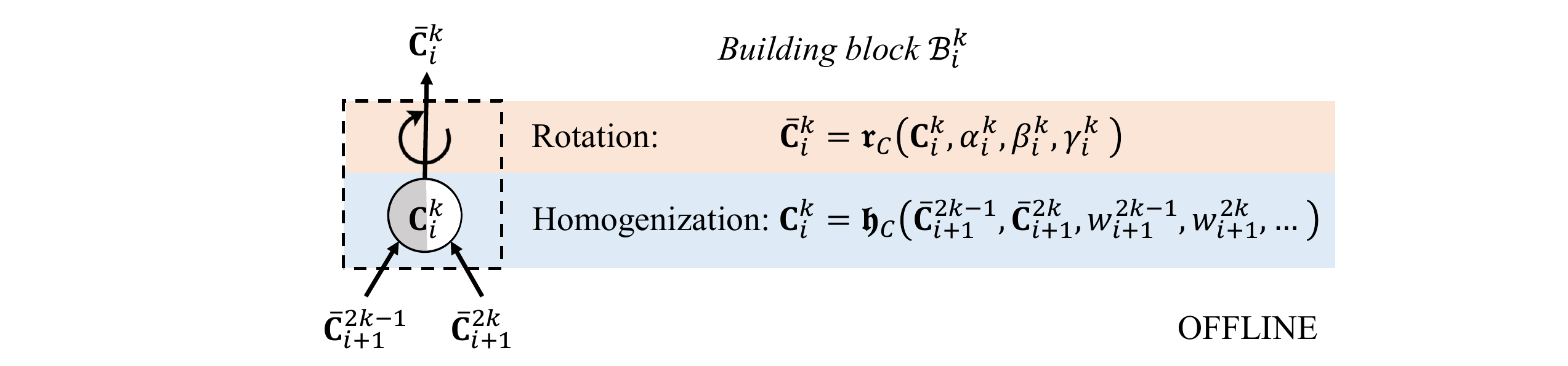}
	\caption{Building block $\mathfrak{B}_i^k$ in the offline stage.}
	\label{fig:offlineBlock}
\end{figure}

Let us zoom into the building block $\mathfrak{B}_i^k$ appeared in the offline stage. As illustrated in Figure \ref{fig:offlineBlock}, the building block has two inputs $\bar{\textbf{C}}_{i+1}^{2k-1}$ and $\bar{\textbf{C}}_{i+1}^{2k}$, and there are two operations within each building block: Homogenization and rotation. The homogenization operation is defined by the transformation function $\boldsymbol{\mathfrak{h}}_c$, which is determined by the underlying physical structure and takes geometric descriptors (e.g. the weights $w$) as parameters. The rotation operation, described by the function $\boldsymbol{\mathfrak{r}}_c$, takes the stiffness matrix $\textbf{C}_i^k$ generated by the homogenization operation and rotates it based on the angles $\alpha_i^k, \beta_i^k$ and $\gamma_i^k$. The result after this 3D rotation is $\bar{\textbf{C}}_{i}^{k}$. Explicit expressions of $\boldsymbol{\mathfrak{h}}_c$ and $\boldsymbol{\mathfrak{r}}_c$  for a generic two-layer structure can be found in Section \ref{sec:linearDeri}. Note that a building block at the bottom layer is slightly different from ones in other layers, since it only takes one input and the homogenization operation is ignored as depicted in Figure \ref{fig:dataFlow} (a).

Each building block generates a more flexible material model than its inputs. In this way, a building block in the upper layer can capture material responses at a more complex level than one in the lower layer. Based on Eq. (\ref{eq:input}), (\ref{eq:output}) and (\ref{eq:wsum}), the overall stiffness matrix of the multi-layer network  $\bar{\textbf{C}}^{rve}$  for a two-phase RVE can be written as a function of the stiffness matrix from each material phase ($\textbf{C}^{p1}$ and $\textbf{C}^{p2}$) and the fitting parameters ($z, \alpha, \beta, \gamma$), 
\begin{equation}\label{eq:final}
\underbrace{\bar{\textbf{C}}^{rve}}_\text{Output}=\textbf{f}_2(\underbrace{\textbf{C}^{p1},\textbf{C}^{p2}}_\text{Inputs}, \overbrace{z^{j=1,2,...,2^{N-1}},\alpha_{i=1,...,N}^{k=1,2,...,2^{i-1}},\beta_{i=1,2,...,N}^{k=1,2,...,2^{i-1}},\gamma  _{i=1,2,...,N}^{k=1,2,...,2^{i-1}}}^\text{Fitting parameters}).
\end{equation}
Similarly, the overall function of DMN for a single-phase RVE becomes
\begin{equation}\label{eq:final1}
\underbrace{\bar{\textbf{C}}^{rve}}_\text{Output}=\textbf{f}_1(\underbrace{\textbf{C}^{p1}}_\text{Input}, \overbrace{z^{j=1,2,...,2^{N-1}},\alpha_{i=1,...,N}^{k=1,2,...,2^{i-1}},\beta_{i=1,2,...,N}^{k=1,2,...,2^{i-1}},\gamma  _{i=1,2,...,N}^{k=1,2,...,2^{i-1}}}^\text{Fitting parameters}).
\end{equation}
For a network with depth $N$, there are totally $(7\times 2^{N-1}-3)$ fitting parameters for a 3D problem.  However, the actual number of active fitting parameters will be diminished on the fly during the training process due to nodal deactivation and model compression.

\textbf{Cost function.}
A cost function based on the mean square error (MSE) and a regularization term \cite{liu2019deep} is formulated to characterize how close is the model prediction $\bar{\textbf{C}}^{rve}$ to the training reference $\bar{\textbf{C}}^{dns}$:
\begin{equation}\label{eq:cost}
J(z,\alpha,\beta,\gamma) = \dfrac{1}{2N_s}\sum_sJ_s(z,\alpha,\beta,\gamma) + \lambda\left(\sum_j a(z^j)-2^{N-2}\right)^2，
\end{equation}
with
\begin{equation}\label{eq:mse}
J_s(z,\alpha,\beta,\gamma)= \dfrac{||\bar{\textbf{C}}^{dns}_s-\textbf{f}_2(\textbf{C}^{p1}_s,\textbf{C}^{p2}_s,z,\alpha,\beta,\gamma)||^2}{||\bar{\textbf{C}}^{dns}_s||^2}\quad \text{for a two-phase RVE},
\end{equation}
or
\begin{equation}\label{eq:mse2}
J_s(z,\alpha,\beta,\gamma)= \dfrac{||\bar{\textbf{C}}^{dns}_s-\textbf{f}_1(\textbf{C}^{p1}_s,z,\alpha,\beta,\gamma)||^2}{||\bar{\textbf{C}}^{dns}_s||^2}\quad \text{for a single-phase RVE}.
\end{equation}
Here $s$ is the index of sample (or data point), and $N_s$ is the total number of training samples. In Eq. (\ref{eq:mse}) and (\ref{eq:mse2}) , the error function is normalized by the squared norm of $\bar{\textbf{C}}^{dns}_s$ to remove the scaling effect. The operator $||...||$ denotes the Frobenius matrix norm.  The regularization term, scaled by a positive hyper-parameter $\lambda$, is introduced to make the optimization problem well-posed. Although $\lambda$  will not alter the optimum fitting parameters, in practice, it should be set appropriately to expedite the gradient descent algorithm in the training process.

\textbf{Design of experiments} 
As mentioned in Section \ref{sec:pre}, the training and test datasets can be generated through high-fidelity DNS of an RVE (e.g. FEM and FFT-based methods). Since the input and output stiffness matrices for DMN can also be measured from experiments, it is also promising to incorporate both numerical and experimental data in the training dataset following the same format. When generating the input data, the components in the compliance matrices $\textbf{D}^{p1}$ and $ \textbf{D}^{p2}$ are first sampled through design of experiments, and then inverted to give the stiffness matrices. 

In general, an anisotropic compliance matrix should be used in order to uniquely determine the material orientation for the bottom-layer nodes. However, we consider both material phases to be orthotropically elastic to reduce the dimension of the sampling space and ease the data generation process. Since the material models used in the online stage have orthotropic symmetry, the non-uniqueness of orientations will not affect the predicted results. In Mandel notation, the components in $\textbf{D}^{p1}$ and $ \textbf{D}^{p2}$ are
\begin{equation}
\textbf{D}^{pi}=
\begin{Bmatrix}
1/E_{11}^{pi}&-\nu_{12}^{pi}/E_{22}^{pi}&-\nu_{31}^{pi}/E_{11}^{pi}&&&\\
&1/E_{22}^{pi}&-\nu_{23}^{pi}/E_{33}^{pi}&&&\\
&&1/E_{33}^{pi}&&&\\
&&&1/(2G_{23}^{pi})&&\\
&&&&1/(2G_{31}^{pi})&\\
&&&&&1/(2G_{12}^{pi})\\
\end{Bmatrix}\quad
\text{with } i=1,2.
\end{equation}
To initiate material anisotropy, the tension moduli of phase $i$ are first randomly assigned as
\begin{equation}
\log_{10}(E_{11}^{pi}),\enskip \log_{10}(E_{22}^{pi}),\enskip \log_{10}(E_{33}^{pi}) \in U[-1, 1],
\end{equation}
where $U$ stands for uniform distribution. The moduli of each material phase will be rescaled to create overall phase contrasts, and the rescaling factor $\bar{E}^{pi}$ are sampled as
\begin{equation}
\bar{E}^{p1}=1, \quad \log_{10}(\bar{E}^{p2})\in U[-3, 3].
\end{equation}
Note that the rescaling factor of phase 1 is kept as constant to remove the scaling effect of RVE. We can then update the tension moduli of each phase
\begin{equation}
E_{kk}^{pi} \leftarrow \dfrac{\bar{E}^{pi}}{(E_{11}^{pi}E_{22}^{pi}E_{33}^{pi})^{1/3}}E_{kk}^{pi}\quad \text{with } k=1,2,3.
\end{equation}
After the tension moduli are determined, the shear moduli are sampled as
\begin{equation}
\dfrac{G_{12}^{pi}}{\sqrt{E_{11}^{pi}E_{22}^{pi}}},\enskip \dfrac{G_{23}^{pi}}{\sqrt{E_{22}^{pi}E_{33}^{pi}}},\enskip \dfrac{G_{31}^{pi}}{\sqrt{E_{33}^{pi}E_{11}^{pi}}} \in U[0.25, 0.5].
\end{equation}
To guarantee the compliance matrices are positive definite, the Poisson's ratios are sampled as
\begin{equation*}
\dfrac{\nu_{12}^{pi}}{\sqrt{E_{22}^{pi}/E_{11}^{pi}}}\in U(0.0,0.5),\enskip \dfrac{\nu_{23}^{pi}}{\sqrt{E_{33}^{pi}/E_{22}^{pi}}}\in U(0.0,0.5),\enskip
\dfrac{\nu_{31}^{pi}}{\sqrt{E_{11}^{pi}/E_{33}^{pi}}}\in U(0.0,0.5).
\end{equation*}

DoE based on the Monte Carlo sampling is performed to explore the input space. For each material system, we generate 500 DNS samples. The first 400 samples are selected as the training dataset, and the remaining 100 samples become the test dataset.  The size of the test set is selected to be large enough so that the distribution of test error is sampled adequately. Ideally, It is always better to increase the size of the training set for covering more possible cases, however, this will put more loads on the data generation process using DNS. In our study, the number of training samples is determined when the average test error (of a trained model) asymptotes sufficiently close to the average training error \cite{Goodfellow-et-al-2016}.

\begin{figure} [!t]
	\centering
	\graphicspath{{Figures/}}
	\subfigure[Anisotropy of phase 1.]{\includegraphics[clip=true,trim = 0.0cm 0.0cm 1.0cm 0.5cm,width=0.44\textwidth]{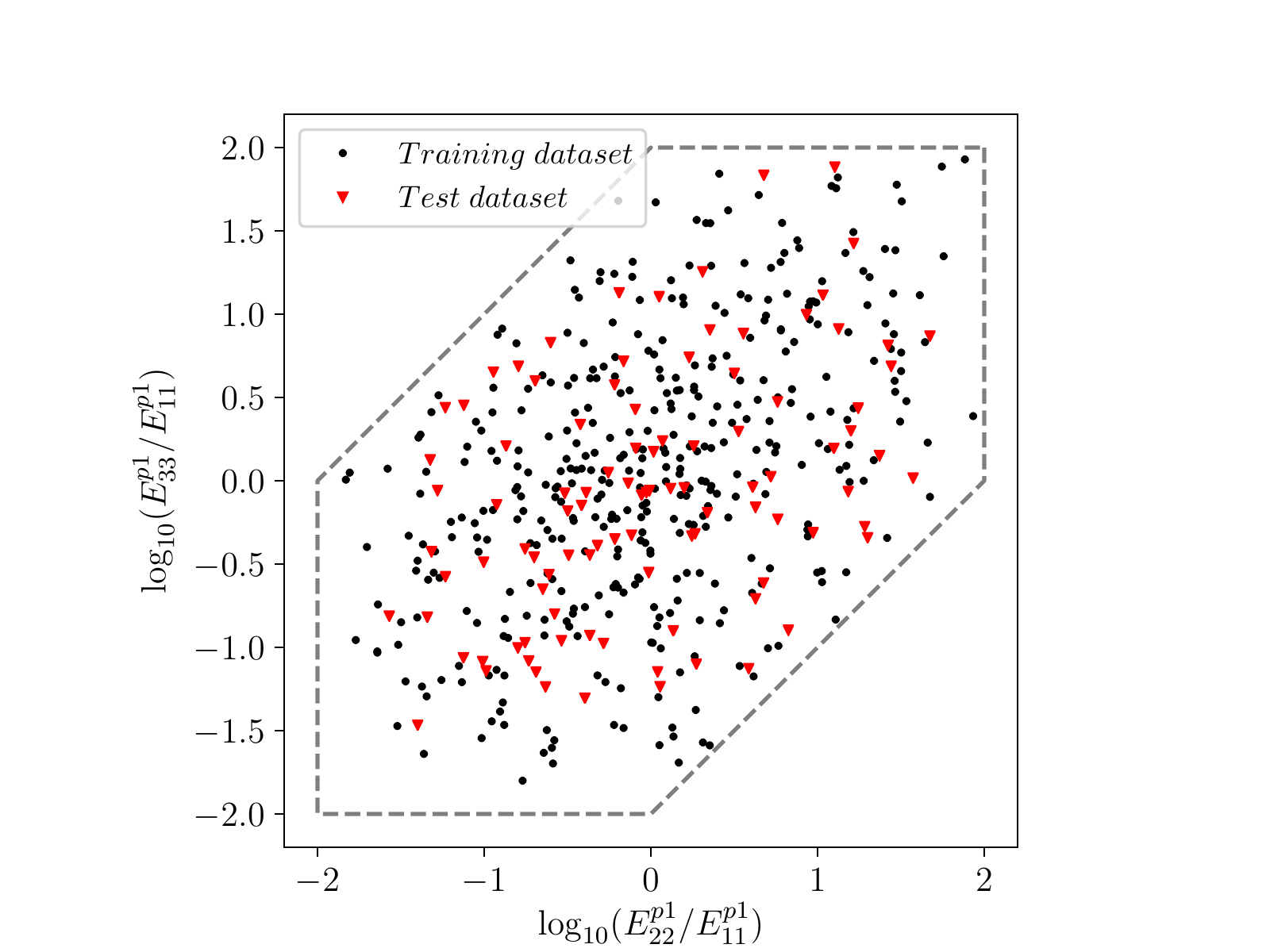}}
	\subfigure[Contrasts of moduli between phase 1 and 2.]{\includegraphics[clip=true,trim = 0.0cm 0.0cm 1.0cm 0.5cm,width=0.44\textwidth]{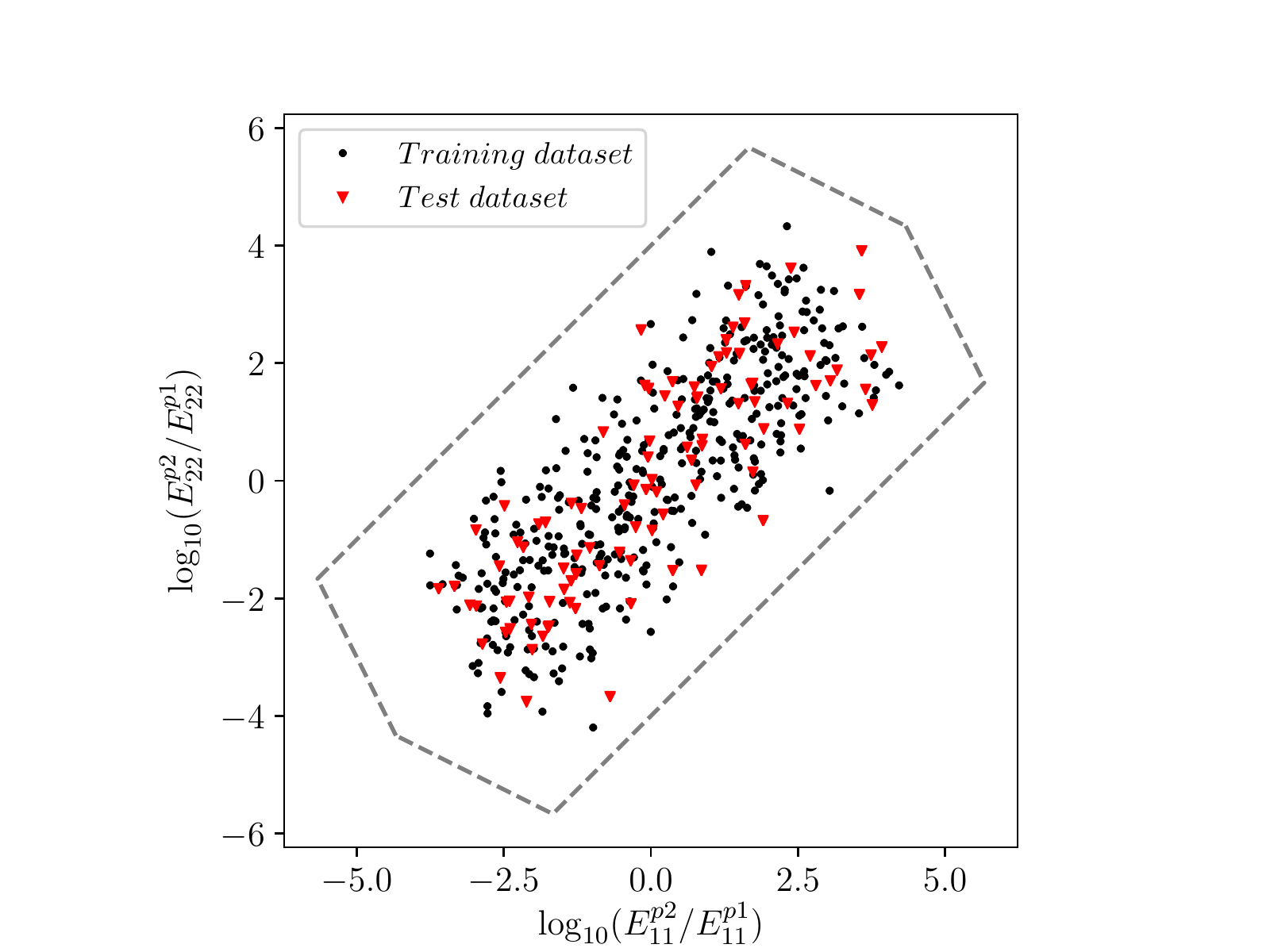}}
	\caption{Distributions of tension moduli in the training and test datasets for the particle-reinforced RVE. Plots for other material systems are similar, and there are 400 training samples ($\bullet$) and 100 test samples ($\blacktriangledown$). The theoretical bounds are shown as the dashed lines.}
	\label{fig:doe}
\end{figure}

As an example, we demonstrate the distributions of training and test samples for the particle-reinforced RVE in Figure \ref{fig:doe}. The DoE procedures for all the material systems are same in terms of the material models and sampling bounds. As a result, their distribution plots generated by random sampling are similar. It can be seen from Figure \ref{fig:doe} (a) that the ratios between tension moduli along different directions for a single phase range from 0.01 to 100. Meanwhile, the contrasts of tension moduli between the two phases can go up to 10000 as shown in Figure \ref{fig:doe} (b). Due to the strong material anisotropy and high phase contrasts in the training dataset, it will be challenging for methods relying on micromechanics assumptions (e.g. analytical mean-field methods \cite{mori1973average,hill1965self}, SCA \cite{liu2016self}) to capture the behaviors of all the samples. By preserving the physics in the building blocks, a DMN can cover the training dataset accurately with sufficient number of layers. In Section \ref{sec:application}, we will validate its capability for various material systems.

\textbf{SGD with model compression.}
Since the building block is designed to have analytical solutions for $\boldsymbol{\mathfrak{h}}_C$ and $\boldsymbol{\mathfrak{r}}_C$, gradient-based optimization methods can be utilized to update the fitting parameters and minimize the cost function, and the gradient vector of the cost function $\nabla J$ is computed by the back-propagation algorithm. To accelerate the training speed, we use the stochastic gradient descent (SGD) to train the network. Instead of computing and averaging the gradients over all the samples at each step, the original dataset is divided into several mini-batches which will be used in a sequence of learning steps. Each time the algorithm has processed all the mini-batches and seen all the samples in the original dataset (or one ``epoch"), the dataset will be randomly shuffled to minimize the sample bias. All fitting parameters are initialized randomly following a uniform distribution at the beginning of the training,
\begin{equation}\label{eq:initialfit}
z^j{}^{(0)}\sim U(0.2,0.8)\quad\text{and}\quad \alpha^k_i{}^{(0)},\beta^k_i{}^{(0)},\gamma^k_i{}^{(0)}\sim U(-\pi/2,\pi/2).
\end{equation}

Similar to 2D DMN \cite{liu2019deep}, the 3D DMN training process can be accelerated through two model compression methods:  1) Deletion of the parent node with only one child node; 2) subtree merging based on the similarity search. In similarity search, both volume fractions and rotation matrices  are compared layer-by-layer between two subtree structures. The representation of 3D rotation using three angles may encounter the loss of one degree of freedom, known as the ``Gimbal lock", in which case two different sets of angles may represent the same 3D rotation. Therefore, we propose to use the rotation matrix $\textbf{R}$ in the similarity search algorithm. Mathematically, two rotations, denoted by $\textbf{R}_1$ and $\textbf{R}_2$, are said to be the same if
\begin{equation}
(\textbf{R}_1)^{-1}\textbf{R}_2 = \textbf{I}.
\end{equation}
However, due to the use of orthotropic material in the offline training data, the similarity condition between two arbitrary rotation matrices $\textbf{R}_1$ and $\textbf{R}_2$ at the bottom layer is relaxed: All the eigenvalues $\lambda_i$ of the matrix $(\textbf{R}_1)^{-1}\textbf{R}_2$ should satisfy
\begin{equation}
||\mathbb{R}(\lambda_i)|-1| < tol,
\end{equation} 
where $tol$ is the tolerance and $\mathbb{R}(\lambda_i)$ denotes the real part of $\lambda_i$. If two subtrees are found similar to each other, they will be merged to one single branch. Each time before the similarity search and subtree merging, the whole material network will be reordered based on the nodal weights. To save the training time, the model compression operations are performed every 10 epochs in our study.

\subsection{Online prediction}\label{sec:online}
\begin{figure}[!t]
	\centering
	\graphicspath{{Figures/}}
	\includegraphics[clip=true,trim = 4cm 6.2cm 4cm 6.2cm, width = 0.75\textwidth]{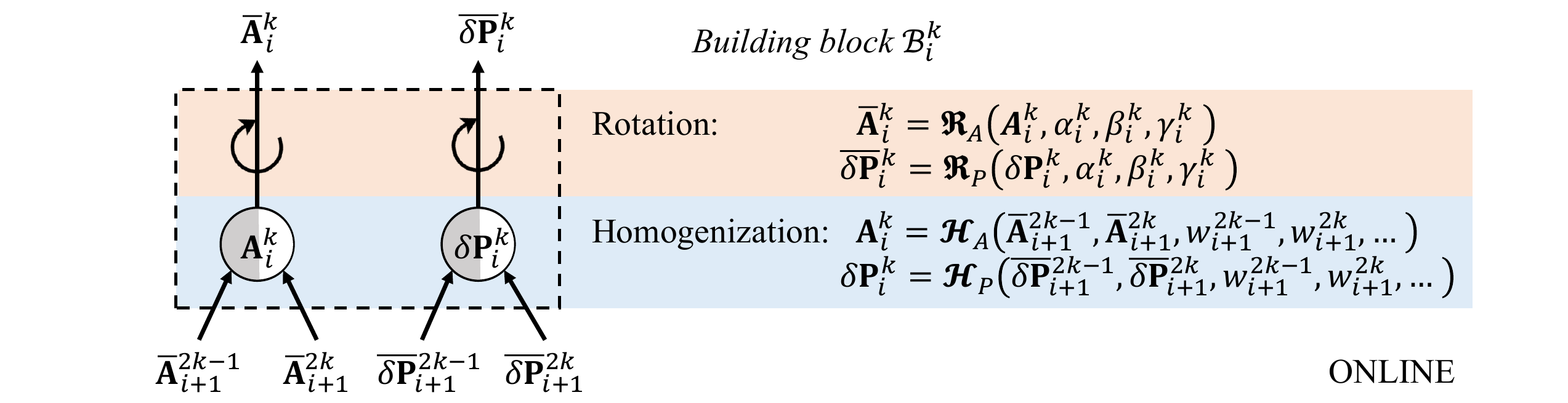}
	\caption{Building block $\mathfrak{B}_i^k$ in the online stage.}
	\label{fig:onlineBlock}
\end{figure}

The online stage is intended to extrapolate the trained DMN to unknown material and loading spaces. Each active node in the bottom layer is treated as an independent material point, with its own loading path and internal variables (if any). For a general finite-strain problem, we will choose the deformation gradient $\textbf{F}$ and the first Piola-Kirchhoff stress $\textbf{P}$ as the strain and stress measures (see Section \ref{sec:nonlinearDeri} for more details). With both material and geometric nonlinearities taken into account, the mechanical data now contains two parts: The tangent stiffness matrix $\textbf{A}$ and the residual stress $\delta \textbf{P}$. As presented in Figure \ref{fig:onlineBlock}, the inputs of the building block are $\bar{\textbf{A}}_{i+1}^{2k-1}, \bar{\textbf{A}}_{i+1}^{2k}, \delta\bar{\textbf{P}}_{i+1}^{2k-1}$ and $\delta\bar{\textbf{P}}_{i+1}^{2k}$ from lower layers. Two homogenization functions $\boldsymbol{\mathcal{H}}_A$ and $\boldsymbol{\mathcal{H}}_P$ are introduced separately to map the inputs to a new stiffness matrix ${\textbf{A}}_{i}^{k}$ and residual stress $\delta {\textbf{P}}_{i}^{k}$, respectively. Then, they are rotated by functions $\boldsymbol{\mathfrak{R}}_A$ and $\boldsymbol{\mathfrak{R}}_P$, and the outputs of building block $\mathfrak{B}_i^k$ are $\bar{\textbf{A}}_i^{k}$ and $\delta\bar{\textbf{P}}_{i}^{k}$.

The nonlinear RVE problem is solved via Newton's method. Each Newton iteration contains one forward homogenization process and one backward de-homogenization process, and the mechanical data in the forward and backward propagations are shown as below,
\begin{equation*}
\text{layer } N \xrightarrow[\text{homogenization}]{\text{forward } (\textbf{A},\delta\textbf{P})} \text{layer } 1 \text{ (macroscale)} \xrightarrow[\text{de-homogenization}]{\text{backward }(\Delta\textbf{F},\Delta\textbf{P})} \text{layer } N.
\end{equation*}
At the start of each iteration, the increment of deformation gradient $\Delta\boldsymbol{F}_N^{j}$ is prescribed on each active node in the bottom layer $N$, and the values of $\bar{\textbf{A}}_N^j$ and $\delta\bar{\textbf{P}}_N^j$ are obtained by evaluating the local constitutive law. Then the data of stiffness matrix and residual stress are feed forward to the top layer through the homogenization and rotation functions at each building block. The overall $\bar{\textbf{A}}^{rve}$ and $\delta\bar{\textbf{P}}^{rve}$ of the RVE are obtained as the results of one forward homogenization process. Given the macroscopic loading condition, we can then compute the overall increments of deformation gradient $\Delta\bar{\textbf{F}}^{rve}$ and stress $\Delta\bar{\textbf{P}}^{rve}$. In the de-homogenization process, the increments of deformation gradients and stress are feed backward from the top layer to the bottom layer, using the stress-strain relations at each building block. Finally, we return to the bottom layer with the new increment of deformation $\Delta\boldsymbol{F}_N^{j (new)}$ at each active node. The convergence condition is evaluated by checking the relative difference between $\Delta\boldsymbol{F}_N^{j}$ and  $\Delta\boldsymbol{F}_N^{j (new)}$. If convergence is not yet achieved, $\Delta\boldsymbol{F}_N^{j}$ will be updated by $\Delta\boldsymbol{F}_N^{j (new)}$ and used for the next iteration. Upon convergence, the internal variables at each active node in the bottom layer will be updated, and the analysis moves on to the next loading step.

An important feature of DMN is that the total number of operations in one Newton's iteration is proportional to the number of active nodes in the bottom layer, and we will study the computational cost of DMN for various 3D RVE problems in Section \ref{sec:time}. Another interesting feature of DMN is that different networks can be easily concatenated to simulate a material system with more than two scales, and the new integrated network can be solved in the online stage without altering the algorithms. In Section \ref{sec:cfrp}, as an example, we will demonstrate a three-scale nonlinear homogenization of CFRP system by attaching networks of the microscale unidirectional fiber composite to the bottom-layer nodes in a network of the mesoscale woven composite.

\section{Mechanistic building block in 3D}\label{sec:theory}
\subsection{Fundamentals  for linear elasticity}\label{sec:linearDeri}
The theory of 3D building block is first discussed under small-strain assumption, which is mainly used in the offline training stage. The stress and strain measures are the Cauchy stress $\utilde{\boldsymbol{\sigma}}$ and the infinitesimal strain $\utilde{\boldsymbol{\varepsilon}}$, and they are related by a fourth-order stiffness tensor $\utilde{{\textbf{C}}}$. The overall stress-strain relation of the building block can be expressed as
\begin{equation}
\utilde{\bar{\boldsymbol{\sigma}}}=\utilde{\bar{\textbf{C}}}:\utilde{\bar{\boldsymbol{\varepsilon}}}.
\end{equation}
For a building block with two materials 1 and 2, we have
\begin{equation}
\utilde{\boldsymbol{\sigma}}^1=\utilde{\bar{\textbf{C}}}^1:\utilde{\boldsymbol{\varepsilon}}^1, \quad \utilde{\boldsymbol{\sigma}}^2=\utilde{\bar{\textbf{C}}}^2:\utilde{\boldsymbol{\varepsilon}}^2.
\end{equation}
Here a bar is placed above each individual stiffness tensor (e.g. $\utilde{\bar{\textbf{C}}}^1$) to denote that it is from the averaging of a previous building block in the network. With the weights $w^1$ and $w^2$ given, the volume fractions $f_1$ and $f_2$ can be computed by
\begin{equation}\label{eq:vf}
f_1 = \dfrac{w^1}{w^1+w^2}, \quad f_2 = 1-f_1.
\end{equation}
In Mandel notation, the stress and strain can be written as
\begin{equation}
\boldsymbol{\sigma}=\{\utilde{\sigma}_{11},\utilde{\sigma}_{22},\utilde{\sigma}_{33},\sqrt{2}\utilde{\sigma}_{23},\sqrt{2}\utilde{\sigma}_{13},\sqrt{2}\utilde{\sigma}_{12}\}^T=\{\sigma_{1},\sigma_{2},\sigma_{3},\sigma_{4},\sigma_{5},\sigma_{6}\}^T 
\end{equation}
and
\begin{equation*}
\boldsymbol{\varepsilon}=\{\utilde{\varepsilon}_{11},\utilde{\varepsilon}_{22},\utilde{\varepsilon}_{33},\sqrt{2}\utilde{\varepsilon}_{23},\sqrt{2}\utilde{\varepsilon}_{13},\sqrt{2}\utilde{\varepsilon}_{12}\}^T=\{\varepsilon_{1},\varepsilon_{2},\varepsilon_{3},\varepsilon_{4},\varepsilon_{5},\varepsilon_{6}\}^T,
\end{equation*}
where the subscripts 3, 4 and 5 denote the shear directions, and the corresponding stiffness matrix $\textbf{C}$ has 21 independent components.

\begin{figure}[!t]
	\centering
	\graphicspath{{Figures/}}
	\includegraphics[clip=true,trim = 4cm 6cm 4cm 6.2cm, width = 0.7\textwidth]{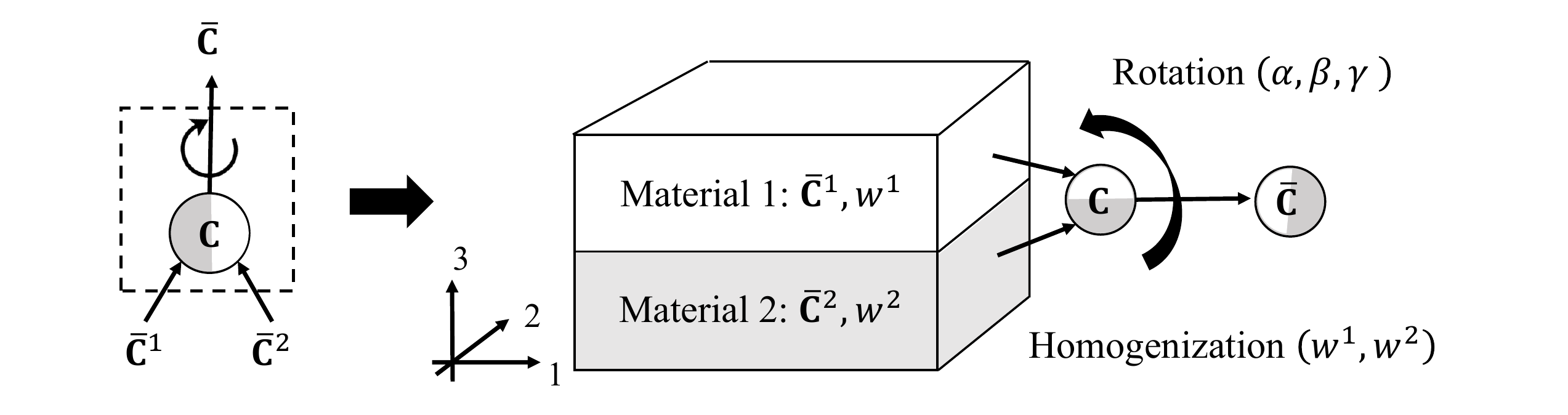}
	\caption{Illustration of the two-layer 3D building block. The stiffness matrix after the homogenization operation is $\textbf{C}$, and the one after the rotation operation is $\bar{\textbf{C}}$.}
	\label{fig:twolayer}
\end{figure}
Similar to the building block in 2D \cite{liu2019deep}, a two-layer structure is proposed in this paper, as shown in Figure \ref{fig:twolayer}. Due to its simplicity, it is possible to derive the analytical forms of its averaging functions, which are essential for calculating the gradients during the network training process. There are two operations evolved in the averaging of building block: 1) Homogenization given by the function  $\boldsymbol{\mathfrak{h}}_C$,
\begin{equation}
\textbf{C} = \boldsymbol{\mathfrak{h}}_C\left(\bar{\textbf{C}}^1,\bar{\textbf{C}}^2,w^1,w^2\right);
\end{equation}
and 2) rotation of the whole two-layer structure given by $\boldsymbol{\mathfrak{r}}_C$,
\begin{equation}
\bar{\textbf{C}} = \boldsymbol{\mathfrak{r}}_C\left(\textbf{C},\alpha,\beta,\gamma\right).
\end{equation}

\textbf{Homogenization function $\boldsymbol{\mathfrak{h}}_C$.} 
The analytical form of homogenization function $\boldsymbol{\mathfrak{h}}_C$ is derived based on the interfacial equilibrium conditions
\begin{equation}\label{eq:con1}
\sigma_{3}^1 = \sigma_{3}^2,\quad \sigma_4^1 = \sigma_4^2,\quad \sigma_5^1 = \sigma_5^2,
\end{equation}
and the kinematic constraints
\begin{equation}\label{eq:con2}
\varepsilon_1^1 = \varepsilon_1^2,\quad \varepsilon_2^1 = \varepsilon_2^2,\quad \varepsilon_6^1 = \varepsilon_6^2.
\end{equation}
First, assume an arbitrary overall strain after homogenization
\begin{equation}
{\boldsymbol{\varepsilon}}=\{{\varepsilon}_{1},{\varepsilon}_{2},{\varepsilon}_{3},{\varepsilon}_{4},{\varepsilon}_{5},{\varepsilon}_{6}\}^T.
\end{equation}
Rewrite the kinematic constraints in Eq. (\ref{eq:con2}),
\begin{equation}\label{eq:strain1}
\varepsilon_1^1 = \varepsilon_1^2={\varepsilon}_{1},\quad \varepsilon_2^1 = \varepsilon_2^2={\varepsilon}_{2},\quad \varepsilon_6^1 = \varepsilon_6^2={\varepsilon}_{6},
\end{equation}
and apply the averaging equation for strain $\boldsymbol{\varepsilon}=f_1\boldsymbol{\varepsilon}^1+f_2\boldsymbol{\varepsilon}^2$,
\begin{equation}\label{eq:strain2}
\varepsilon_3^2 = \dfrac{1}{f_2}{\varepsilon}_3-\dfrac{f_1}{f_2}\varepsilon_3^1,\quad \varepsilon_4^2 = \dfrac{1}{f_2}{\varepsilon}_4-\dfrac{f_1}{f_2}\varepsilon_4^1,\quad \varepsilon_5^2 = \dfrac{1}{f_2}{\varepsilon}_5-\dfrac{f_1}{f_2}\varepsilon_5^1.
\end{equation}
Utilizing the constitutive law for each material and substituting Eq. (\ref{eq:strain1}) and (\ref{eq:strain2}) into the equilibrium condition in Eq. (\ref{eq:con1}) yield the following equation for the strain components of material 1,
\begin{equation}
\begin{Bmatrix}
\hat{C}_{33}&\hat{C}_{34}&\hat{C}_{35}\\
\hat{C}_{34}&\hat{C}_{44}&\hat{C}_{45}\\
\hat{C}_{35}&\hat{C}_{45}&\hat{C}_{55}\\
\end{Bmatrix}
\begin{Bmatrix}
\varepsilon_{3}^1\\
\varepsilon_{4}^1\\
\varepsilon_{5}^1\\
\end{Bmatrix}=
\begin{Bmatrix}
f_2(\Delta C_{13}{\varepsilon}_{1}+\Delta C_{23}{\varepsilon}_{2}+\Delta C_{36}{\varepsilon}_{6})+\left( \bar{C}_{33}^2{\varepsilon}_{3}+\bar{C}_{34}^2{\varepsilon}_{4}+\bar{C}_{35}^2{\varepsilon}_{5}\right)\\
f_2(\Delta C_{14}{\varepsilon}_{1}+\Delta C_{24}{\varepsilon}_{2}+\Delta C_{46}{\varepsilon}_{6})+\left( \bar{C}_{34}^2{\varepsilon}_{3}+\bar{C}_{44}^2{\varepsilon}_{4}+\bar{C}_{45}^2{\varepsilon}_{5}\right)\\
f_2(\Delta C_{15}{\varepsilon}_{1}+\Delta C_{25}{\varepsilon}_{2}+\Delta C_{56}{\varepsilon}_{6})+\left( \bar{C}_{35}^2{\varepsilon}_{3}+\bar{C}_{45}^2{\varepsilon}_{4}+\bar{C}_{55}^2{\varepsilon}_{5}\right)\\
\end{Bmatrix},
\end{equation}
where
\begin{equation}
\hat{\textbf{C}}=f_2\bar{\textbf{C}}^1+f_1\bar{\textbf{C}}^2\quad\text{and}\quad \Delta\textbf{C}=\bar{\textbf{C}}^2-\bar{\textbf{C}}^1.
\end{equation}
After solving the linear system, the unknown strain components $\varepsilon_{3}^1$, $\varepsilon_{4}^1$ and $\varepsilon_{5}^1$ in material 1 can be expressed as a function of the overall strain,
\begin{equation}\label{eq:c345}
\begin{Bmatrix}
\varepsilon_{3}^1\\
\varepsilon_{4}^1\\
\varepsilon_{5}^1\\
\end{Bmatrix}=\left(\hat{\textbf{C}}_{345}\right)^{-1}
\begin{Bmatrix}
f_2\Delta C_{13}&f_2\Delta C_{23}&\bar{C}_{33}^2&\bar{C}_{34}^2&\bar{C}_{35}^2&f_2\Delta C_{36}\\
f_2\Delta C_{14}&f_2\Delta C_{24}&\bar{C}_{34}^2&\bar{C}_{44}^2&\bar{C}_{45}^2&f_2\Delta C_{46}\\
f_2\Delta C_{15}&f_2\Delta C_{25}&\bar{C}_{35}^2&\bar{C}_{45}^2&\bar{C}_{55}^2&f_2\Delta C_{56}\\
\end{Bmatrix}{\boldsymbol{\varepsilon}}=\textbf{s}_{3\times 6}\cdot \boldsymbol{\varepsilon}
\end{equation}
with
\begin{equation}
\hat{\textbf{C}}_{345}=\begin{Bmatrix}
\hat{C}_{33}&\hat{C}_{34}&\hat{C}_{35}\\
\hat{C}_{34}&\hat{C}_{44}&\hat{C}_{45}\\
\hat{C}_{35}&\hat{C}_{45}&\hat{C}_{55}\\
\end{Bmatrix}.
\end{equation}
Combining Eq. (\ref{eq:strain1}) and (\ref{eq:c345}) gives the definition of strain concentration tensor of material 1 $\textbf{s}^1$,
\begin{equation}\label{eq:A1}
\boldsymbol{\varepsilon}^1 = \textbf{s}^1{\boldsymbol{\varepsilon}},\quad\text{with}\quad {s}^1_{11}={s}^1_{22}={s}^1_{66}=1, \quad \textbf{s}^1_{([3,4,5],:)}=\textbf{s}_{3\times 6}.
\end{equation}
Rewrite the averaging equation of the stress based on the definition of $\textbf{s}^1$:
\begin{equation}\label{eq:stress1}
{\boldsymbol{\sigma}}=f_1\boldsymbol{\sigma}^1 + f_2\boldsymbol{\sigma}^2=f_1\bar{\textbf{C}}^1\boldsymbol{\varepsilon}^1+f_2\bar{\textbf{C}}^2\boldsymbol{\varepsilon}^2=\bar{\textbf{C}}^2{\boldsymbol{\varepsilon}}-f_1\Delta\textbf{C}\boldsymbol{\varepsilon}^1=\left(\bar{\textbf{C}}^2-f_1\Delta\textbf{C}\textbf{s}^1\right){\boldsymbol{\varepsilon}}.
\end{equation}
Finally, the homogenized stiffness matrix before the rotation operation is obtained:
\begin{equation}
\textbf{C}=\boldsymbol{\mathfrak{h}}_C\left(\bar{\textbf{C}}^1,\bar{\textbf{C}}^2,w^1,w^2\right)=\bar{\textbf{C}}^2-f_1\Delta\textbf{C}\textbf{s}^1.
\end{equation}

\textbf{Rotation function $\boldsymbol{\mathfrak{r}}_C$.} 
Here we use the Tait-Bryan angles $(\alpha, \beta, \gamma)$ to parameterize the 3D rotation. Any rotation can be achieved by composing three element rotations, in other words, a rotation matrix $\textbf{R}$ can be decomposed as a product of three element rotation matrices,
\begin{equation}\label{eq:smallR}
\textbf{R}(\alpha, \beta, \gamma) = \textbf{X}(\alpha)\textbf{Y}(\beta)\textbf{Z}(\gamma).
\end{equation}
The expressions of elementary rotation matrices are given in \ref{ap:ap1}.
The rotation function $\boldsymbol{\mathfrak{r}}_C$ takes the following form:
\begin{equation}
\bar{\textbf{C}}= \boldsymbol{\mathfrak{r}}_C\left(\textbf{C},\alpha,\beta,\gamma\right)=\textbf{R}^{-1}(\alpha, \beta, \gamma){\textbf{C}}\textbf{R}(\alpha, \beta, \gamma).
\end{equation}
In practice, the 3D rotation is decomposed into 3 rotation steps, characterized by $\alpha$, $\beta$ and $\gamma$,
\begin{equation}\label{eq:r3}
{\textbf{C}}^\alpha = \textbf{X}(-\alpha){\textbf{C}}\textbf{X}(\alpha),\quad
{\textbf{C}}^{\alpha\beta} = \textbf{Y}(-\beta){\textbf{C}}^\alpha\textbf{Y}(\beta),\quad
\bar{\textbf{C}}=\textbf{Z}(-\gamma){\textbf{C}}^{\alpha\beta}\textbf{Z}(\gamma).
\end{equation}
More details on computing the gradients for training are provided in \ref{ap:ap2}.

\subsection{Extension to material and geometric nonlinearities}\label{sec:nonlinearDeri}
In this section, we extend the solutions of two-layer building block from linear elasticity to problems with both material and geometric nonlinearities, as to appear in the online extrapolation stage. Note that the same procedure can be degenerated to problems with only material nonlinearity (e.g. small-strain plasticity) or geometric nonlinearity (e.g. hyperelasticity), so that we will only focus on the general situation.

For finite-strain problem, the deformation gradient $\utilde{\textbf{F}}$ and first Piola-Kirchhoff stress $\utilde{\textbf{P}}$ are chosen as the strain and stress measures, respectively. Here, we defined the tangent elasticity tensor $\utilde{\textbf{A}}$ in the rate form as
\begin{equation}
\utilde{\dot{\textbf{P}}}=\utilde{\textbf{A}}:\utilde{\dot{\textbf{F}}} \quad\text{ or}\quad \utilde{\dot{P}}_{ij} = \utilde{A}_{ijkl}\utilde{\dot{F}}_{kl}.
\end{equation}
The first elasticity tensor has major symmetry, $\utilde{A}_{ijkl}=\utilde{A}_{klij}$, but does not have minor symmetries. As a result, Voigt notation can not be applied in this case, instead, we vectorize the deformation gradient and first Piola-Kirchhoff stress in 3D based on all the 9 components in each tensor:
\begin{equation}\label{eq:finiteFP}
\textbf{F} = \{\utilde{F}_{11},\utilde{F}_{22},\utilde{F}_{33},\utilde{F}_{23},\utilde{F}_{32},\utilde{F}_{13},\utilde{F}_{31},\utilde{F}_{12},\utilde{F}_{21}\}^T=\{F_{1},F_{2},F_{3},...,F_{9}\}^T,
\end{equation}
\begin{equation*}
\textbf{P} = \{\utilde{P}_{11},\utilde{P}_{22},\utilde{P}_{33},\utilde{P}_{23},\utilde{P}_{32},\utilde{P}_{13},\utilde{P}_{31},\utilde{P}_{12},\utilde{P}_{21}\}^T=\{P_{1},P_{2},P_{3},...,P_{9}\}^T.
\end{equation*}
Under the above notations, the stiffness matrix $\textbf{A}$ is defined as
\begin{equation}\label{eq:finiteA}
{\dot{\textbf{P}}}={\textbf{A}}{\dot{\textbf{F}}}.
\end{equation}
Due to major symmetry, $\textbf{A}$ is symmetric, and it has 45 independent components. For a general loading step, the stress-strain relation takes the following form,
\begin{equation}
\Delta\textbf{P} = \textbf{A}\Delta \textbf{F}+\delta\textbf{P},
\end{equation}
where $\Delta\textbf{P}$ and $\Delta \textbf{F}$ are the increments of stress and deformation gradient. Due to the existence of nonlinearity, the residual first Piola-Kirchoff stress $\delta\textbf{P}$ needs to be included in the relation. 

In the two-layer building block, the constitutive laws of materials 1 and 2 are
\begin{equation}
\Delta{\textbf{P}}^1 = \bar{\textbf{A}}^1\Delta \textbf{F}^1+\delta\bar{\textbf{P}}^1,\quad \Delta{\textbf{P}}^2 = \bar{\textbf{A}}^2\Delta \textbf{F}^2+\delta\bar{\textbf{P}}^2.
\end{equation}
Again, we place the bar notations above the stiffness matrix and residual stress of each material (e.g.$\bar{\textbf{A}}^1$, $\delta\bar{\textbf{P}}^1$) to keep the fact that these quantities should come from the averaging of previous building blocks in the network. As illustrated in Figure \ref{fig:onlineBlock}, two steps are involved in the averaging of the building block with nonlinearities: 1) Homogenization given by the functions  $\boldsymbol{\mathcal{H}}_A$ and $\boldsymbol{\mathcal{H}}_P$
\begin{equation}
\textbf{A} = \boldsymbol{\mathcal{H}}_A\left(\bar{\textbf{A}}^1,\bar{\textbf{A}}^2,w^1,w^2\right), \quad
\delta \textbf{P} = \boldsymbol{\mathcal{H}}_P\left(\delta\bar{\textbf{P}}^1,\delta\bar{\textbf{P}}^2,w^1,w^2\right);
\end{equation}
and 2) rotation of the homogenized two-layer structure given by the functions $\boldsymbol{\mathfrak{R}}_A$ and $\boldsymbol{\mathfrak{R}}_P$,
\begin{equation}
\bar{\textbf{A}} = \boldsymbol{\mathfrak{R}}_A\left(\textbf{A},\alpha,\beta,\gamma\right), \quad
\delta\bar{\textbf{P}} = \boldsymbol{\mathfrak{R}}_P\left(\delta\textbf{P},\alpha,\beta,\gamma\right).
\end{equation}
The intermediate constitutive law right after homogenization can be written as
\begin{equation}\label{eq:finite1}
\Delta{{\textbf{P}}} = {\textbf{A}}\Delta {\textbf{F}}+\delta{\textbf{P}},
\end{equation}
and the overall constitutive law after rotation is
\begin{equation}
\Delta\bar{{\textbf{P}}} = \bar{\textbf{A}}\Delta \bar{\textbf{F}}+\delta\bar{\textbf{P}}.
\end{equation}

\textbf{Homogenization functions $\boldsymbol{\mathcal{H}}_A$ and $\boldsymbol{\mathcal{H}}_P$.}
Similar to the linear elastic case discussed in Section \ref{sec:linearDeri}, analytical forms of the homogenization functions $\boldsymbol{\mathcal{H}}_A$ and $\boldsymbol{\mathcal{H}}_P$ can be derived based on the equilibrium conditions
\begin{equation}\label{eq:con3}
P_{3}^1 = P_{3}^2, \quad P_{4}^1 = P_{4}^2,  \quad P_{6}^1 = P_{6}^2,
\end{equation}
and the kinematic constraints
\begin{equation}\label{eq:con4}
F_{1}^1 = F_{1}^2, \quad F_{2}^1 = F_{2}^2,\quad F_{5}^1 = F_{5}^2, \quad F_{7}^1 = F_{7}^2, \quad F_{8}^1 = F_{8}^2, \quad F_{9}^1 = F_{9}^2.
\end{equation}
In the derivation of the stiffness tensor, we first consider the situation with no residual stress and solve for the unknown components of the increment of deformation gradient $\Delta \textbf{F}^1$ of material 1 in the 3, 4, 6 directions. Then the strain concentration tensor $\textbf{S}^{1}$ of material 1, defined by $\Delta\textbf{F}^1=\textbf{S}^{1}\Delta \textbf{F}$, is found to be
\begin{equation}
S^{1}_{([3,4,6],:)}=\left(\hat{\textbf{A}}_{346}\right)^{-1}
\begin{Bmatrix}
f_2\Delta A_{13}&f_2\Delta A_{23}&\bar{A}_{33}^2&\bar{A}_{34}^2&f_2\Delta A_{35}&\bar{A}_{36}^2&f_2\Delta A_{37}&f_2\Delta A_{38}&f_2\Delta A_{39}\\
f_2\Delta A_{14}&f_2\Delta A_{24}&\bar{A}_{34}^2&\bar{A}_{44}^2&f_2\Delta A_{45}&\bar{A}_{46}^2&f_2\Delta A_{47}&f_2\Delta A_{48}&f_2\Delta A_{49}\\
f_2\Delta A_{16}&f_2\Delta A_{26}&\bar{A}_{35}^2&\bar{A}_{46}^2&f_2\Delta A_{56}&\bar{A}_{66}^2&f_2\Delta A_{67}&f_2\Delta A_{68}&f_2\Delta A_{69}\\
\end{Bmatrix},
\end{equation}
\begin{equation*}
S^{1}_{11}=S^{1}_{22}=S^{1}_{55}=S^{1}_{77}=S^{1}_{88}=S^{1}_{99}=1
\end{equation*}
with
\begin{equation}
\hat{\textbf{A}}_{346}=\begin{Bmatrix}
\hat{A}_{33}&\hat{A}_{34}&\hat{A}_{36}\\
\hat{A}_{34}&\hat{A}_{44}&\hat{A}_{46}\\
\hat{A}_{36}&\hat{A}_{46}&\hat{A}_{66}\\
\end{Bmatrix},\quad 
\hat{\textbf{A}}=f_2\bar{\textbf{A}}^1+f_1\bar{\textbf{A}}^2\quad \text{and}\quad \Delta\textbf{A}=\bar{\textbf{A}}^2-\bar{\textbf{A}}^1,
\end{equation}
where definitions of $f_1$ and $f_2$ are given in Eq. (\ref{eq:vf}). The homogenization function $\boldsymbol{\mathcal{H}}_A$ of the stiffness matrix ${\textbf{A}}$ is derived to be
\begin{equation}
\textbf{A} = \boldsymbol{\mathcal{H}}_A\left(\bar{\textbf{A}}^1,\bar{\textbf{A}}^2,w^1,w^2\right)=\bar{\textbf{A}}^2-f_1\Delta\textbf{A}\textbf{S}^{1}.
\end{equation}

To derive the homogenization function $\boldsymbol{\mathcal{H}}_P$ for the residual stress $\delta \textbf{P}$, we first set the average strain after the homogenization $\Delta \textbf{F}$ in Eq. (\ref{eq:finite1}) to zero:
\begin{equation}
\Delta\textbf{F}=\textbf{0},\quad \text{so that } \Delta\textbf{P} = \delta\textbf{P}.
\end{equation}
With the kinematic constraints defined in Eq. (\ref{eq:con4}) and the averaging conditions, we have
\begin{equation}
F_{1}^1 = F_{1}^2=F_{2}^1 = F_{2}^2=F_{5}^1 = F_{5}^2=F_{7}^1 = F_{7}^2=F_{8}^1 = F_{8}^2=F_{9}^1 = F_{9}^2=0,
\end{equation}
\begin{equation*}
F_3^2 = -\dfrac{f_1}{f_2}F_3^1,\quad F_4^2 = -\dfrac{f_1}{f_2}F_4^1,\quad F_6^2 = -\dfrac{f_1}{f_2}F_6^1.
\end{equation*}
Substituting the above equations to the equilibrium condition defined in Eq. (\ref{eq:con3}) yields
\begin{equation}
\hat{\textbf{A}}_{346}
\begin{Bmatrix}
F_{3}^1\\
F_{4}^1\\
F_{6}^1\\
\end{Bmatrix}=f_2
\begin{Bmatrix}
\delta\bar{P}_{3}^2-\delta\bar{P}_{3}^1\\
\delta\bar{P}_{4}^2-\delta\bar{P}_{4}^1\\
\delta\bar{P}_{6}^2-\delta\bar{P}_{6}^1\\
\end{Bmatrix},\quad
\begin{Bmatrix}
F_{3}^1\\
F_{4}^1\\
F_{6}^1\\
\end{Bmatrix}=f_2\left(\hat{\textbf{A}}_{346}\right)^{-1}
\begin{Bmatrix}
\delta\bar{P}_{3}^2-\delta\bar{P}_{3}^1\\
\delta\bar{P}_{4}^2-\delta\bar{P}_{4}^1\\
\delta\bar{P}_{6}^2-\delta\bar{P}_{6}^1\\
\end{Bmatrix}.
\end{equation}
After solving for $\Delta{\textbf{F}}^1$ in material 1, the homogenized residual stress is given by the average stress increment $\Delta \textbf{P}$ in the structure,
\begin{equation}
\delta\textbf{P} = \Delta\textbf{P}=f_1\Delta\textbf{P}^1+f_2\Delta\textbf{P}^2.
\end{equation}
Therefore, the homogenization function $\boldsymbol{\mathcal{H}}_P$ is derived to be
\begin{equation}
\delta\textbf{P}=\boldsymbol{\mathcal{H}}_P\left(\delta\bar{\textbf{P}}^1,\delta\bar{\textbf{P}}^2,w^1,w^2\right)=-f_1f_2\Delta\textbf{A}_{(:,[3,4,6])}\left(\hat{\textbf{A}}_{346}\right)^{-1}
\begin{Bmatrix}
\delta\bar{P}_{3}^2-\delta\bar{P}_{3}^1\\
\delta\bar{P}_{4}^2-\delta\bar{P}_{4}^1\\
\delta\bar{P}_{6}^2-\delta\bar{P}_{6}^1\\
\end{Bmatrix}+f_1\delta\bar{\textbf{P}}^1+f_2\delta\bar{\textbf{P}}^2.
\end{equation}

\textbf{Rotation functions $\boldsymbol{\mathfrak{R}}_A$ and $\boldsymbol{\mathfrak{R}}_P$.}
The rotation matrices are also required to be modified under finite-strain notation in Eq. (\ref{eq:finiteFP}). Based on the Tait-Bryan angles $\alpha,\beta,\gamma$, the new rotation matrix $\textbf{R}^f$ can be decomposed as a product of three element rotation matrices,
\begin{equation}\label{eq:finiteR}
\textbf{R}^f(\alpha,\beta,\gamma) = \textbf{X}^f(\alpha)\textbf{Y}^f(\beta)\textbf{Z}^f(\gamma).
\end{equation}
The expressions of finite-strain element rotation matrices $\textbf{X}^f$, $\textbf{Y}^f$ and $\textbf{Z}^f$ are given in \ref{ap:ap1}.
With the definition of the rotation matrix $\textbf{R}^f$, the rotation functions can be expressed as
\begin{equation}
\bar{\textbf{A}} = \boldsymbol{\mathfrak{R}}_A\left(\textbf{A},\alpha,\beta,\gamma\right)=\left[\textbf{R}^{f}\left(\alpha,\beta,\gamma\right)\right]^{-1}\textbf{A}\textbf{R}^{f}\left(\alpha,\beta,\gamma\right),
\end{equation}
and
\begin{equation}
\delta\bar{\textbf{P}} = \boldsymbol{\mathfrak{R}}_P\left(\delta\textbf{P},\alpha,\beta,\gamma\right)=\left[\textbf{R}^{f}\left(\alpha,\beta,\gamma\right)\right]^{-1}\delta\textbf{P}.
\end{equation}

\section{Applications }\label{sec:application}
In this paper, we explore the 3D architectures of DMN for three typical multiscale material systems which are popular in both academic research and industrial applications: 1) Particle-reinforced rubber composite with Mullins effect in Section \ref{sec:particle}; 2) polycrystalline materials in Section \ref{sec:poly}; 3) carbon-fiber reinforced polymer (CFRP) composites in Section \ref{sec:cfrp}. In particular, three scales are considered in the CFRP example, where the corresponding two RVEs are the microscale unidirectional fiber (UD) composite and the mesoscale woven composite. We will first show the DMN training results based upon orthotropic linear elasticity. Then we will extrapolate the trained networks to nonlinear materials, including Mooney-Rivlin hyperelasticity with Mullins effect, isotropic von Mises plasticity and rate-dependent crystal plasticity. In the SGD algorithm, the mini-batch size is chosen to be 20, and there are 20 learning steps in each epoch (for 400 training samples).  For all the cases, the coefficient $\lambda$ is set to 0.001, so that the regularization term will not affect the training at early stage with large MSE, but it will eventually limit the magnitudes of activations when the overall cost $J$ becomes smaller. Here we use $N_a$ to denote the number of active nodes in the bottom layer, so that $N_a\leq2^{N-1}$. Relative errors are computed at the end of training to evaluate the accuracy of DMN. For a sample $s$, its error $e_s$ is defined as
\begin{equation}
e_s=\dfrac{||\bar{\textbf{C}}^{dns}_s-\bar{\textbf{C}}^{rve}_s||}{||\bar{\textbf{C}}^{dns}_s||}.
\end{equation}
The average error of a whole dataset with S samples is
\begin{equation}
\bar{e}=\dfrac{1}{S}\sum_s e_s.
\end{equation}
The average errors of training and test datasets are denoted by $\bar{e}^{tr}$ and $\bar{e}^{te}$, respectively. By default, finite-strain formulation with geometric nonlinearity is adopted in the online stages of all the RVEs.

\subsection{Hyperelastic polymer composite with Mullins effect}\label{sec:particle}
\begin{figure}[!t]
	\centering
	\graphicspath{{Figures/}}
	\subfigure[Geometry and mesh.]{\includegraphics[clip=true,trim = 8.0cm 2.5cm 8.0cm 3.6cm,width=0.44\textwidth]{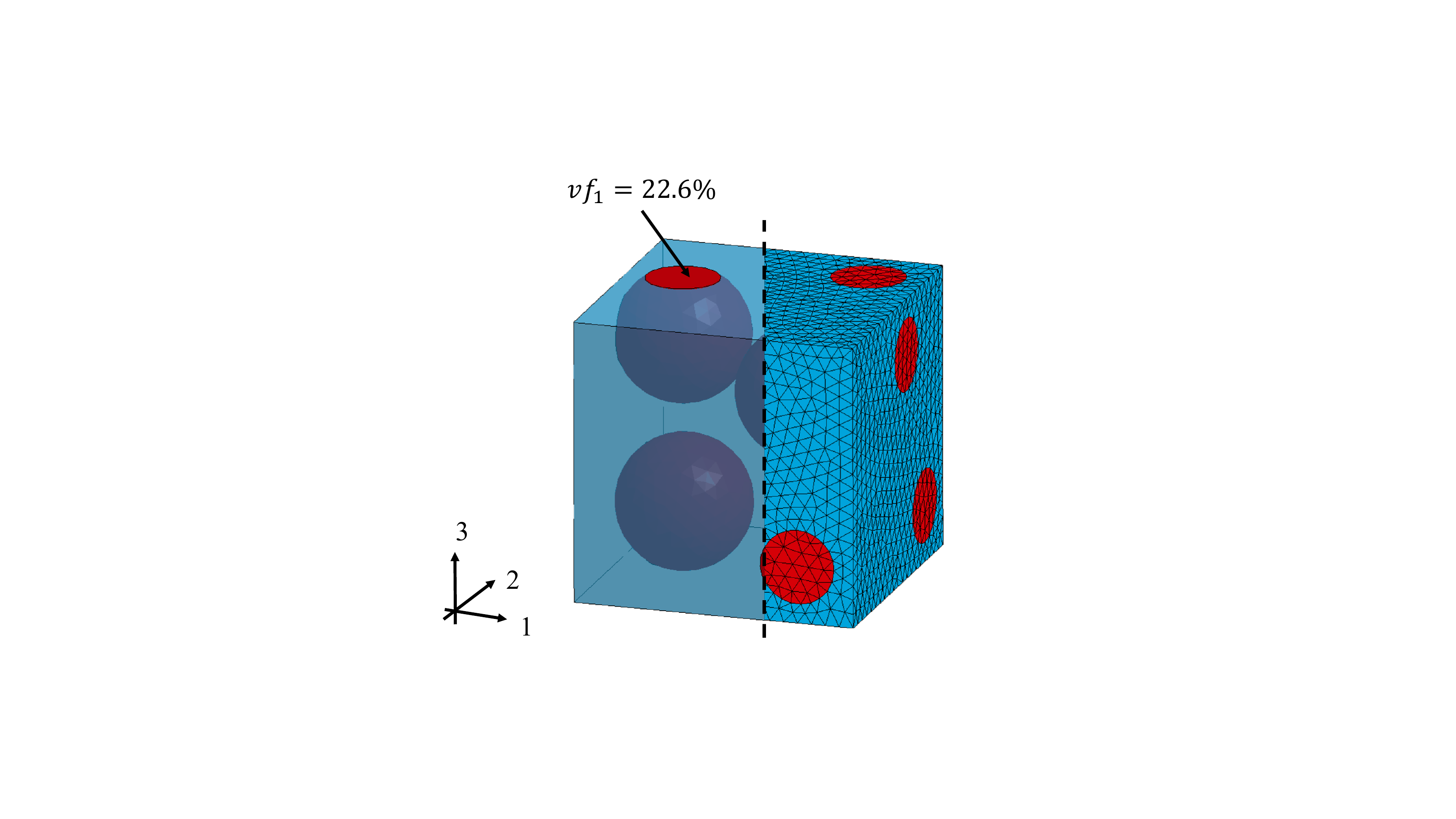}}
	\subfigure[Uniaxial responses of individual phases (online).]{\includegraphics[clip=true,trim = 0.0cm 0.0cm 1.0cm 0.5cm,width=0.44\textwidth]{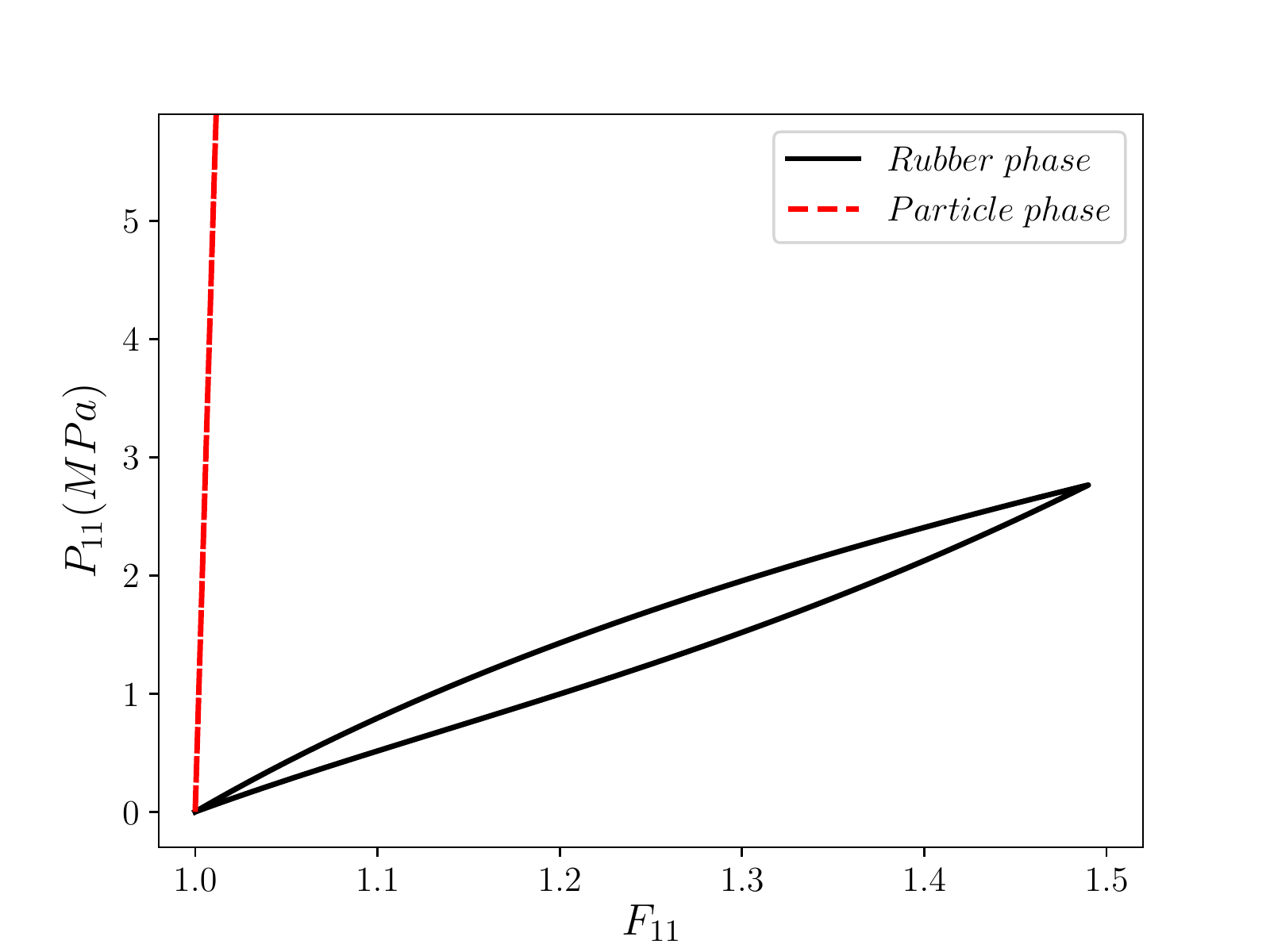}}
	\caption{Particle reinforced RVE: (a) The volume fraction of the particle phase is 22.6\% and the FE model has 84693 nodes and 59628 10-node tetrahedron elements; (b) In the online extrapolation stage, the matrix phase is considered as a Mooney-Rivlin hyperelastic rubber with Mullins effect and the particle is a Neo-Hookean material which is 100 times harder than the matrix.}
	\label{fig:np}
\end{figure}

The first RVE example is a two-phase particle-reinforced polymer composite. The geometry of the RVE is presented in Figure \ref{fig:np}, together with the uniaxial responses of material models that will be studied in the online stage. Phase index of the particle material is 1, and there are four identical spherical particles embedded in the matrix. Volume fraction of the particle phase is 22.6\%. The DNS RVE is discretized by 84693 nodes and 59628 10-node tetrahedron finite elements.

\subsubsection{Offline evaluation}
\begin{figure} [!t]
	\centering
	\graphicspath{{Figures/}}
	\subfigure[Training histories.]{\includegraphics[clip=true,trim = 0.0cm 0.0cm 1.0cm 0.5cm,width=0.44\textwidth]{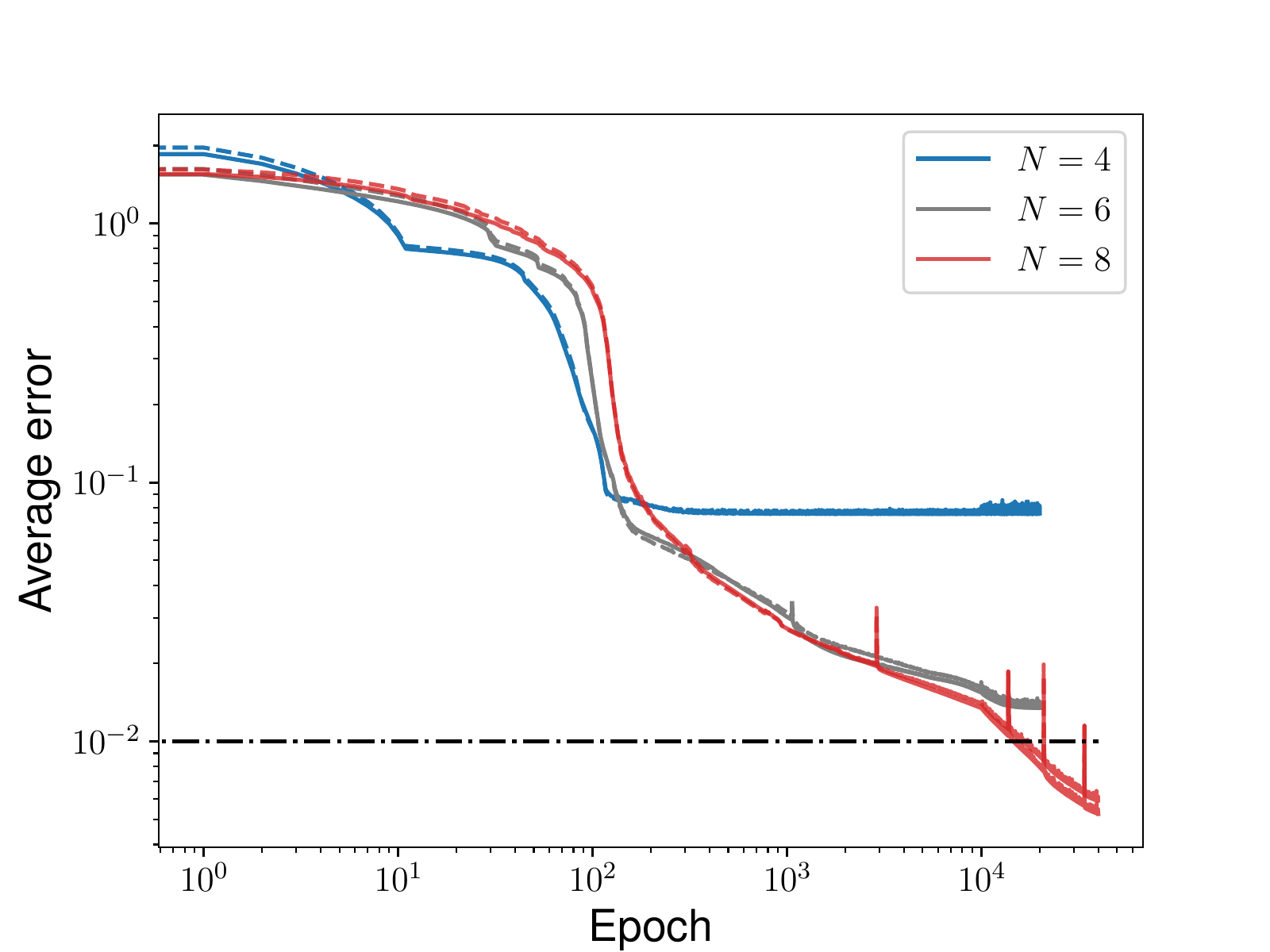}}
    \subfigure[Distributions of test error $e^{te}$.]{\includegraphics[clip=true,trim = 0.0cm 0.0cm 1.0cm 0.5cm,width=0.44\textwidth]{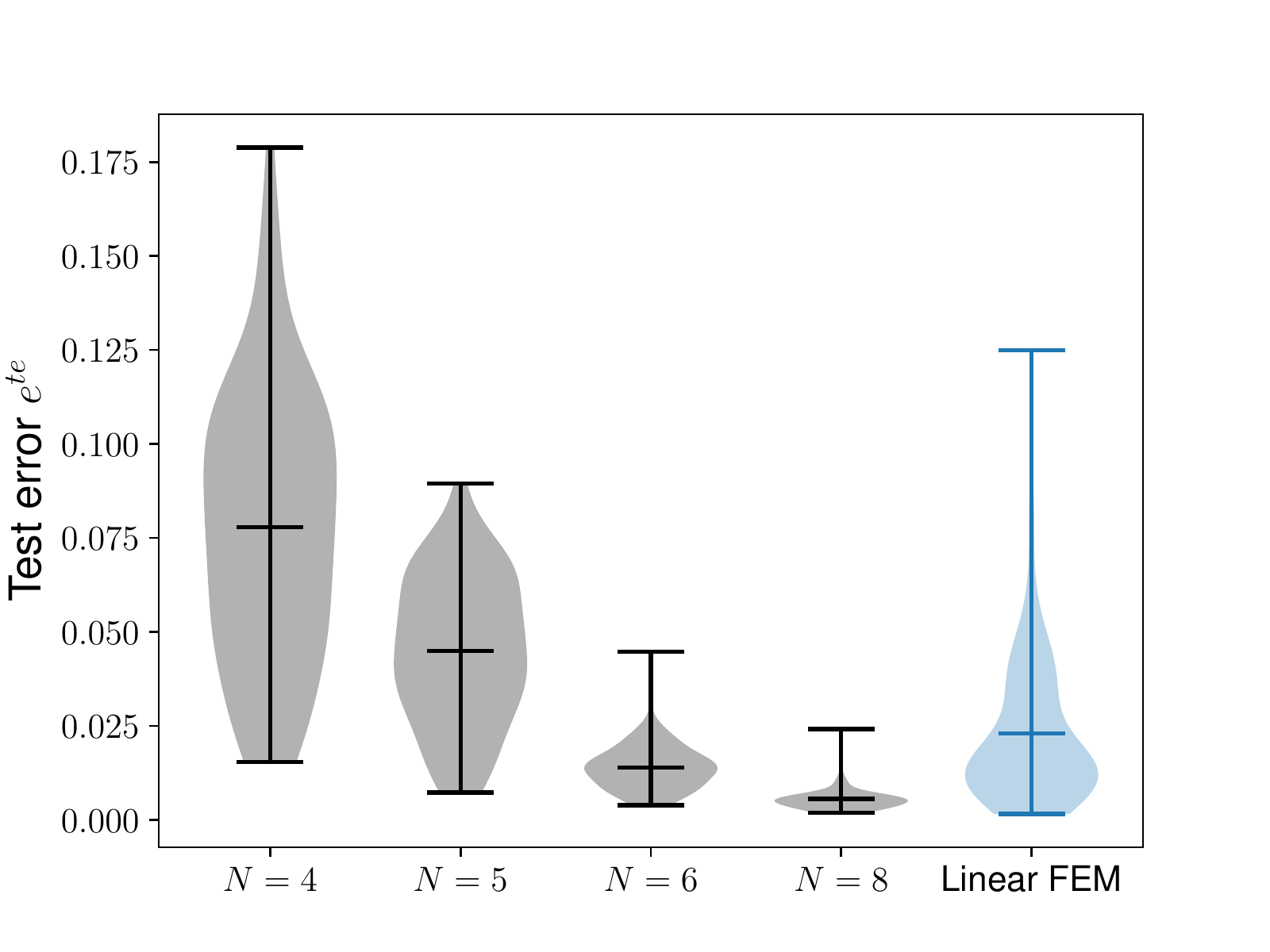}}
	\caption{Error histories and distribution of DMN for the particle reinforced composite. In (a), the histories of the average training and test errors are denoted by solid and dashed lines, respectively. In (b), the distributions of test error are shown for trained DMNs with various depths (black), and the test result of the linear FE model (blue) is also provided for comparison.}
	\label{fig:npHist}
\end{figure}

Histories of average training and test errors are plotted in Figure \ref{fig:npHist} for network depth $N=4, 5, 6, 8$, and the errors at the end of training are also listed in Table \ref{table:trnp}. Networks with $N= 4,$ 5 and $6$ are trained for 20000 epochs. While for $N=8$, we observed that the errors still decrease rapidly after 20000 epochs of training, so we choose to restart the learning process by doubling the learning rate,  and train the network for another 20000 epochs. As we can see from the figure, a network with $N=4$ is not sufficient to capture the RVE responses, and its average training error stopped decreasing around 7.6\%. Good accuracy can be achieved by the deeper networks with $N=6$ and $N=8$. Due to the use of mechanistic building block, the difference between the training and test errors is almost negligible, indicating no over-fitting issue. For $N=8$, the average training error $\bar{e}^{tr}$ was reduced to 0.53\% after 40000 epochs of training, and the maximum error among all the samples in the test dataset is only 2.41\%. Note that the contrast of stiffness between the two material phases can reach up to 1000 in the offline sampling space, indicating that the trained DMN discovers a good reduced representation of the original 3D RVE microstructure.
\begin{table}[!t]
	\captionabove{Training results of the particle-reinforced composite. Average training error $\bar{e}^{tr}$, average test error $\bar{e}^{tr}$, maximum test error and predicted volume fraction $vf_1$  are provided for each DMN. Test errors of the linear FEM model are also shown.} 
	\centering 
	\label{table:trnp} 
	{\tabulinesep=1.0mm
		\begin{tabu}{c c c c c c} 
			\hline 
			&Epochs & Training $\bar{e}^{tr}$ & Test $\bar{e}^{te}$ & Maximum $e_s^{te}$ &  $vf_1$\\
			\hline
			$N=4$&20000&7.61\%&7.79\%&17.9\%&0.211 (-6.63\%)\\ 
			$N=5$ &20000&4.47\%&4.49\%&8.94\%&0.220 (-2.65\%)\\
			$N=6$ &20000&1.34\%&1.39\%&4.46\%&0.220 (-2.65\%)\\
			$N=8$ &40000&0.53\%&0.59\%&2.41\%&0.224 (-0.88\%) \\
			Linear FEM& \textbackslash&\textbackslash&2.30\%&12.5\%&\textbackslash\\
			\hline
	\end{tabu}}
\end{table}
\begin{figure} [!t]
	\centering
	\graphicspath{{Figures/}}
	\subfigure[$N=4, N_a =4, vf_1=0.211$]{\includegraphics[clip=true,trim = 0.0cm 0.0cm 1.0cm 0.5cm,width=0.28\textwidth]{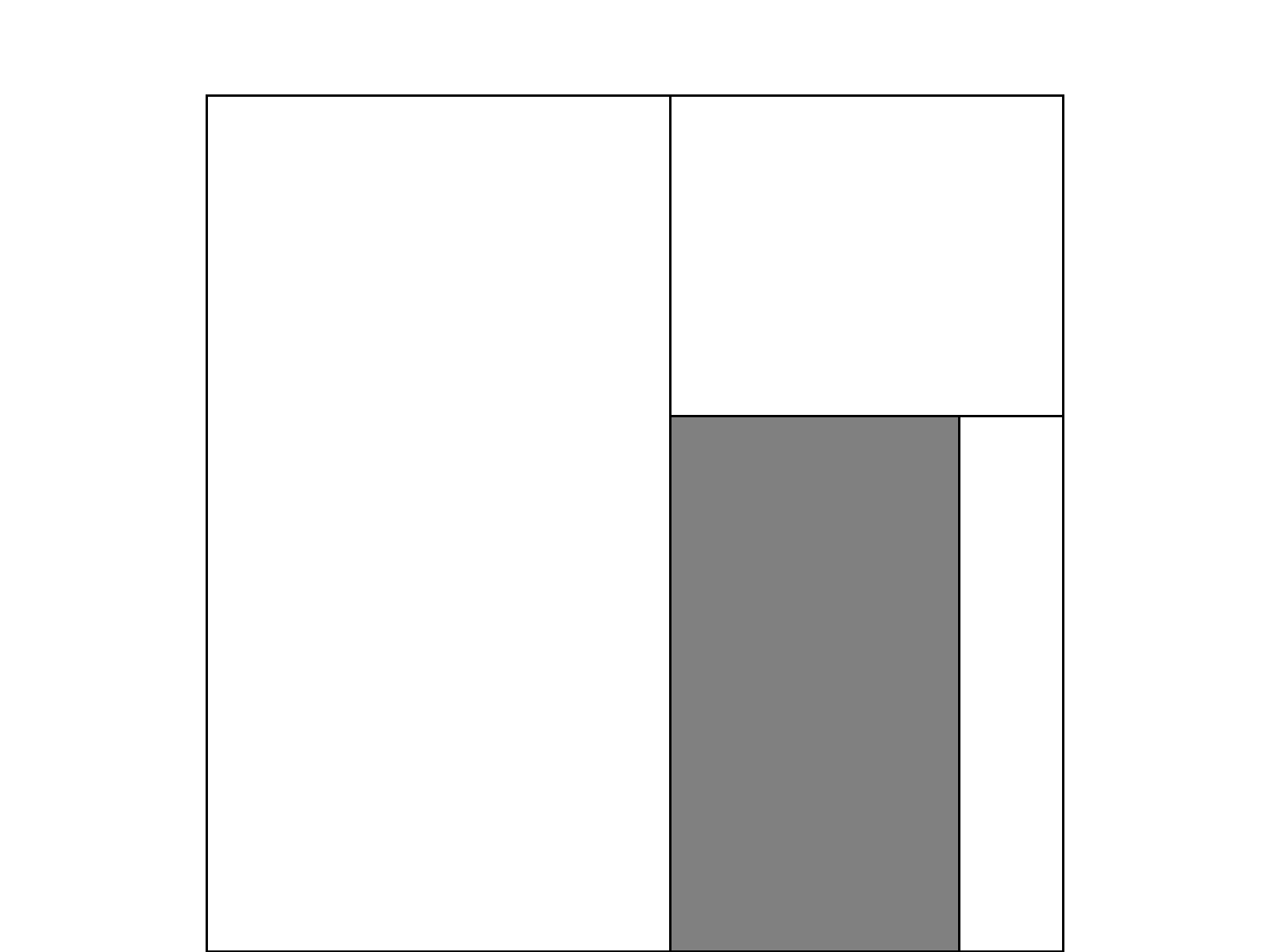}}
	\subfigure[$N=6, N_a =13, vf_1=0.220$]{\includegraphics[clip=true,trim = 0.0cm 0.0cm 1.0cm 0.5cm,width=0.28\textwidth]{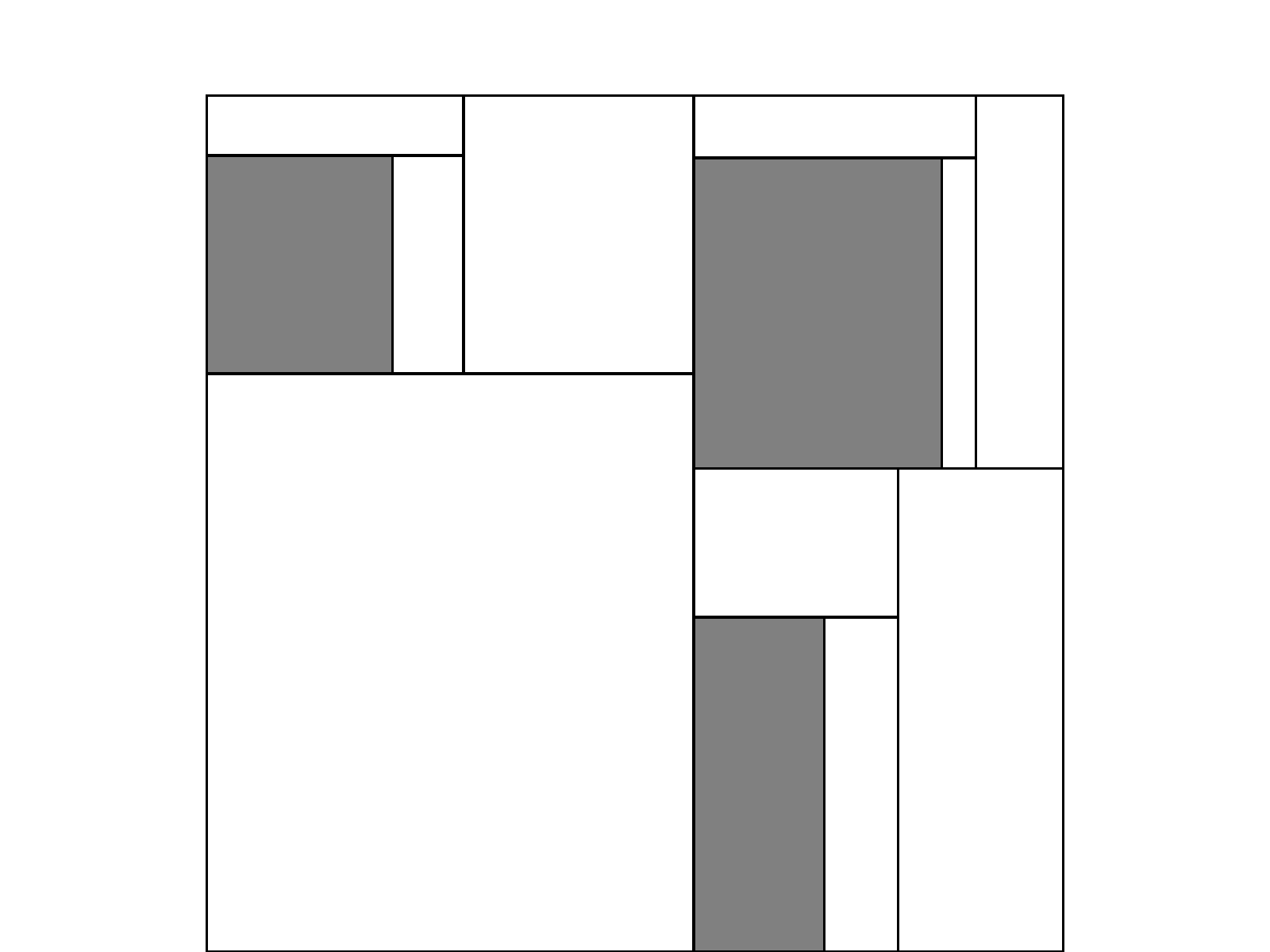}}
	\subfigure[$N=8, N_a =28, vf_1=0.224$]{\includegraphics[clip=true,trim = 0.0cm 0.0cm 1.0cm 0.5cm,width=0.28\textwidth]{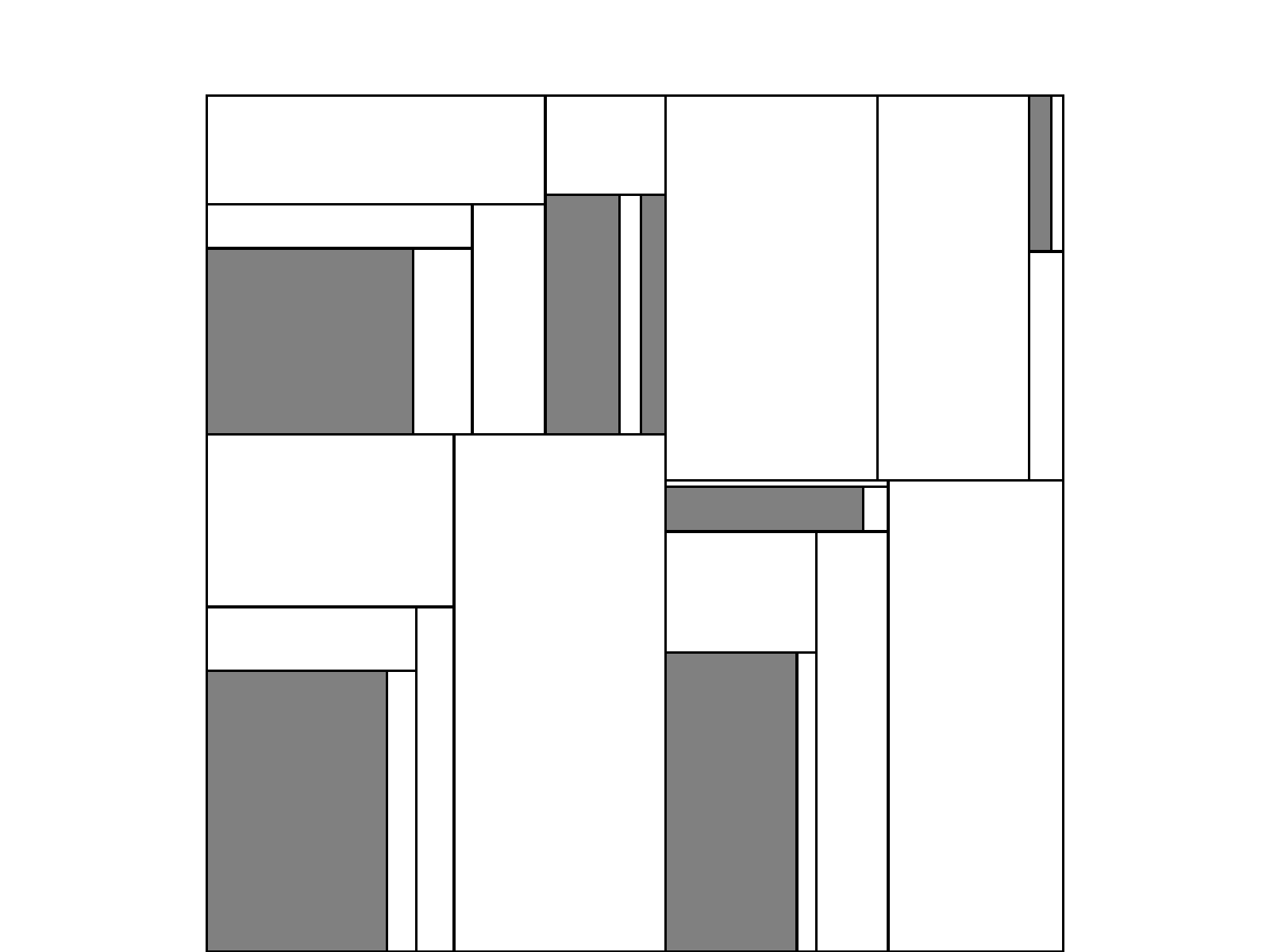}}
	\caption{Treemaps of DMN for the particle-reinforced composite. The network depths $N$ are (a) 4, (b) 6 and (c) 8. The number of active nodes in the bottom layer $N_a$ and the predicted volume fraction of phase 1 $vf_1$ are also shown under each plot.}
	\label{fig:npTree}
\end{figure}

To evaluate the accuracy and efficiency of DMN objectively, we also introduce another FE model with linear elements as a reference and measure its performance on the test dataset with 100 samples. As shown in Table \ref{table:trnp} and Figure \ref{fig:npHist} (b), the average test error $\bar{e}^{te}$ of the linear FE model is 2.30\%, and the maximum test error is 12.5\%. If we use $\bar{e}^{te}$ as the measure, the linear FE model performs comparably to the DMN model with $N=6$ and $N_a=13$. Moreover, it has only 11236 nodes, hence, is more efficient than the original DNS model. The computational time of the linear FE model for the particle-reinforced RVE will be given in Section \ref{sec:time}.

Treemaps of DMN at the end of training are provided in Figure \ref{fig:npTree} to show the fraction of each active bottom-layer node within the whole network. It should be noted that no rotation information is reflected in these treemaps. At the beginning of training, all the nodes in the bottom layer are active, so that we have $N_a=8,32,128$ for $N=4, 6, 8$, respectively. At the end of training, these numbers are reduced to $N_a = 4, 13, 28$ due to model compression. Volume fractions predicted by DMN based on the weights can be found in Table \ref{table:trnp} and Figure \ref{fig:npTree}. For $N=8$, the relative difference between $vf_1$ predicted by DMN and the reference one of the full-field DNS model is only 0.88\%. Although the training dataset only contains information of the overall RVE stiffness matrices, the trained DMN is capable of extracting essential geometric information of the RVE accurately from these mechanical data. The treemaps also provide some physical insights on the RVE microstructure: the particle block (dark) is always surrounded by multiple matrix blocks (light), indicating that the matrix phase dominates the overall RVE properties. 

\subsubsection{Online extrapolation}\label{sec:npOnline}
We start from a Mooney-Rivlin hyperelastic rubber model. To model the rubber as an unconstrained material, a hydrostatic work term scaled by the bulk modulus $K$ is included in the strain energy functional which is function of the relative volume $J$ \cite{ogden1997non}:
\begin{equation}\label{eq:generalhyper}
W(J_1,J_2,J)=W_d(J_1,J_2) + W_h(J) = C_{10}(J_1-3) + C_{01}(J_2-3)+K(J-1-\ln J).
\end{equation}
In order to prevent volumetric work from contributing to the hydrostatic work, the first and second invariants are modified,
\begin{equation}
J_1 = I_1I_3^{-1/3}, \quad J_2 = I_2I_3^{-2/3},
\end{equation}
where $I_1,I_2,I_3$ are the invariants of right Cauchy-Green tensor $C=F^TF$. The independent parameters in the hyperelastic material model are chosen as $C_{10}, C_{01}$ and Poisson's ratio $\nu$. The shear and bulk moduli can be computed by
\begin{equation}
\mu = 2(C_{10}+C_{01}),\quad K = \dfrac{4(C_{10}+C_{01})(1+\nu)}{3(1-2\nu)}.
\end{equation} 
\begin{table}[t!]
	\captionabove{Material parameters for the hyperelastic particle-reinforced rubber composite with Mullins effect.} 
	\centering 
	\label{table:hyperpara} 
	{\tabulinesep=1.0mm
		\begin{tabu}{c c c c c c c} 
			\hline\hline
			\multirow{2}{*}{Particle}&$C_{10}$ (MPa) & $C_{01}$ (MPa) & $\nu$ &&&\\
			&100 & 0 & 0.3 &&& \\ 
			\hline\hline
			\multirow{2}{*}{Matrix}&$C_{10}$ (MPa) & $C_{01}$ (MPa) & $\nu$ &$\eta$&a (J$\cdot \text{cm}^{-3}$)&b\\
			&1.0 & 0.5 & 0.495 & 0.8 & 1.0 & 1.0 \\ 
			\hline
	\end{tabu}}
\end{table}
\begin{figure} [t!]
	\centering
	\graphicspath{{Figures/}}
	\subfigure[Uniaxial tension.]{\includegraphics[clip=true,trim = 0.0cm 0.0cm 1.0cm 0.5cm,width=0.44\textwidth]{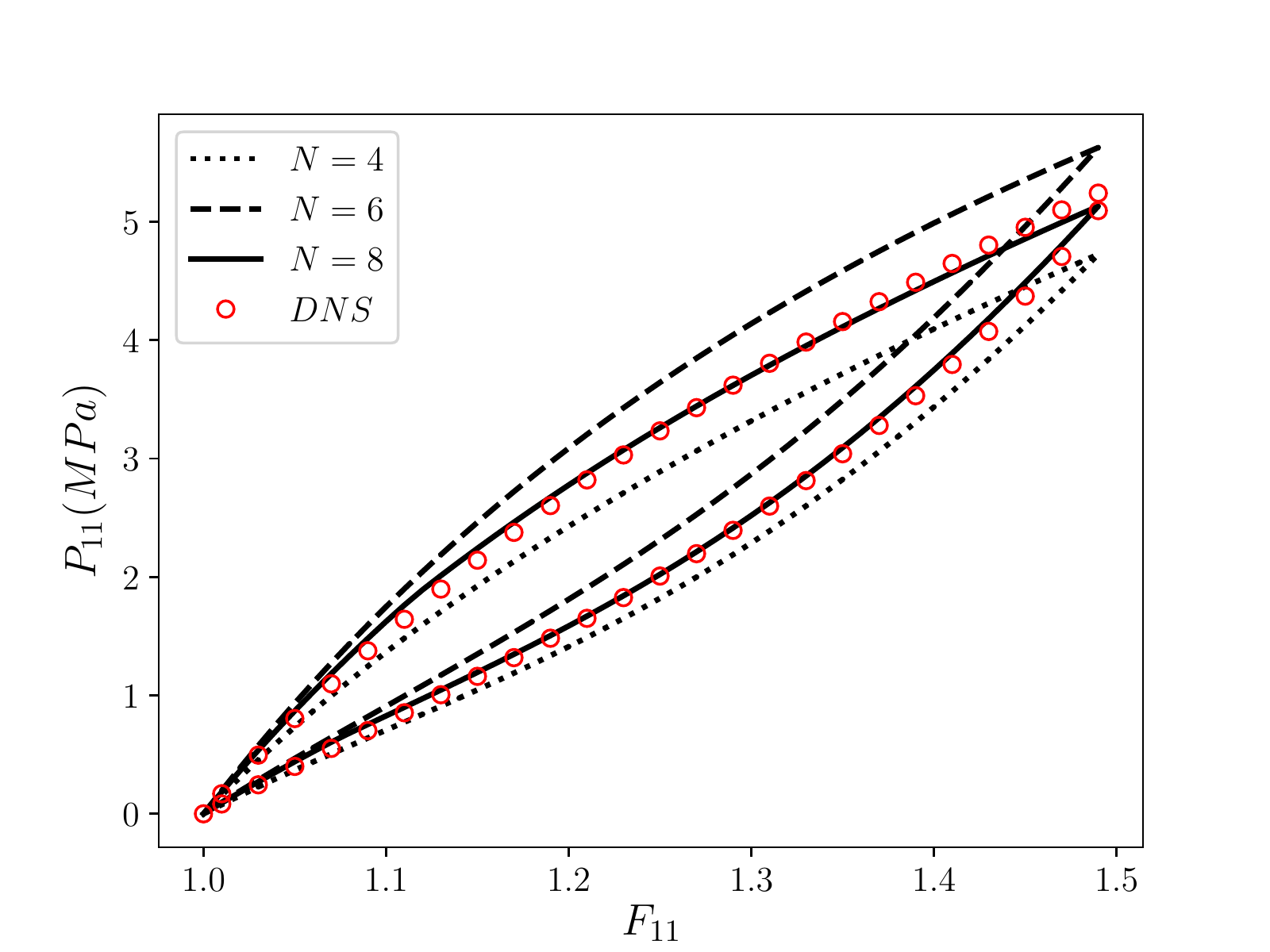}}
	\subfigure[Shear.]{\includegraphics[clip=true,trim = 0.0cm 0.0cm 1.0cm 0.5cm,width=0.44\textwidth]{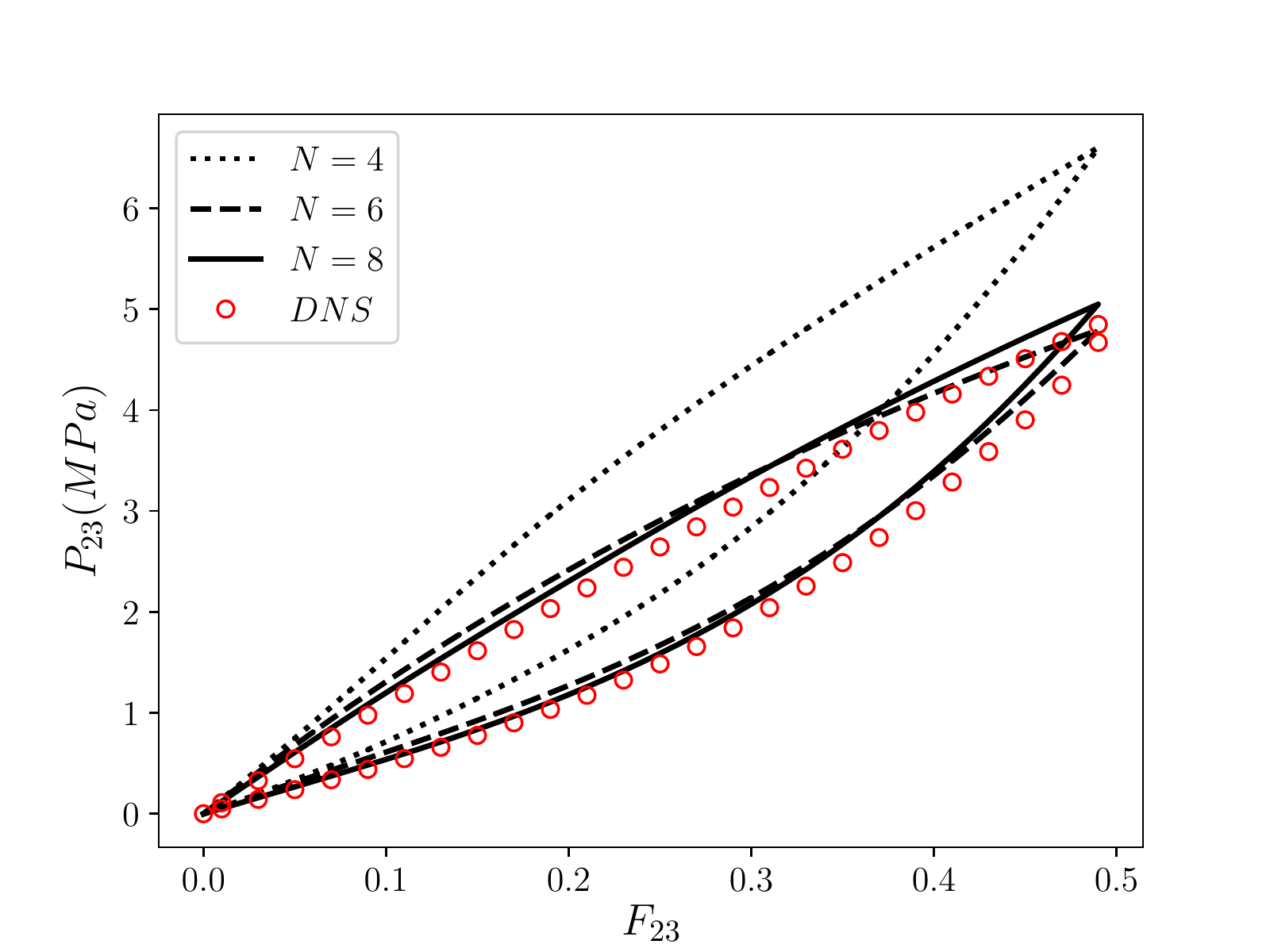}}
	\caption{Stress-strain curves from DMN and DNS for hyperelastic particle-reinforced rubber composite with Mullins effect under (a) uniaxial tension and (b) shear loading conditions. Both loading and unloading are considered. The network depth are $N=3$ (dotted), 5 (dashed) and 7 (solid).}
	\label{fig:mullin}
\end{figure}

Mullins effect, a kind of stress softening phenomena, is considered in the matrix rubber material. As the rubber material is loaded,  the breakage of links between rubber chains happens at different deformation levels, thereby leading to continuous damage in the macroscale. Note that the Mullins effect will cause dispersed softening in the RVE, rather than localized damage or even discrete cracks. For RVE problems with localizations, the current homogenization scheme based on the first-order Hill-Mandel condition is questionable and needs to be revised \cite{geers2010multi}, so does the corresponding DMN framework. To model the Mullins effect, the resulting stress is reduced by a damage factor according to
\begin{equation}
\textbf{S} = D(W_d,W_d^{max})\dfrac{\partial W_d}{\partial \textbf{E}} + \dfrac{\partial W_h}{\partial \textbf{E}},
\end{equation}
where $W_d$ is the current value of the deviatoric strain energy density, and  $W_d^{max}$ is the peak deviatoric strain energy density reached up to this point in time. The damage factor is defined as
\begin{equation}
D(W_d,W_d^{max}) = 1-\eta\;\text{erf}\left(\dfrac{W_d^{max}-W_d}{a+b W_d^{max}}\right),
\end{equation}
where $\text{erf}()$ denotes the Gauss error function, and $\eta, a, b$ are the empirical material constants related to the Mullins effect. In addition, the particle phase is modeled as a Neo-Hookean material which is 100 times harder than the matrix phase in terms of $C_{10}$. All the material parameters are provided in Table \ref{table:hyperpara}.

The stress-strain curves of the particle-reinforced rubber composite under uniaxial tension and shear loadings are shown in Figure \ref{fig:mullin}. For both cases, the RVEs are unloaded to check the Mullins effect due to matrix damage. It can been seen from the figure that material responses for all the cases can be well captured by the DMN with $N=8$, which has 28 active nodes in the bottom layer. As the DNS model contains much more DOFs (84693 nodes), the computation based on DMN is much faster than the DNS. Detailed investigation on the computational cost are provided in Section \ref{sec:time}. Since the active nodes in the bottom layer serve as the microscale material points, the distributions of local quantities (e.g. strain, stress and internal variables) can be recovered by DMN. More results on this subject can be found in \cite{liu2019deep}. It is also possible to relate the bottom-layer nodes directly to parts in the real RVE microstructure using clustering techniques, which will be investigated in our future work.

\subsection{Polycrystalline materials with rate-dependent crystal plasticity}\label{sec:poly}
\begin{figure}[!t]
	\centering
	\graphicspath{{Figures/}}
	\subfigure[Geometry and mesh.]{\includegraphics[clip=true,trim = -5.0cm -1.0cm -5.0cm 0.cm,width=0.44\textwidth]{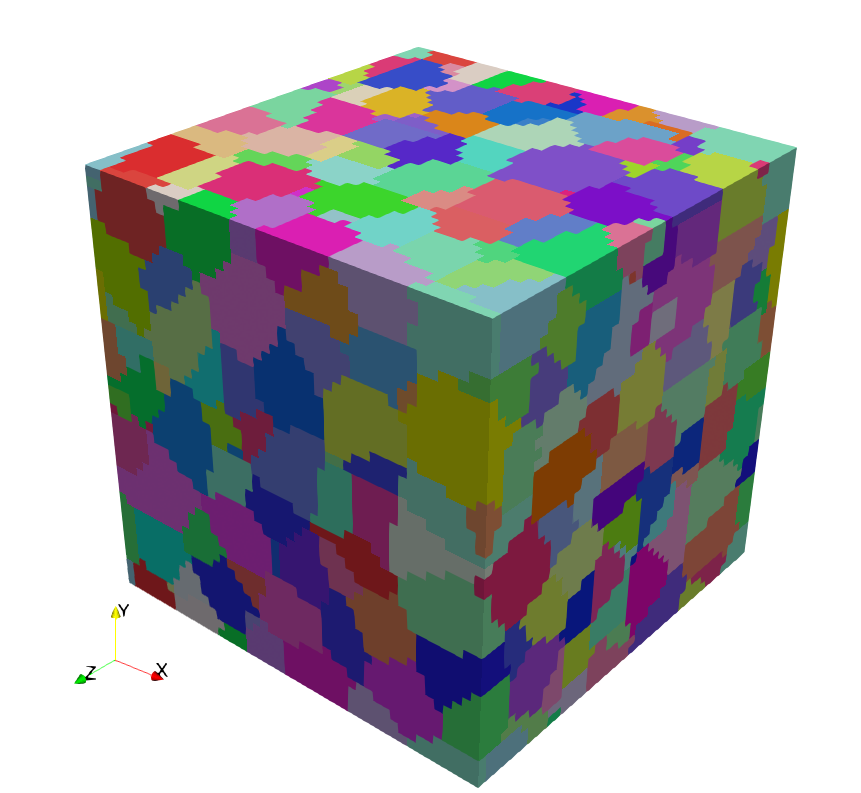}}
	\subfigure[Uniaxial-tension responses of single grains (online).]{\includegraphics[clip=true,trim = 0.0cm 0.0cm 1.0cm 0.5cm,width=0.44\textwidth]{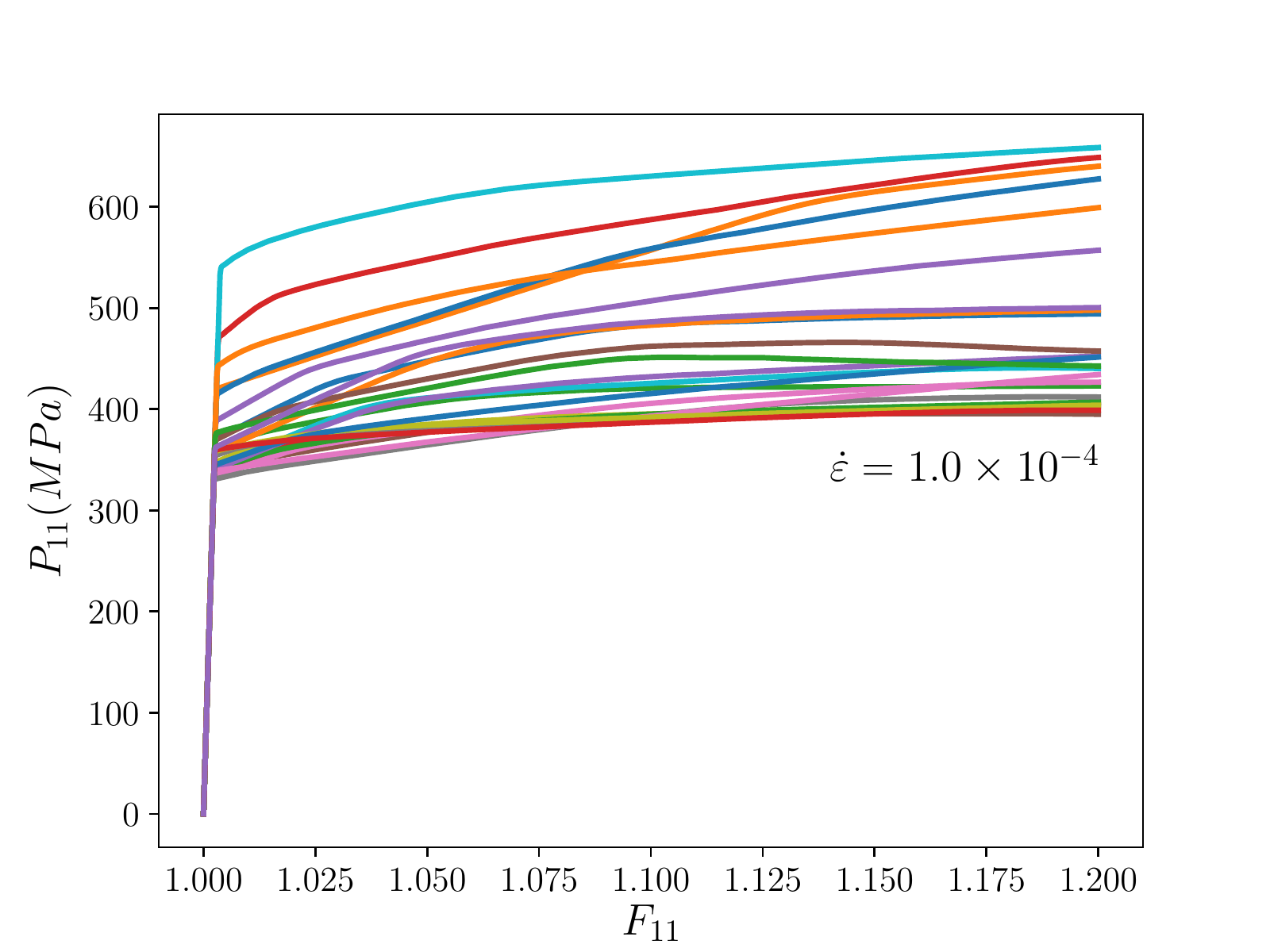}}
	\caption{Geometry and single-crystal responses of the polycrystalline RVE with random ODF. In (a), the RVE of equiaxed grains is generated with nominal number 415 and mesh size $45\times45\times45$. In (b), we selected 25 single grains randomly from the RVE, and pulled them under the crystal plasticity law used in the online stage at strain rate $\dot{\varepsilon}=1.0\times10^{-4}$. }
	\label{fig: polycrystal}
\end{figure}

Another application of the 3D DMN is to evaluate the effective properties of polycrystalline materials, such as Ni-based super-alloy. We used the software package DREAM.3D \cite{groeber2014dream} to generate the polycrystalline RVEs of equiaxed grains with two different crystallographic orientation distribution functions (ODF): 1) Random ODF and 2) textured ODF with preferred grain orientations. For both cases, the nominal number of grains inside the RVE is set to 415. As shown in Figure \ref{fig: polycrystal} (a), the polycrystalline RVE is represented in a $45\times45\times45$ mesh with 91125 material points. Since the mesh is uniform, FFT-based micromechanics method \cite{moulinec1998a} is chosen for the full-field DNS to accelerate the computation. In the online stage, we will extrapolate trained polycrystalline DMNs to finite-strain rate-dependent crystal plasticity, and the uniaxial responses of 25 randomly selected grains from the RVE with random ODF are presented in Figure \ref{fig: polycrystal} (b). 

\subsubsection{Offline evaluation}
The polycrystalline RVE can be treated as a single-phase material. In other words, all the grains shares the same material model, but with different orientations. Hence, the DMN in the offline stage takes only one input $\textbf{C}^{p1}$ as shown in Eq. (\ref{eq:final1}), while other parts of the training process remain the same as before. 

\begin{figure}[!t]
	\centering
	\graphicspath{{Figures/}}
	\subfigure[Random ODF.]{\includegraphics[clip=true,trim = 0.0cm 0.0cm 1.0cm 0.5cm,width=0.44\textwidth]{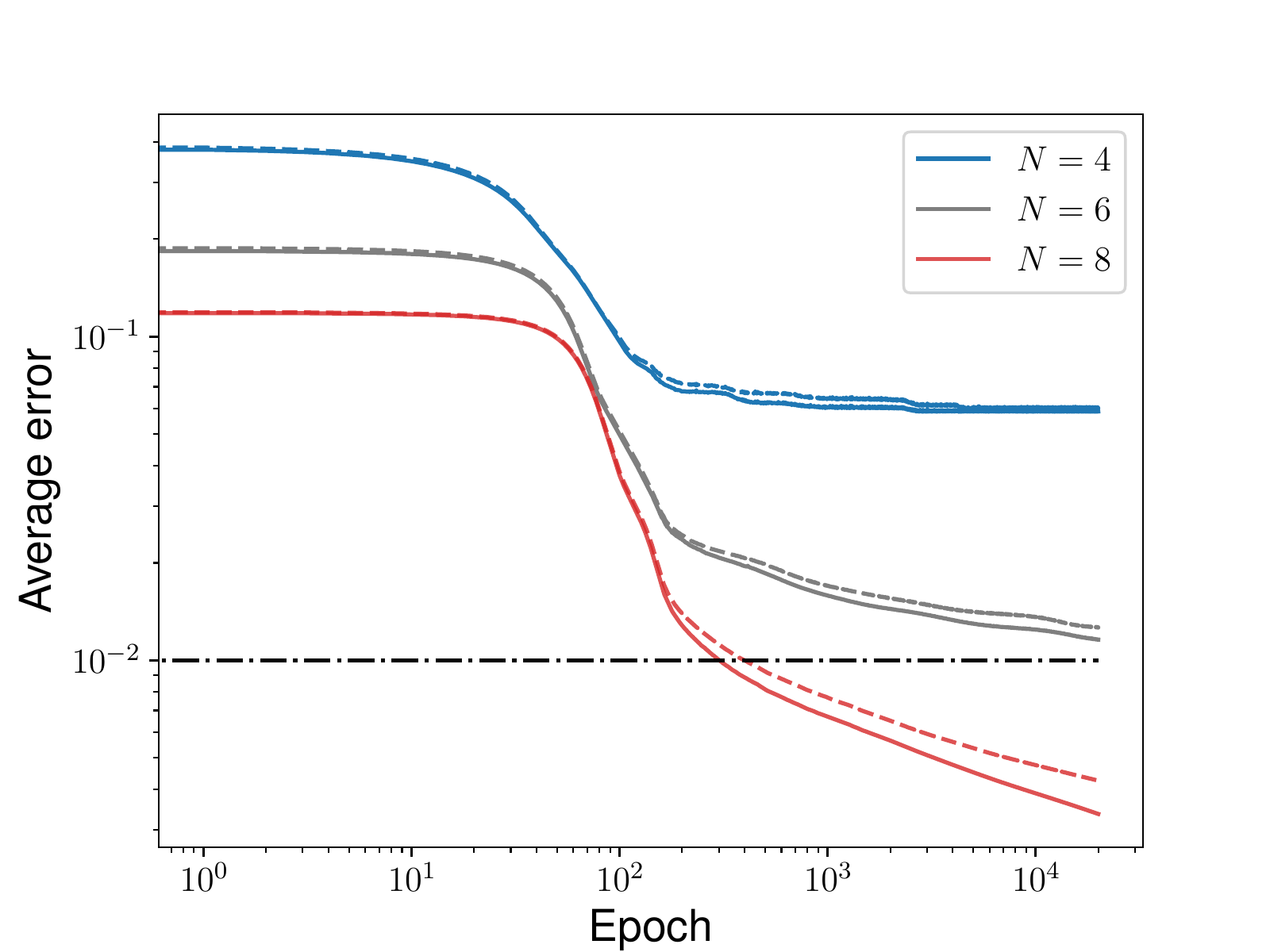}}
	\subfigure[Textured ODF.]{\includegraphics[clip=true,trim = 0.0cm 0.0cm 1.0cm 0.5cm,width=0.44\textwidth]{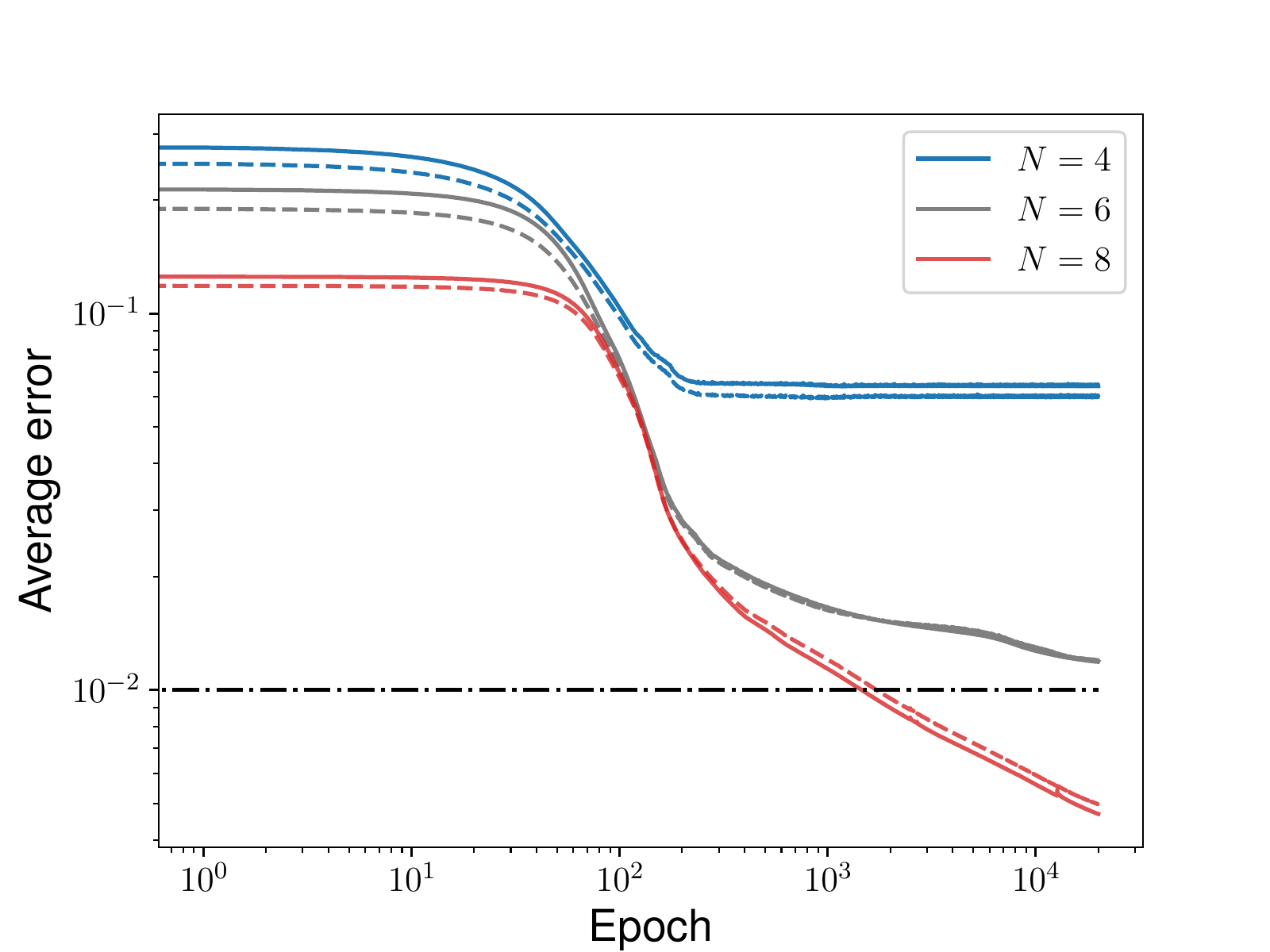}}
	\caption{Training histories of DMN for polycrystalline RVEs with (a) random ODF and (b) textured ODF. The histories of the average training and test errors are denoted by solid and dashed lines, respectively. All the networks are trained for 20000 epochs.}
	\label{fig:cpHist}
\end{figure}

Figure \ref{fig:cpHist} presents the histories of the training and test errors for two RVEs with random and textured ODFs. In general, the training process of polycrystalline RVE is smoother than the one of a two-phase material (see Figure \ref{fig:npHist}), because it has less deactivation or compression operations which may cause abrupt temporary changes in the error. DMNs with three network depths $N=4, 6, 8$ are investigated, and their initial number of active nodes in the bottom layers are $N_a = 8, 32, 128$, respectively. We can see from the figures that the errors of $N=8$ can be reduced to less than 0.5\% for both random and textured ODFs after 20000 epochs. Since the slopes are not saturated yet, further reduction of the errors could be achieved by continuing the training for more epochs. 

Errors of the RVE with random ODF at the end of training are also provided in Table \ref{table:cp}. For $N=6$, the mean training and test errors are close to 1\%, and the maximum test error is only $3.64\%$. In practice, the choice of the network depth $N$ is a trade-off between the desired accuracy and computational cost. For example, if 1.27\% of average error is satisfactory for a certain application, DMN with $N=6$ would be a better candidate for representing the polycrystalline RVE, since it will be around four times faster than the one with $N=8$ in terms of both offline training and online extrapolation.
\begin{table}[!t]
	\captionabove{Training results of the polycrystalline RVE with random ODF. Average training error $\bar{e}^{tr}$, average test error $\bar{e}^{tr}$ and maximum test error  are provided for each DMN.} 
	\centering 
	\label{table:cp} 
	{\tabulinesep=1.0mm
		\begin{tabu}{c c c c c} 
			\hline 
			&Epochs & Training $\bar{e}^{tr}$ & Test $\bar{e}^{te}$ & Maximum $e_s^{te}$\\
			\hline
			$N=4$&20000&5.87\%&6.00\%&15.5\%\\ 
			$N=6$ &20000&1.16\%&1.27\%&3.64\% \\
			$N=8$ &20000&0.36\%&0.43\%&1.80\% \\
			\hline
	\end{tabu}}
\end{table}

\begin{figure} [!t]
	\centering
	\graphicspath{{Figures/}}
	\subfigure[$N=4, N_a =8$]{\includegraphics[clip=true,trim = 0.0cm 0.0cm 1.0cm 0.5cm,width=0.28\textwidth]{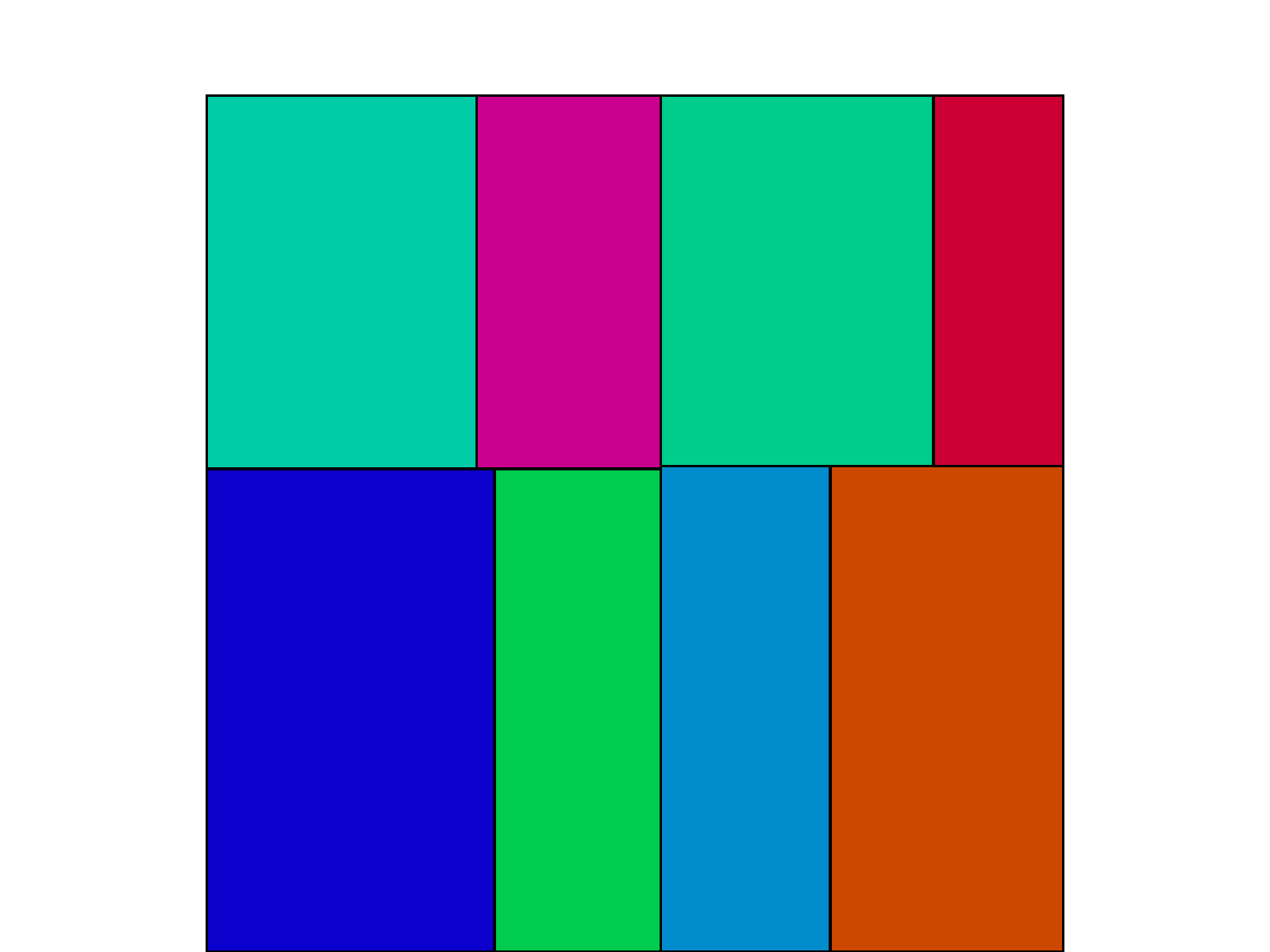}}
	\subfigure[$N=6, N_a =31$]{\includegraphics[clip=true,trim = 0.0cm 0.0cm 1.0cm 0.5cm,width=0.28\textwidth]{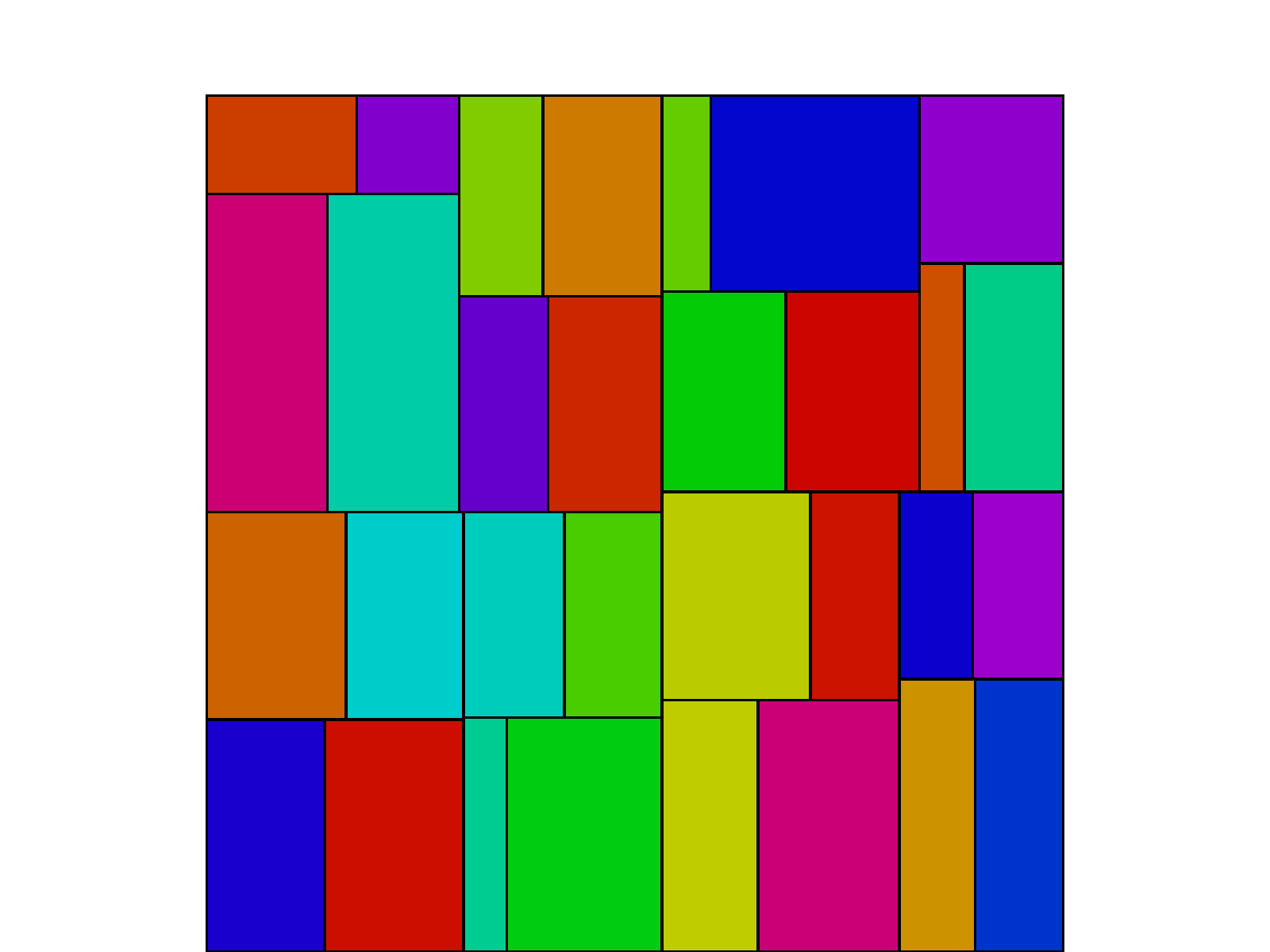}}
	\subfigure[$N=8, N_a =128$]{\includegraphics[clip=true,trim = 0.0cm 0.0cm 1.0cm 0.5cm,width=0.28\textwidth]{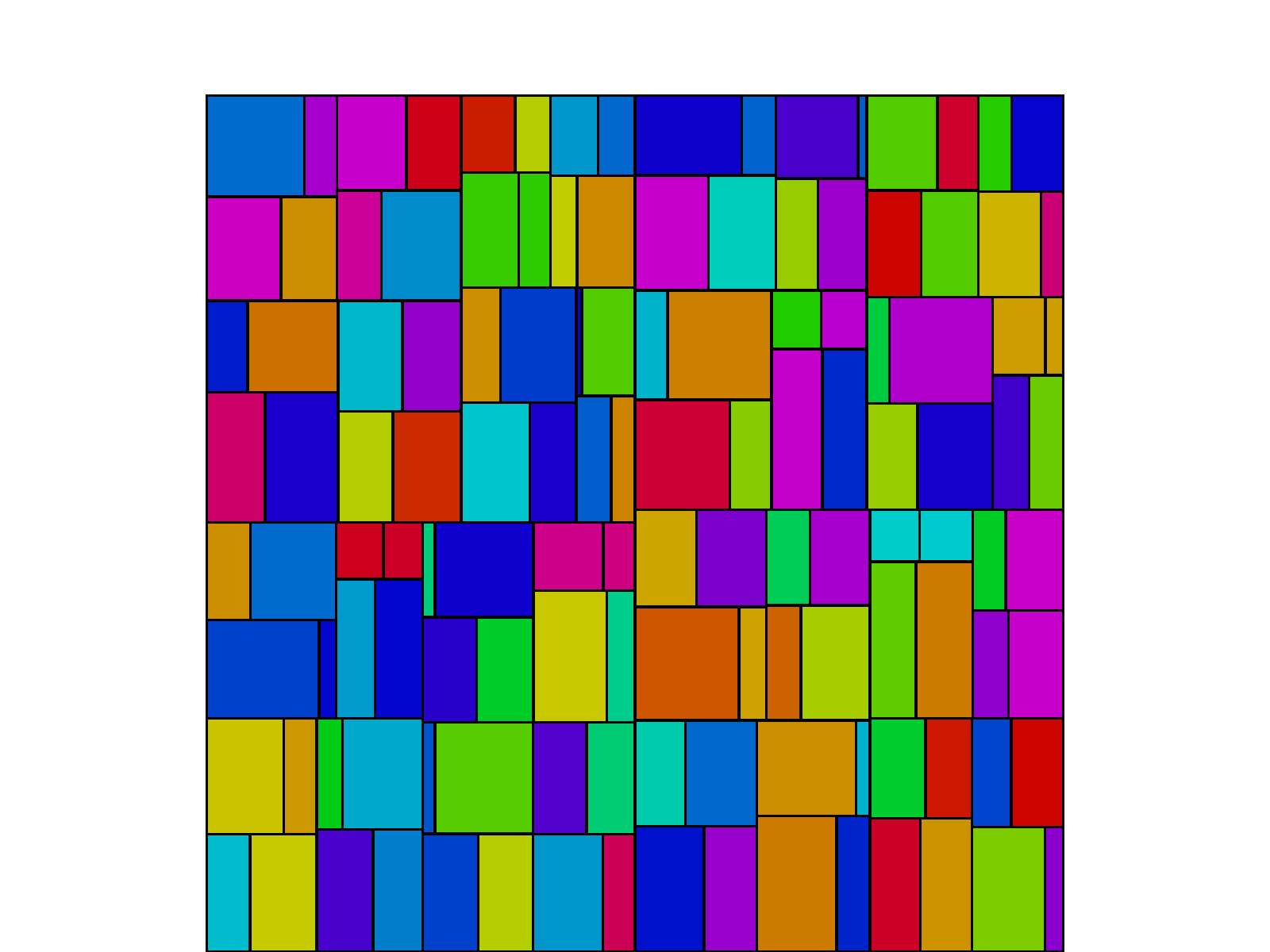}}
	\caption{Treemaps of DMN for the polycrystalline RVE with random ODF. The network depths $N$ are (a) 4, (b) 6 and (c) 8. The number of active nodes in the bottom layer $N_a$ is listed under each plot. The block colors are randomly assigned.}
	\label{fig:cpTree}
\end{figure}

Treemaps of DMN at the end of training are plotted in Figure \ref{fig:cpTree} for the RVE with random ODF. Since there is only one phase in the RVE, the volume fractions cannot be defined, thus, are not shown in the plots. In each polycrystalline DMN, the number of active bottom-layer nodes did not decrease during the training, except for $N=6$ where only one node is deactivated. After 20000 epochs, the weights of bottom-layer nodes are still uniformly distributed. In some sense, this is consistent with the DNS RVE model with equiaxed grains.   
\begin{figure}[!t]
	\centering
	\graphicspath{{Figures/}}
	\subfigure[DNS, 415 grains]{\includegraphics[clip=true,trim = 0.cm 0.0cm 0.0cm 0.cm,width=0.44\textwidth]{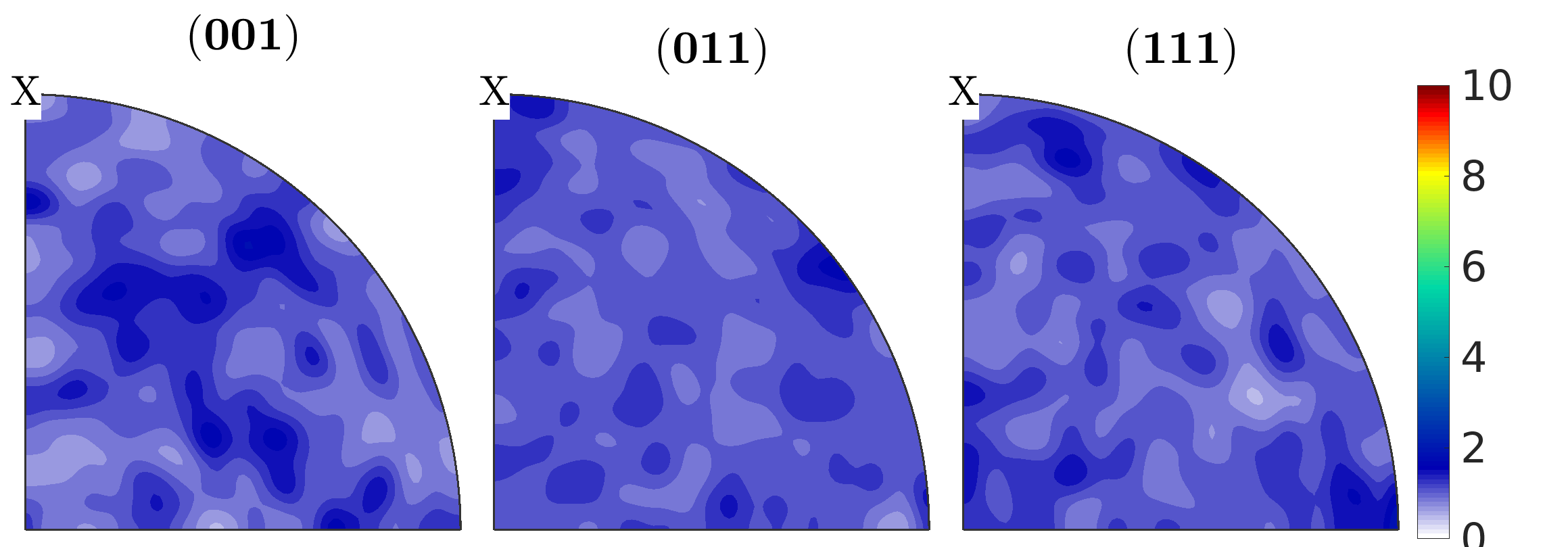}}
	\subfigure[$N=4, N_a = 8$]{\includegraphics[clip=true,trim = 0.cm 0.7cm 0.0cm -0.7cm,width=0.44\textwidth]{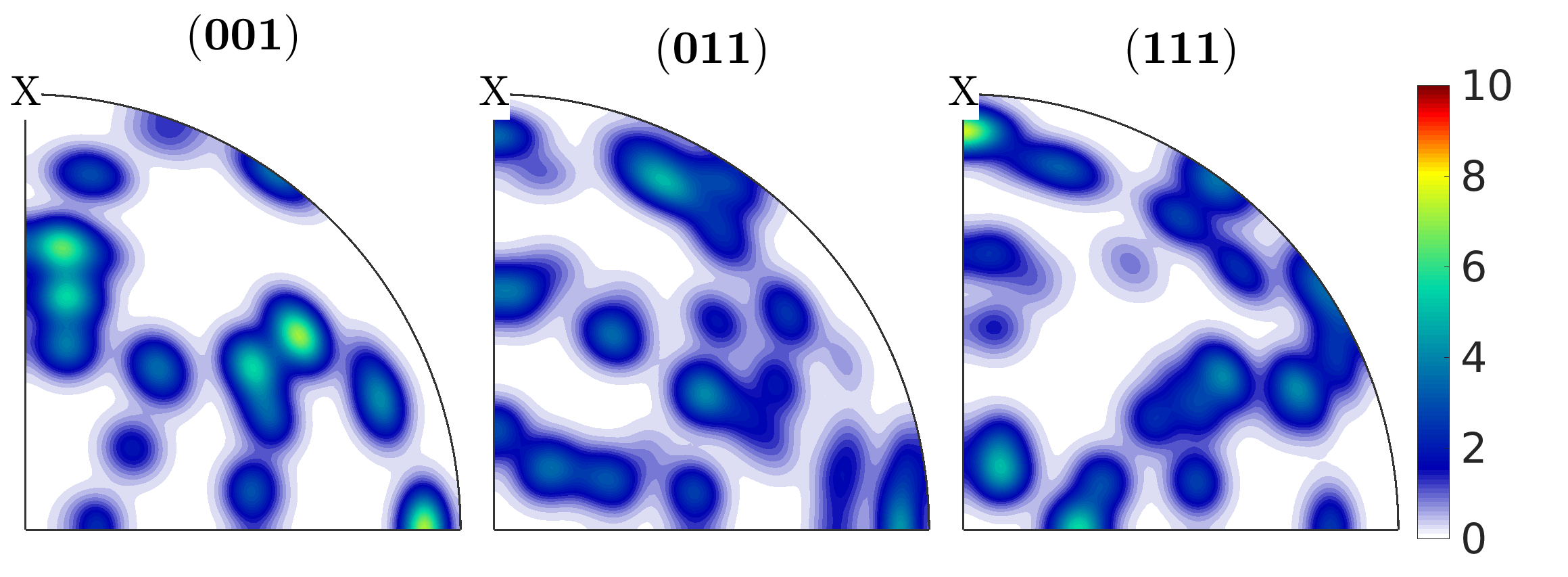}}
	\subfigure[$N=6, N_a = 31$]{\includegraphics[clip=true,trim = 0.cm 0.0cm 0.0cm 0.cm,width=0.44\textwidth]{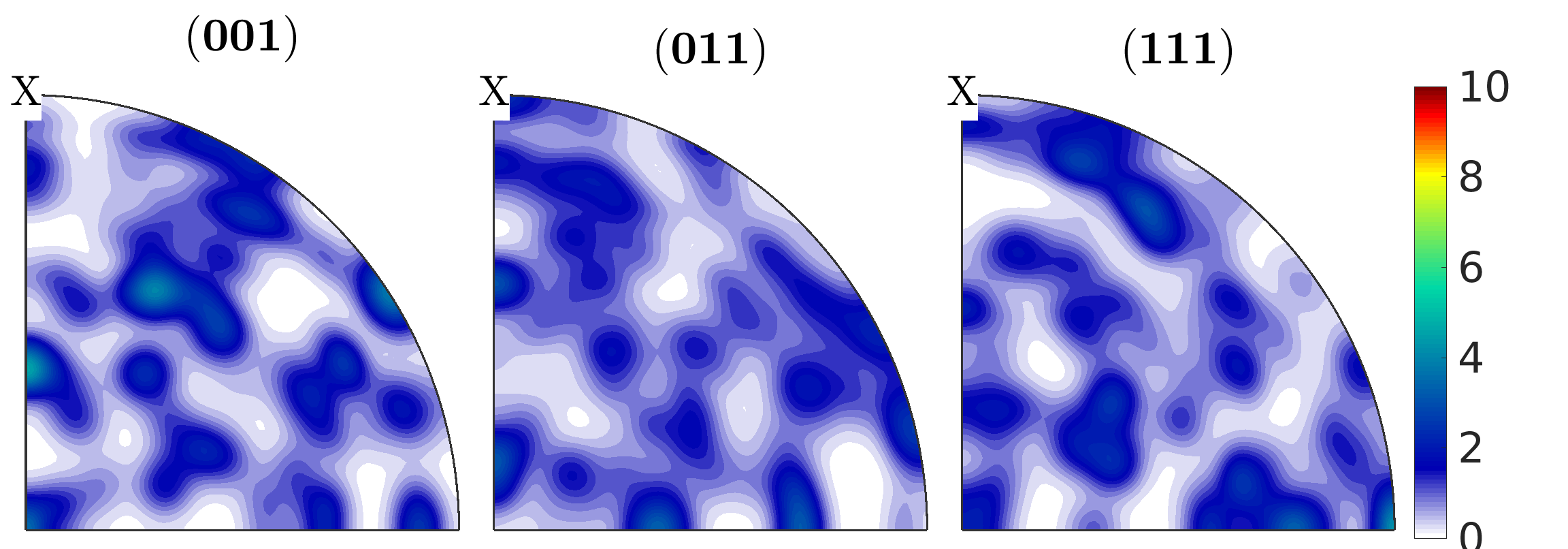}}
	\subfigure[$N=7, N_a = 64$]{\includegraphics[clip=true,trim = 0.cm 0.7cm 0.0cm -0.7cm,width=0.44\textwidth]{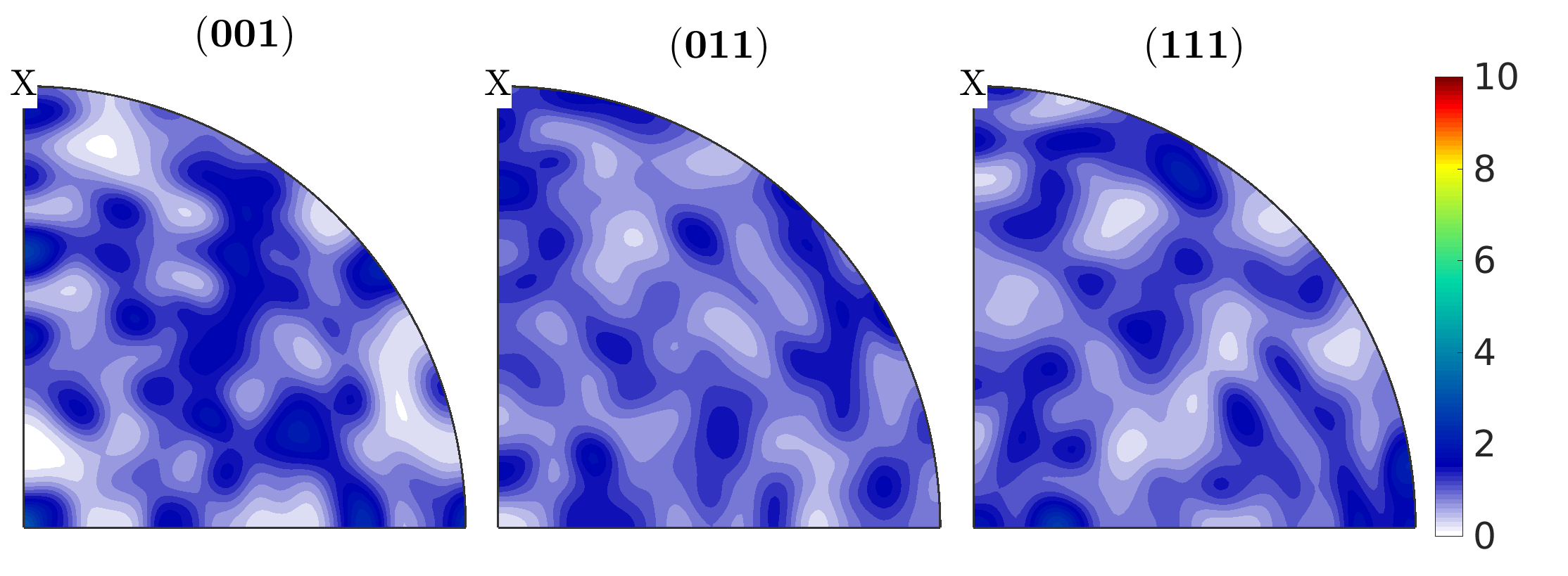}}
	\subfigure[$N=8, N_a = 128$]{\includegraphics[clip=true,trim = 0.cm 0.0cm 0.0cm 0.cm,width=0.44\textwidth]{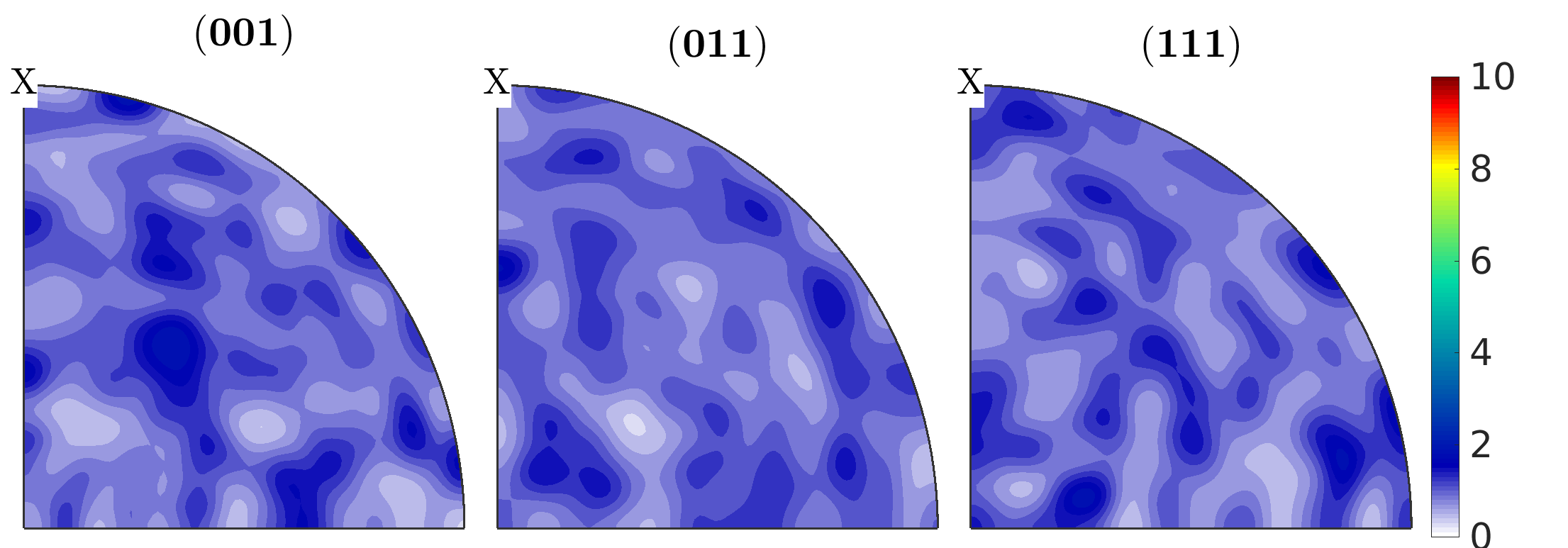}}
	\subfigure[$N=9, N_a = 255$]{\includegraphics[clip=true,trim = 0.cm 0.0cm 0.0cm 0.cm,width=0.44\textwidth]{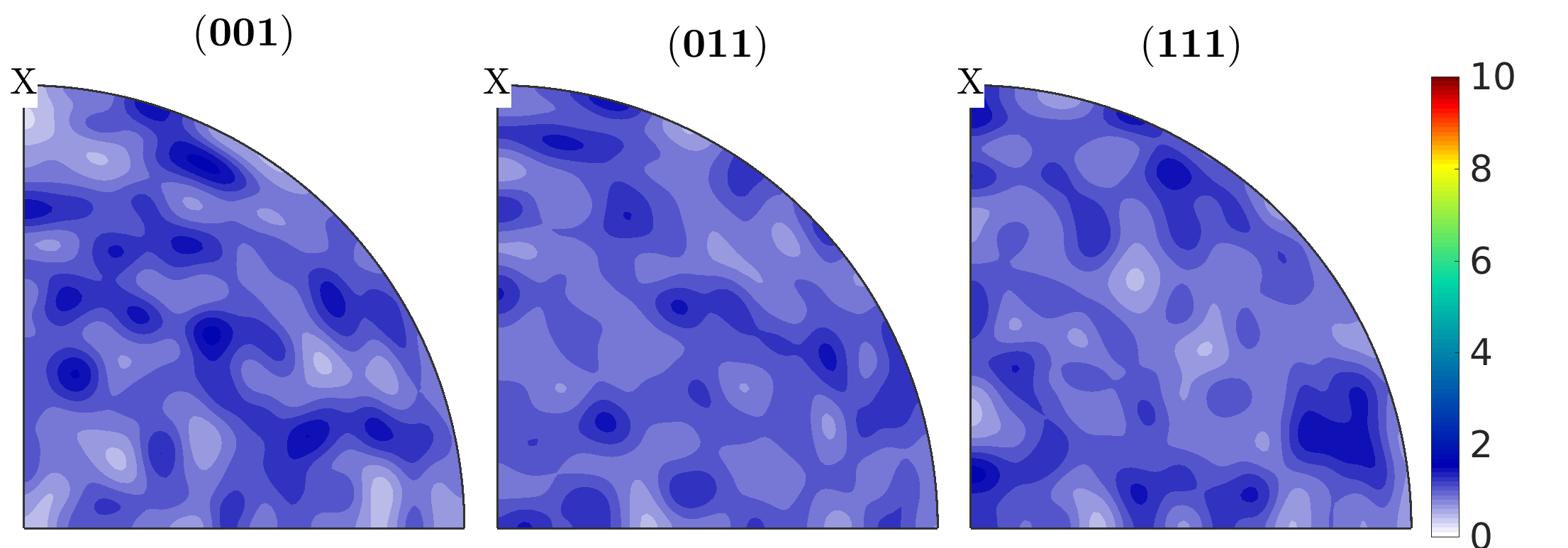}}
	\caption{Pole figures from DNS and predicted by DMN for the polycrystalline RVE with random ODF. The color bar indicates the multiples of random distribution (MRD).}
	\label{fig: polecrystal_case1}
\end{figure}
\begin{figure}[!t]
	\centering
	\graphicspath{{Figures/}}
	\subfigure[DNS, 417 grains]{\includegraphics[clip=true,trim = 0.cm 0.0cm 0.0cm 0.cm,width=0.44\textwidth]{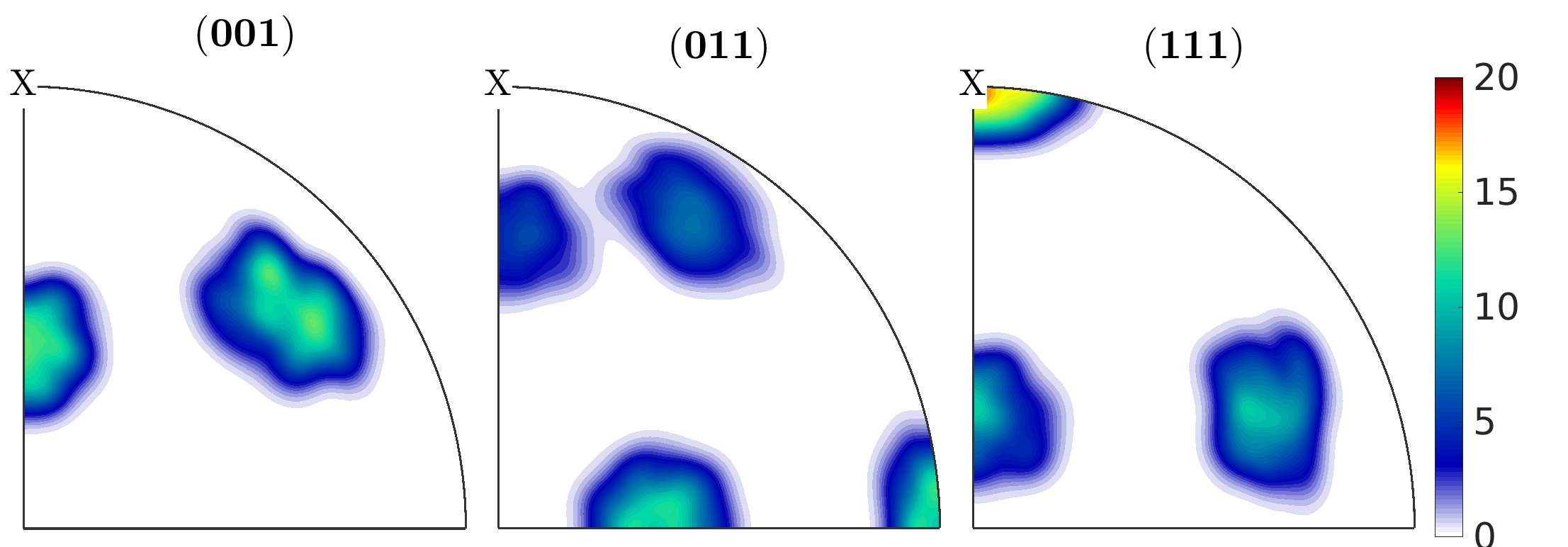}}
	\subfigure[$N=4, N_a = 8$]{\includegraphics[clip=true,trim = 0.cm 0.0cm 0.0cm 0.cm,width=0.44\textwidth]{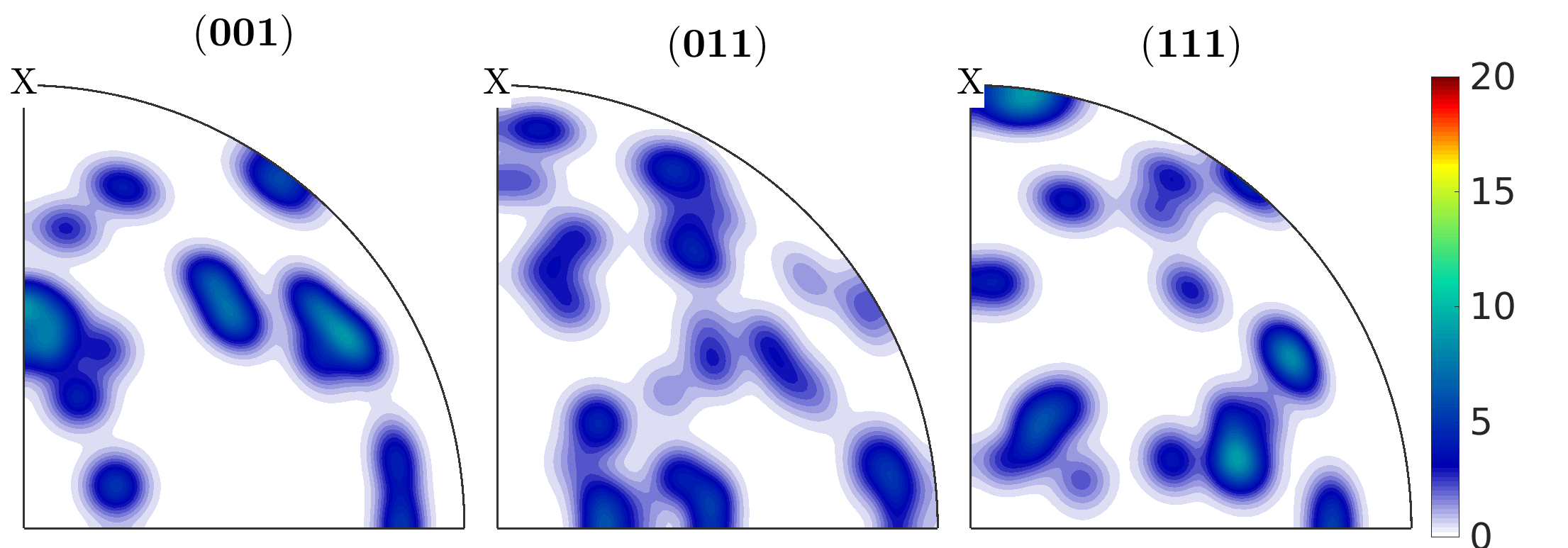}}
	\subfigure[$N=6, N_a = 32$]{\includegraphics[clip=true,trim = 0.cm 0.0cm 0.0cm 0.cm,width=0.44\textwidth]{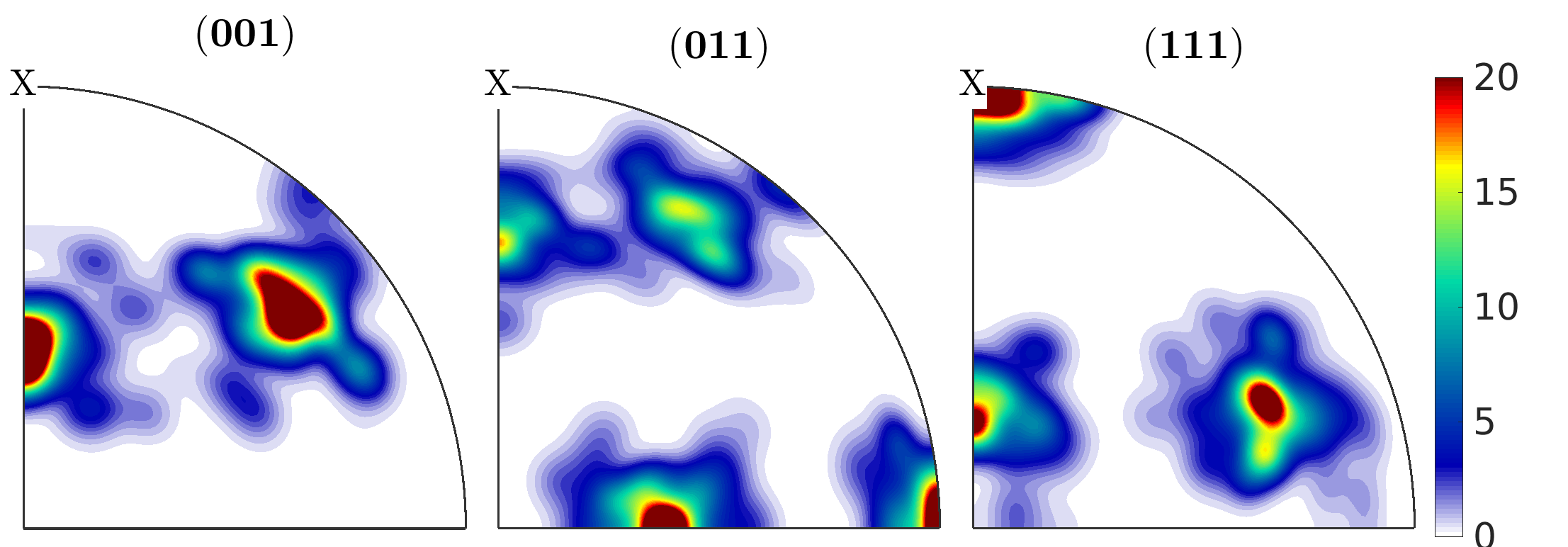}}
	\subfigure[$N=7, N_a = 60$]{\includegraphics[clip=true,trim = 0.cm 0.0cm 0.0cm 0.cm,width=0.44\textwidth]{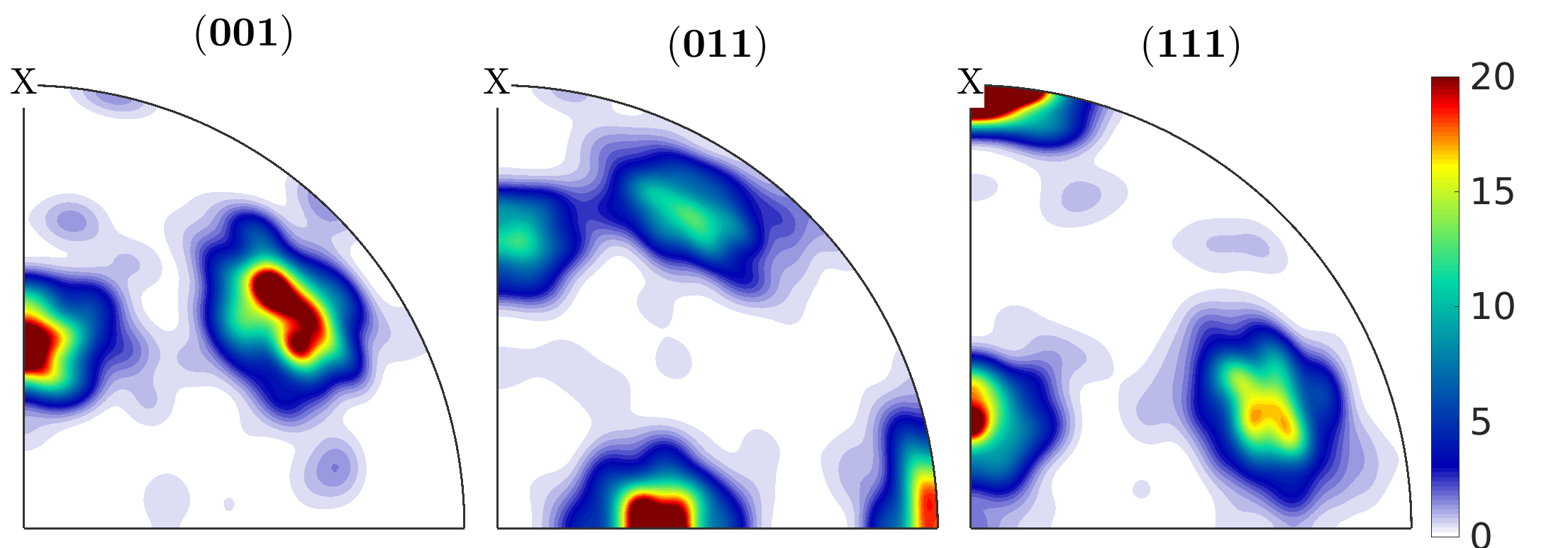}}
	\subfigure[$N=8, N_a = 111$]{\includegraphics[clip=true,trim = 0.cm 0.0cm 0.0cm 0.cm,width=0.44\textwidth]{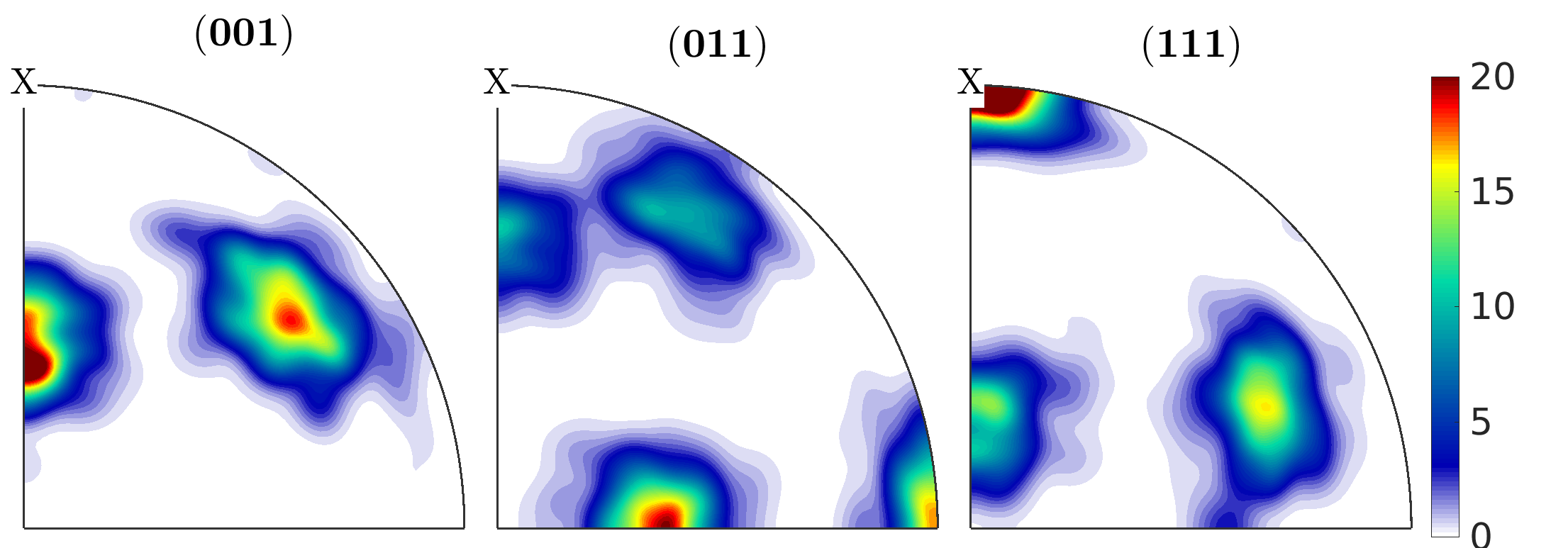}}
	\subfigure[$N=9, N_a = 226$]{\includegraphics[clip=true,trim = 0.cm 0.0cm 0.0cm 0.cm,width=0.44\textwidth]{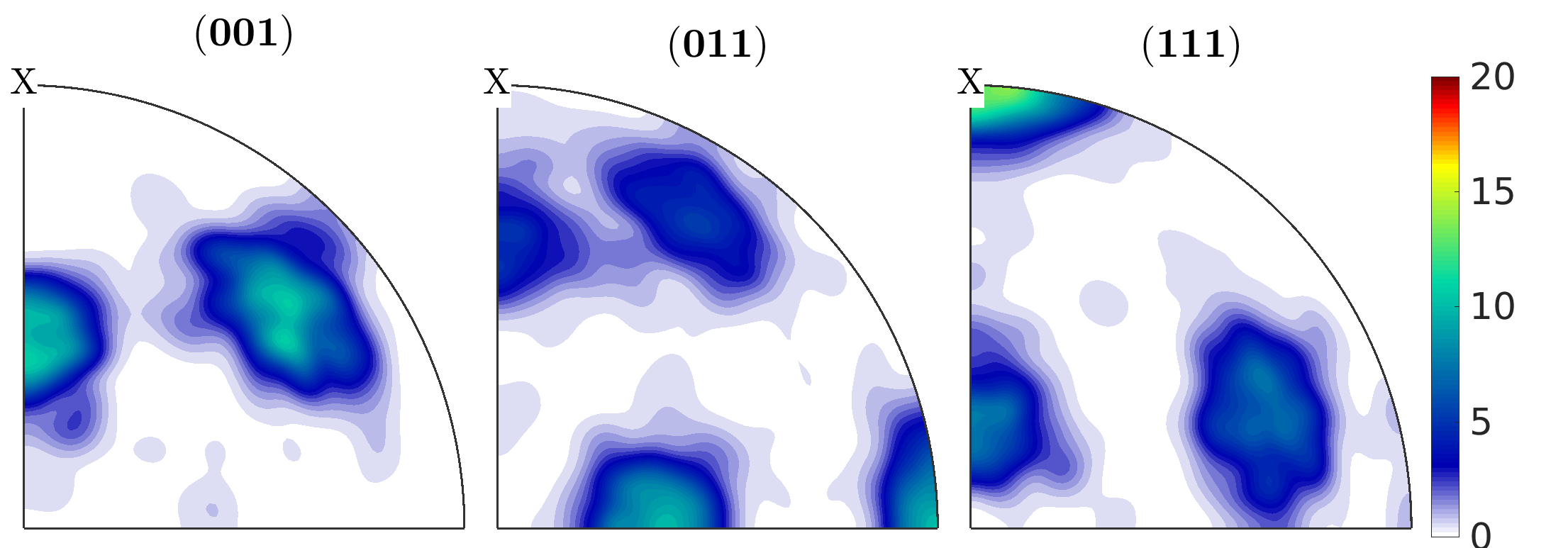}}
	\caption{Pole figures from DNS and predicted by DMN for the polycrystalline RVE with textured ODF. The color bar indicates the multiples of random distribution (MRD).}
	\label{fig: polecrystal_case2}
\end{figure}

As the weights in the network do not change much, the main training effort is devoted to optimizing the angles in each building block. Therefore, it is also interesting to check whether our DMN can learn the distribution of angles or ODF purely from the mechanical data of stiffness matrices. In this work, we follow the Bunge convention of Euler angles ($\phi_1, \Phi, \phi_2$) \cite{bunge2013texture} and use the MTEX package \cite{bachmann2010texture} to generate the pole figures. Figure \ref{fig: polecrystal_case1} collects the (001),(011), (111) pole figures from DNS model and predicted by DMNs with network depth ranging from 4 to 9.  Note that only a quarter of the pole figure is shown due to the symmetry of orthotropic materials used in the offline stage. By looking at these plots, we can conclude that DMN with sufficient number of layers can discover the nature of random ODF hidden in the mechanical dataset. Finding the ODF for the RVE with textured ODF can be more challenging, and the pole figures for this case are listed Figure \ref{fig: polecrystal_case2}. It shows that for $N>4$, the DMN is able to recover the dense regions appearing in the DNS pole figures, and the best match with DNS is achieved by the network with $N=9$. Due to the existence of dense regions in the textured ODF, the compression operations (e.g. subtree merging) are more likely to be triggered during the training process, and the resulting DMN usually has less number of active bottom-layer nodes, or a smaller $N_a$, than the one for random ODF with the same network depth.

\subsubsection{Online extrapolation}
In the online stage, we evaluate the trained DMN on a finite-strain rate-dependent crystal plasticity model \cite{liu2018data,McGinty2001}. To begin with, the deformation gradient $\mathbf{F}$ is decomposed multiplicatively as:
\begin{equation} \label{eq:decompf}
\mathbf{F} = \mathbf{F}^e\mathbf{F}^p,
\end{equation}
where the plastic part $\mathbf{F}^p$ maps points in the reference configuration onto an intermediate one which is then mapped to the current one through the elastic part $\mathbf{F}^e$. 
The effect of dislocation motion is modeled by relating the plastic velocity gradient $\tilde{\mathbf{L}}^p$ in the intermediate configuration to simple shear deformation $\gamma^{(\alpha)}$:
\begin{equation}
\tilde{\mathbf{L}}^p = \sum_{\alpha=1}^{N_\text{slip}} \dot{\gamma}^{(\alpha)}(\tilde{\mathbf{s}}^{(\alpha)} \otimes \tilde{\mathbf{n}}^{(\alpha)})
\end{equation}
where $N_\text{slip}$ is the number of slip systems. For a crystal slip systems $(\alpha)$ in the intermediate configuration, $\dot{\gamma}^{(\alpha)}$ is a shear rate, $\tilde{\mathbf{s}}^{(\alpha)}$ is the slip direction, and $\tilde{\mathbf{n}}^{(\alpha)}$ is the slip plane normal. The relationship between $\tilde{\mathbf{L}}^p$ and $\mathbf{F}^p$ is given by
\begin{equation}\label{eq:pvg}
\tilde{\mathbf{L}}^p = \mathbf{\dot{F}}^p \cdot \big(\mathbf{F}^p\big)^{-1}.
\end{equation}
We choose to formulate the constitutive laws of elasto-plasticity based on the Green strain $\mathbf{E}^e$ and Second Piola-Kirchhoff stress $\mathbf{S}^e$, which are related by:
\begin{equation}
\mathbf{S}^e = \tilde{\mathbf{C}} \cdot \mathbf{E}^e = \frac{1}{2} \tilde{\mathbf{C}} \cdot \big[ (\mathbf{F}^e)^{\text{T}}\mathbf{F}^e - \mathbf{I} \big],
\end{equation}
where the elastic stiffness tensor $\tilde{\mathbf{C}}$ is defined in the intermediate configuration. The plastic shear rate in each slip system given by a phenomenological power law
\begin{equation}\label{eq:shearrate}
\dot{\gamma}^{(\alpha)} = \dot{\gamma}_0 \left | \frac{ {\tau}^{(\alpha)} - a^{(\alpha)} }{ \tau^{(\alpha)}_0 } \right | ^{(m-1)} \bigg( \frac{ {\tau}^{(\alpha)} - a^{(\alpha)} }{ \tau^{(\alpha)}_0 } \bigg),
\end{equation}
where ${\tau}^{(\alpha)}$ is the resolved shear stress, $a^{(\alpha)}$ is a back-stress that describes kinematic hardening, $\dot{\gamma}_0$ is a reference shear rate, $\tau^{(\alpha)}_0$ is a reference shear stress that accounts for isotropic hardening, and the exponent $m$ determines the material strain rate sensitivity. The shear stress ${\tau}^{(\alpha)}$ is resolved onto the slip directions with:
\begin{equation}
{\tau}^{(\alpha)} = \bm{\sigma} : (\mathbf{s}^{(\alpha)} \otimes \mathbf{n}^{(\alpha)}),
\end{equation}
where $\bm{\sigma}$, $\mathbf{s}^{(\alpha)}$ and $\mathbf{n}^{(\alpha)}$ are the Cauchy stress, slip direction and slip plane normal in the current configuration, respectively. Moreover, the evolution equation of the reference shear stress $\tau^{(\alpha)}_0$  is defined as
\begin{equation}
\dot{\tau}^{(\alpha)}_0 = H \sum_ {\beta = 1} ^ {N_\text{slip}} q^{\alpha\beta} \dot{\gamma}^{(\beta)} - R\tau^{(\alpha)}_0 \sum_ {\beta = 1} ^ {N_\text{slip}}  | \dot{\gamma}^{(\beta)} |,
\end{equation}
where $H$ is a direct hardening coefficient and $R$ is a dynamic recovery coefficient and $q^{\alpha\beta}$ is the latent hardening ratio given by:
\begin{equation}
q^{\alpha\beta} = \chi + (1-\chi)\delta_{\alpha\beta}
\end{equation}
where $\chi$ is a latent hardening parameter. The back-stress $a^{(\alpha)}$ in Eq. (\ref{eq:shearrate}) evolves based on the following expression,
\begin{equation} \label{eq:backstress}
\dot{a}^{(\alpha)} = h\dot{\gamma}^{(\alpha)} - ra | \dot{\gamma}^{(\alpha)} |,
\end{equation}
where $h$ and $r$ are direct and dynamic hardening factors respectively. The set of nonlinear equations from Eq. (\ref{eq:decompf}) to Eq. (\ref{eq:backstress}) need to be solved numerically in the crystal plasticity model. In this work, we defined the model for a FCC metal (e.g. Ni-based super-alloy) with 12 slip planes. In the elastic regime, the material behaves isotropically. All the crystal plasticity parameters are given in Table \ref{table:polypara}.
\begin{table}[ht]
	\captionabove{Crystal plasticity parameters for a FCC metal with 12 slip planes \cite{liu2018data,McGinty2001}.} 
	\centering 
	\label{table:polypara} 
	{\tabulinesep=1.0mm
		\begin{tabu}{c c c c c c} 
			\hline\hline
			$C_{1111}$ (GPa) & $C_{1122}$ (GPa) & $C_{2323}$ (GPa) & $\dot{\gamma}_0$ (s$^{-1}$) & $m$ & initial $\tau_0$ (MPa)\\
			196.4 & 84.2 & 56.1 & 0.00242 & 58.8 & 171.85 \\ 
			\hline
			$H$ (MPa) & $R$ (MPa) & $\chi$ & initial $a_0$ (MPa) & $h$ (MPa) &$r$ (MPa)\\
			1.0& 0.0 & 1.0& 0.0 & 500.0 & 0.0  \\
			\hline
	\end{tabu}}
\end{table}
\begin{figure} [!t]
	\centering
	\graphicspath{{Figures/}}
	\subfigure[Random ODF.]{\includegraphics[clip=true,trim = 0.0cm 0.0cm 1.0cm 0.5cm,width=0.44\textwidth]{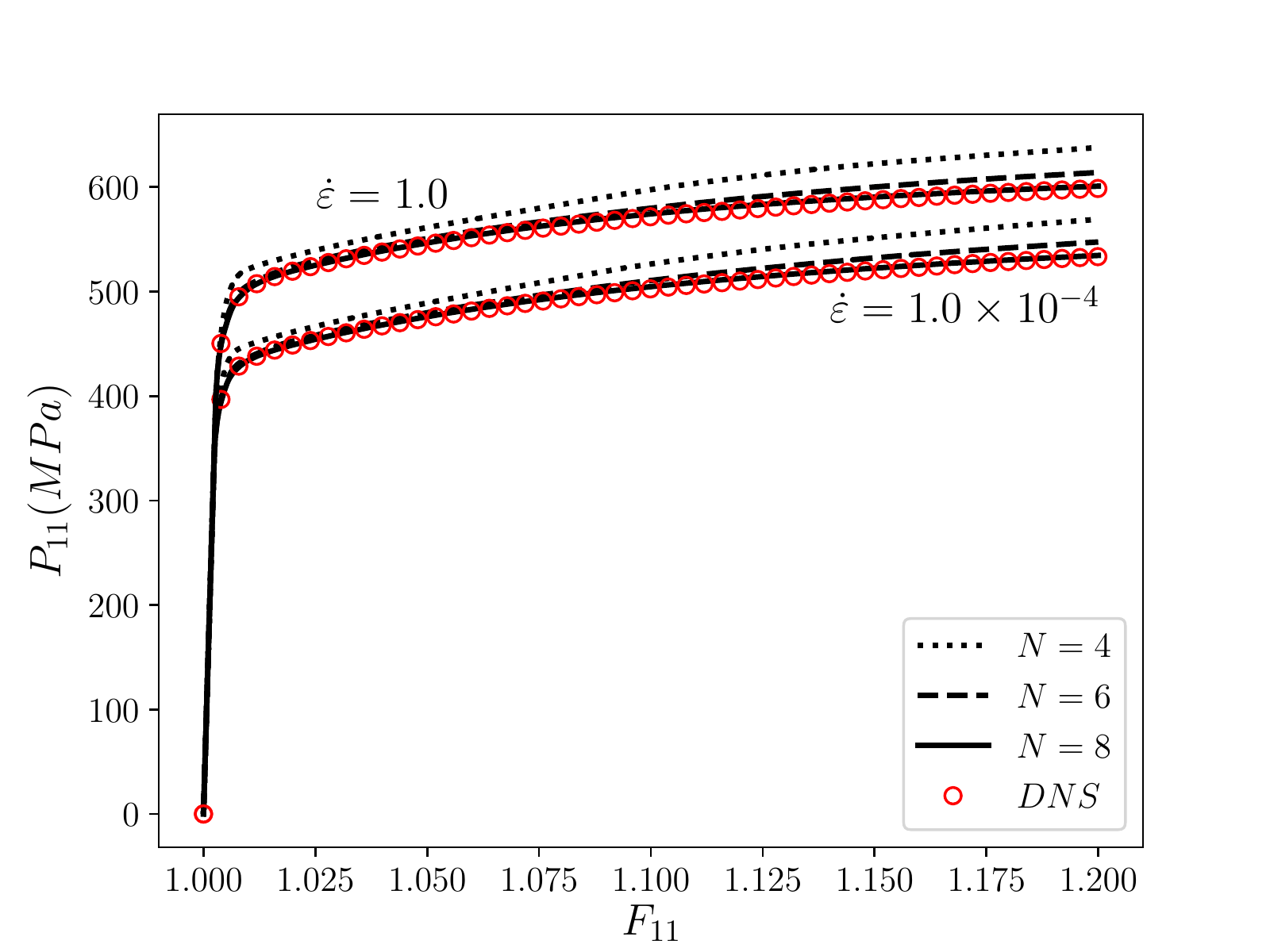}}
	\subfigure[Textured ODF.]{\includegraphics[clip=true,trim = 0.0cm 0.0cm 1.0cm 0.5cm,width=0.44\textwidth]{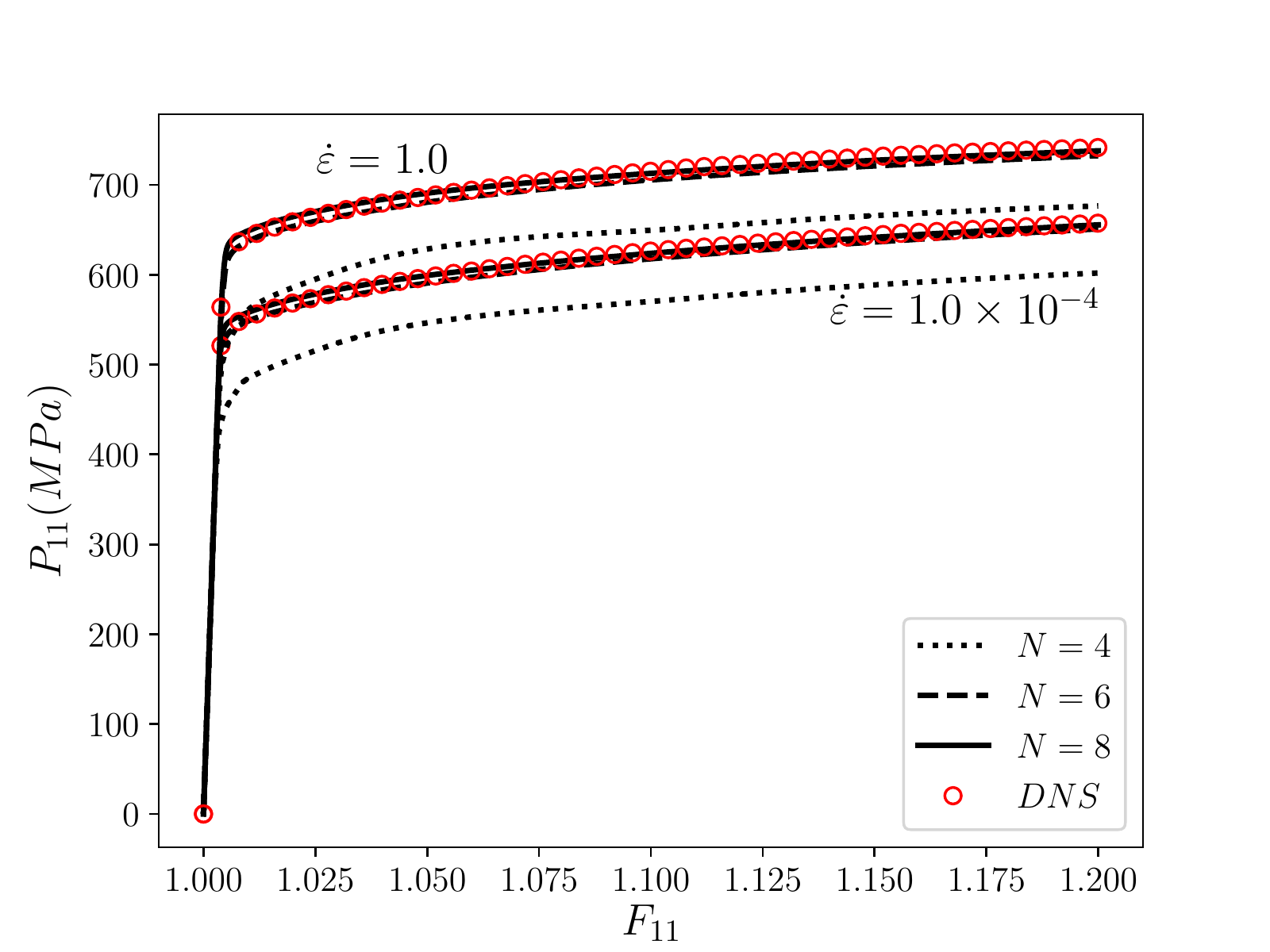}}
	\caption{Uniaxial stress-strain curves predicted by DNS and DMN for the polycrystalline RVEs with (a) random ODF and (b) textured ODF based on finite-strain rate-dependent crystal plasticity. Two strain rates are considered: $\dot{\varepsilon}=1.0\times10^{-4}$ and $\dot{\varepsilon}=1.0$. The network depths are $N=$4 (dotted), 6 (dashed) and 8 (solid).}
	\label{fig:cp}
\end{figure}

To show the rate-dependency in the crystal plastic model, two strain rates were considered: $\dot{\varepsilon}=1.0\times10^{-4}$ and $\dot{\varepsilon}=1.0$. The polycrystalline RVEs with random and textured ODFs are simulated by both DNS and DMN under uniaxial loading up to $F_{11}=1.2$, and the results are compared in Figure \ref{fig:cp}. The stress-strain curves of $N=4$ are shown as the dotted lines, and we can see from the plots that it overestimates the yield stress for random ODF while underestimates the value for textured ODF. Thus, it appears to be too ambitious to use 8 active bottom-layer nodes to represent the a RVE with more than 400 grains subject to different orientations. For $N=6$ and $N=8$, the networks can predict the hardening behavior very well under different strain rates. 

\subsection{Carbon fiber-reinforced polymer with three scales}\label{sec:cfrp}
\begin{figure}[!t]
	\centering
	\graphicspath{{Figures/}}
	\subfigure[Geometries and meshes of CFRPs.]{\includegraphics[clip=true,trim = 8.0cm 2.5cm 8.0cm 3.6cm,width=0.44\textwidth]{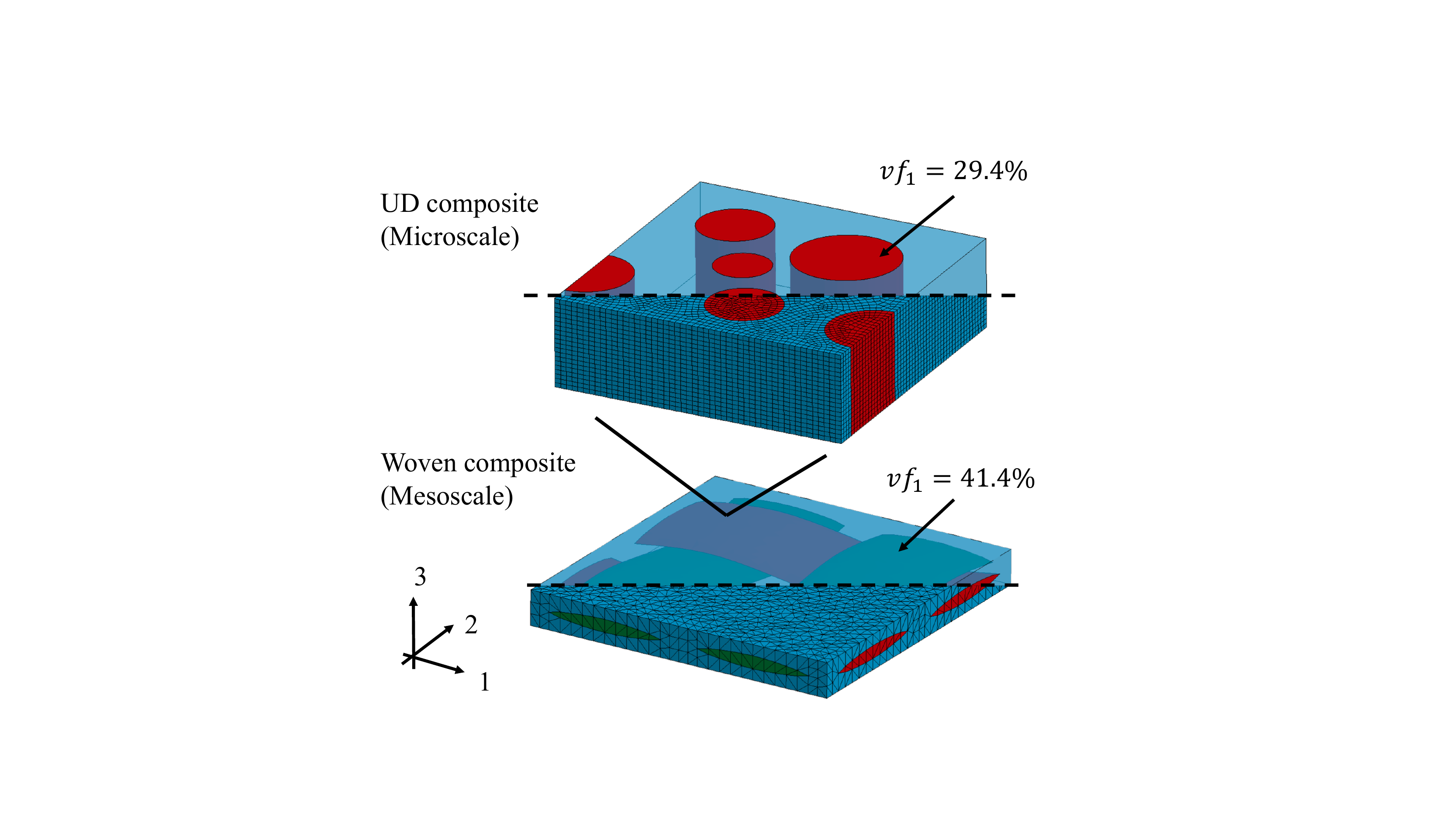}}
	\subfigure[Uniaxial responses of individual phases.]{\includegraphics[clip=true,trim = 0.0cm 0.0cm 1.0cm 0.5cm,width=0.44\textwidth]{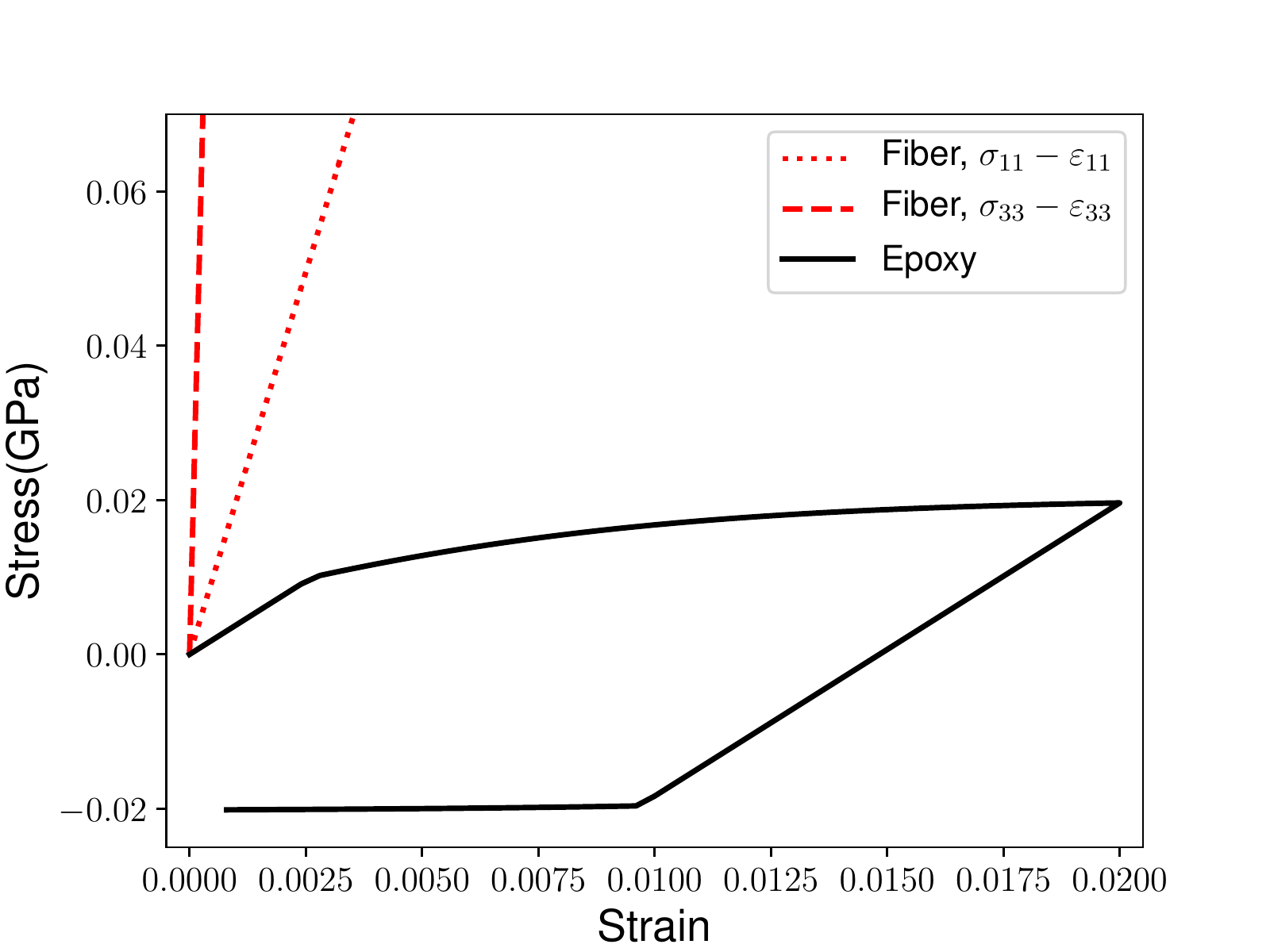}}
	\caption{CFRP composites: (a) In the UD RVE at microscale, the volume fraction of fiber phase is 29.4\%, and its FE model has 69048 nodes and 63720 8-node hexahedron elements. In the woven RVE at mesoscale, the volume fraction of yarn phase is 41.4\%, and its FE model has 43226 nodes and 28475 10-node tetrahedron elements. Material properties of the yarn phase come from the homogenization of the UD RVE; b) In the online stage, the epoxy matrix is considered as an elasto-plastic material, while the carbon fiber is modeled as an orthotropic material with high strength along the fiber direction.}
	\label{fig: cfrp}
\end{figure}

The last material system that we have investigated using DMN is the carbon fiber reinforced polymer (CFRP) composite. A typical example of CFRP system is illustrated in Figure \ref{fig: cfrp} (a). The microscale RVE is a unidirectional (UD) fiber composite with straight fibers penetrating through the matrix in one direction. The UD RVE is then homogenized to provide properties for the yarn phase within the mesoscale RVE of a woven composite, and the macroscale overall material properties are obtained by homogenizing the woven RVE. Compared to the previous two-scale examples, the CFRP system contains three scales with an additional mesoscale RVE, which poses more challenges on the efficiency and robustness of a reduced order model. 

The UD RVE is discretized by 69048 nodes and 63720 8-node hexahedron finite elements, and the volume fraction of the fiber phase is $vf_1=29.5\%$. Additionally, the woven RVE has 43226 nodes and 28475 10-node tetrahedron finite elements, and the volume fraction of its yarn phase is $vf_1=41.4\%$. The uniaxial responses of online material models for the carbon fiber and epoxy matrix are presented in Figure \ref{fig: cfrp} (b).

\subsubsection{Offline evaluation}
\begin{figure} [!t]
	\centering
	\graphicspath{{Figures/}}
	\subfigure[UD composite.]{\includegraphics[clip=true,trim = 0.0cm 0.0cm 1.0cm 0.5cm,width=0.44\textwidth]{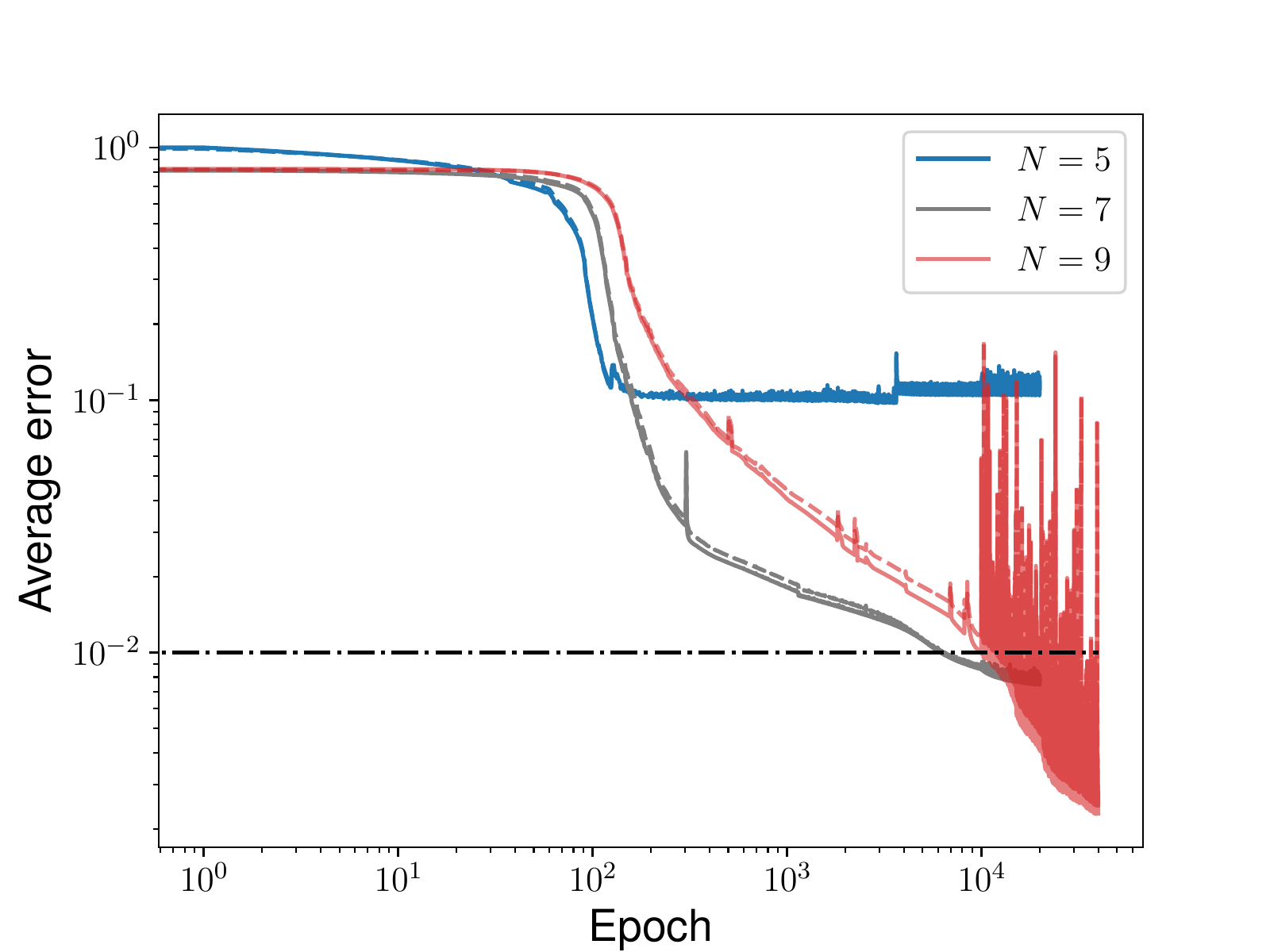}}
	\subfigure[Woven composite.]{\includegraphics[clip=true,trim = 0.0cm 0.0cm 1.0cm 0.5cm,width=0.44\textwidth]{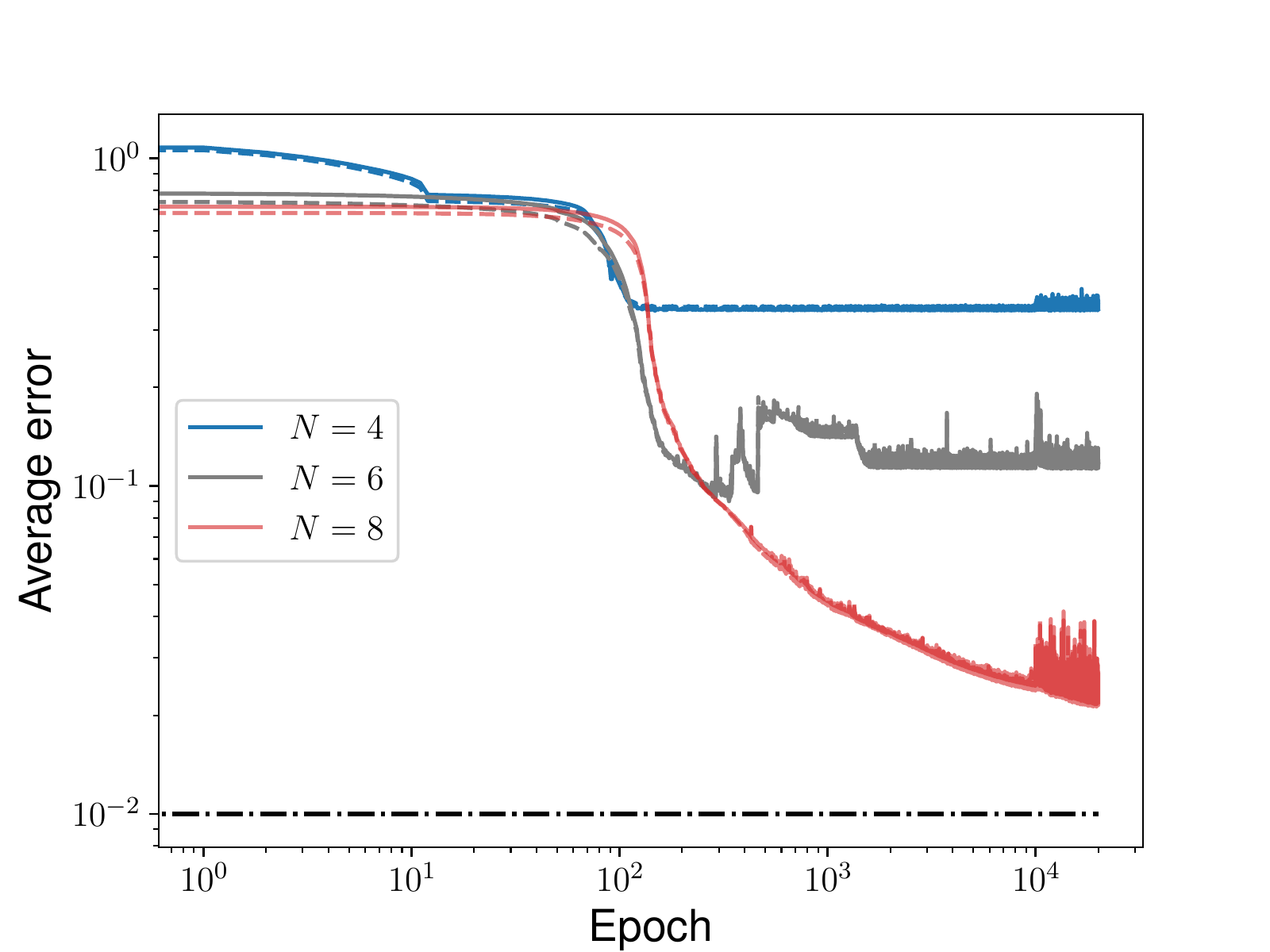}}
	\caption{Training histories of DMN for (a) UD composite and (b) woven composite. The histories of the average training and test errors are denoted by solid and dashed lines, respectively.}
	\label{fig:cfrpHist}
\end{figure}

In the offline stage, the microscale UD RVE and the mesoscale woven RVE are trained separately to find their own DMN representations. Histories of the average training and test errors of UD and woven DMN are given in Figure \ref{fig:cfrpHist}.  For all the cases, we doubled the learning rate of SGD algorithm after 10000 epochs to accelerate the training speed, which also causes more oscillations in the error histories. According to our study, the bottom-layer nodes in a DMN of UD composite is more often deactivated than the other RVE examples during the training, mainly due to its strongly anisotropic microstructure.  A network with $N=4$ may unintentionally lose all its nodes for the fiber phase. Therefore, we choose to show the DMN results of UD composite from $N=5$ to 9. After 20000 epochs of training, errors for $N=7$ and 9 both reduced to be less than 1\%. 

Finding the optimum DMN that well represents the woven RVE appears to be a more challenging task. The lowest training error we achieved is 2.19\% with $N=8$, and the training process tends to be saturated in the end. This is somehow expected since the morphology of woven microstructure is more complex than the other cases. For example, it has penetrating phases along two different directions, as well as the entanglement of yarns. The current level of accuracy achieved by DMN is satisfactory in many practical applications, however, further reduction of errors may require more advanced learning techniques.
\begin{figure} [!t]
	\centering
	\graphicspath{{Figures/}}
	\subfigure[$N=5, N_a =4, vf_1 =0.303$]{\includegraphics[clip=true,trim = 0.0cm 0.0cm 1.0cm 0.5cm,width=0.28\textwidth]{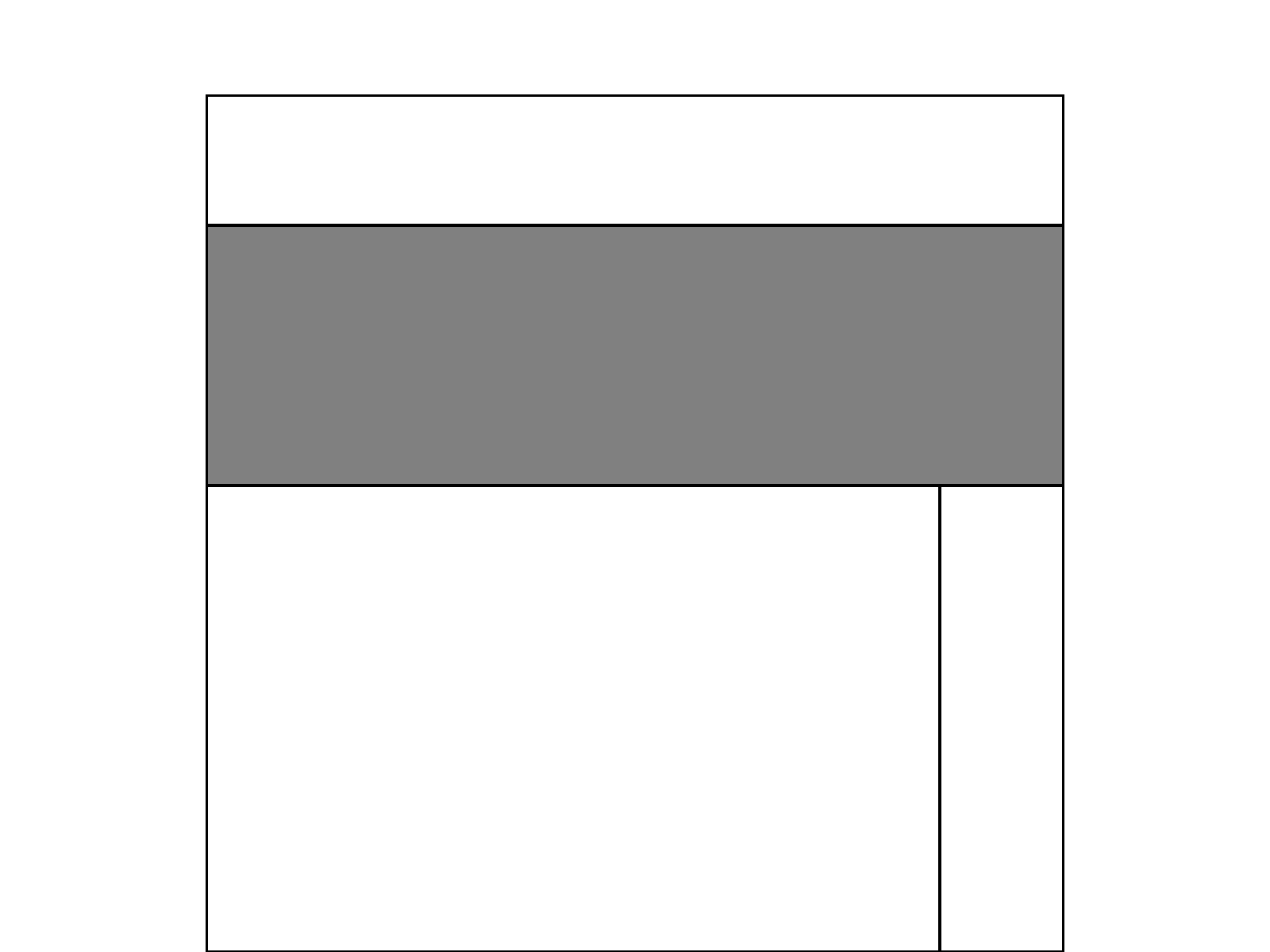}}
	\subfigure[$N=7, N_a =14, vf_1 =0.294$]{\includegraphics[clip=true,trim = 0.0cm 0.0cm 1.0cm 0.5cm,width=0.28\textwidth]{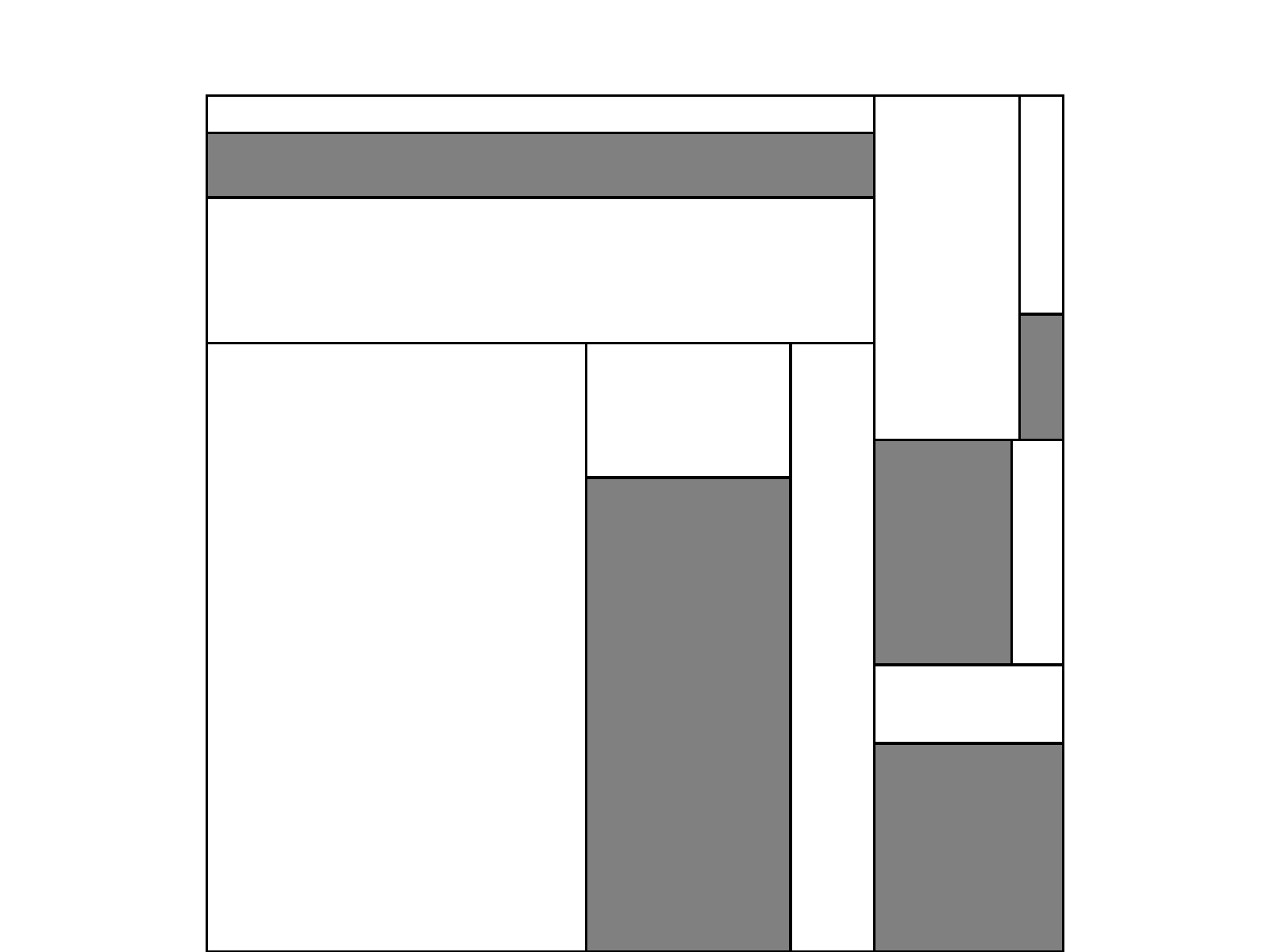}}
	\subfigure[$N=9, N_a =60, vf_1 =0.295$]{\includegraphics[clip=true,trim = 0.0cm 0.0cm 1.0cm 0.5cm,width=0.28\textwidth]{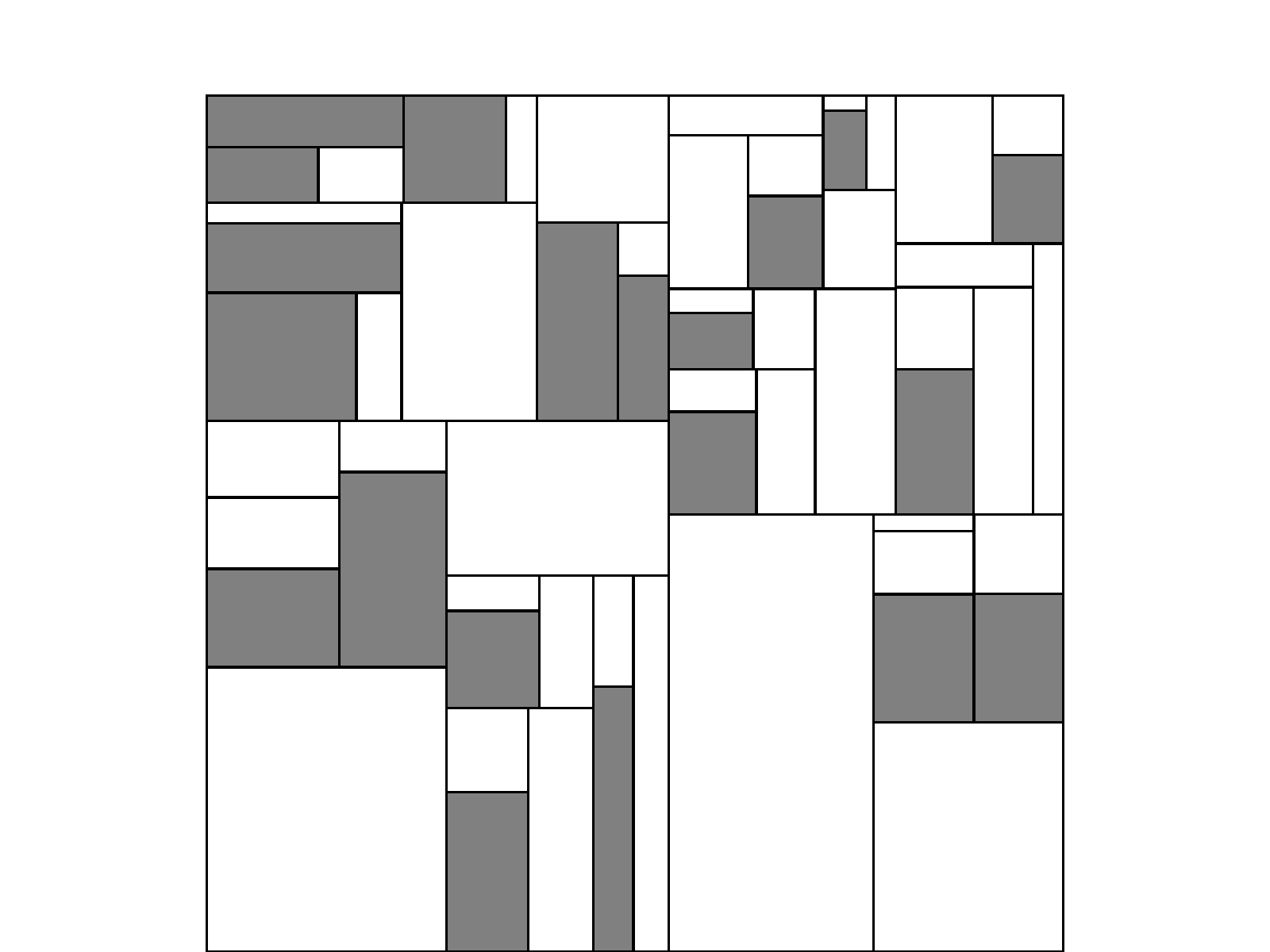}}
	\subfigure[$N=4, N_a =3, vf_1 =0.301$]{\includegraphics[clip=true,trim = 0.0cm 0.0cm 1.0cm 0.5cm,width=0.28\textwidth]{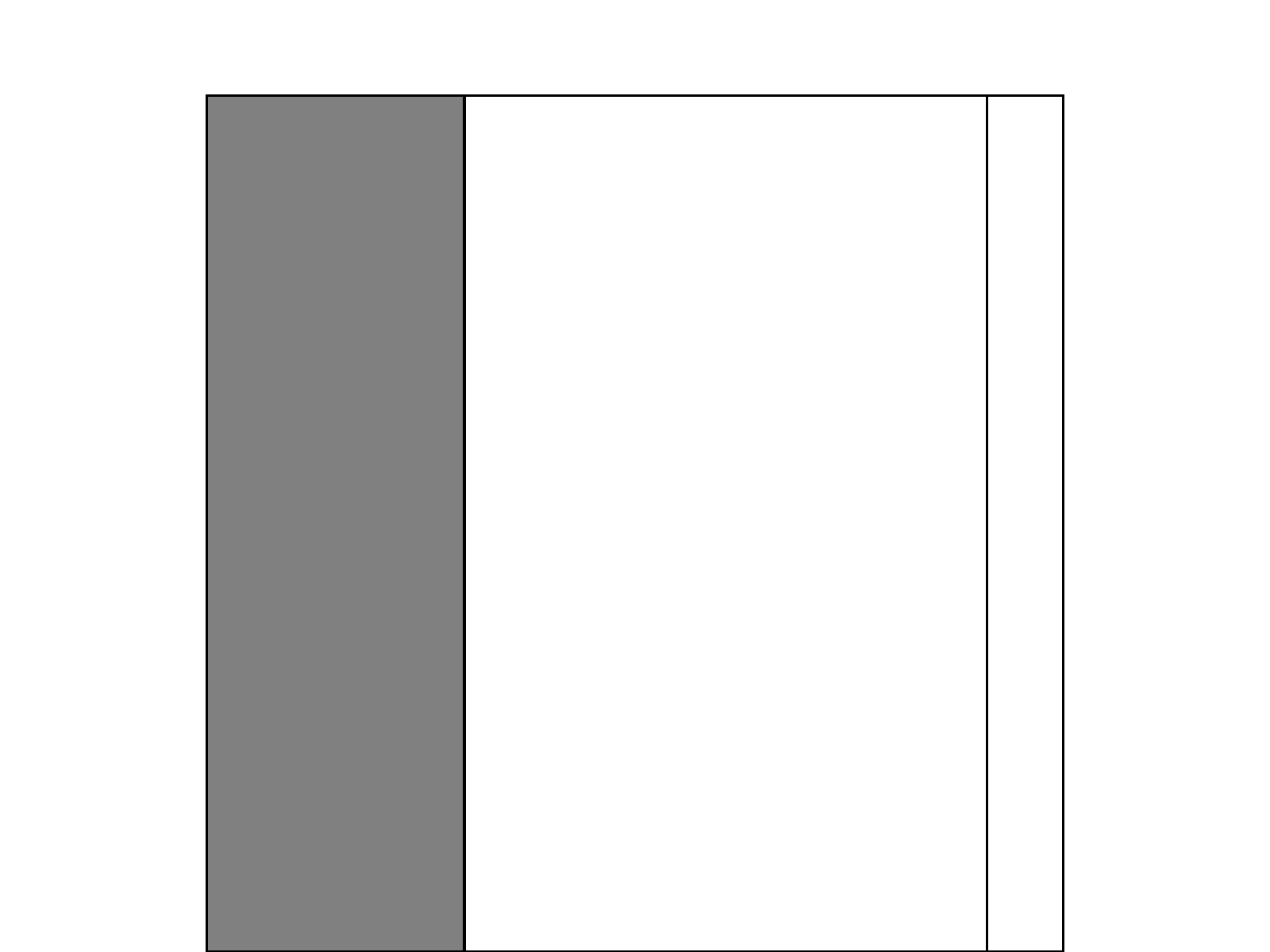}}
	\subfigure[$N=6, N_a =7, vf_1 =0.348$]{\includegraphics[clip=true,trim = 0.0cm 0.0cm 1.0cm 0.5cm,width=0.28\textwidth]{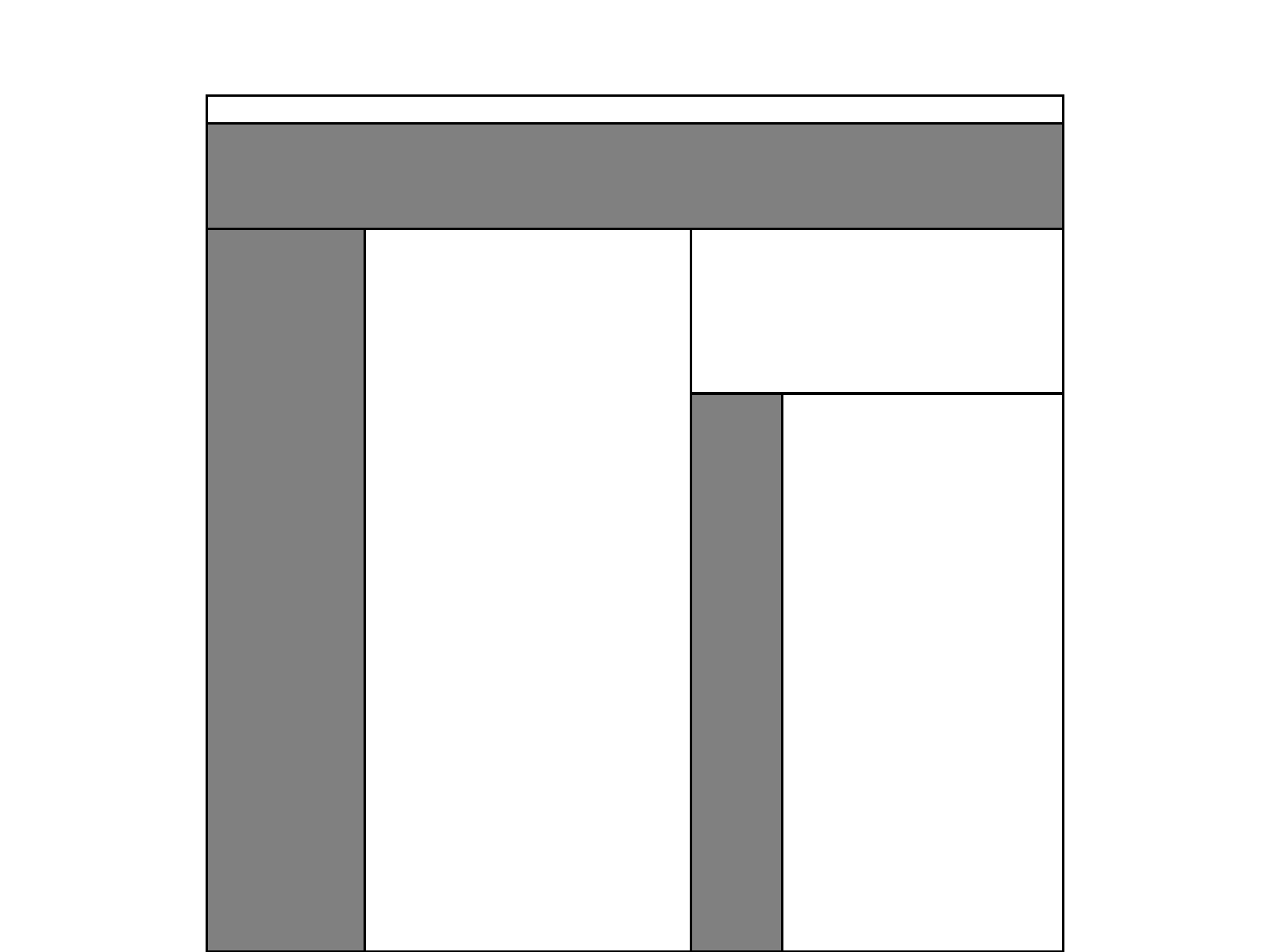}}
	\subfigure[$N=8, N_a =38, vf_1 =0.420$]{\includegraphics[clip=true,trim = 0.0cm 0.0cm 1.0cm 0.5cm,width=0.28\textwidth]{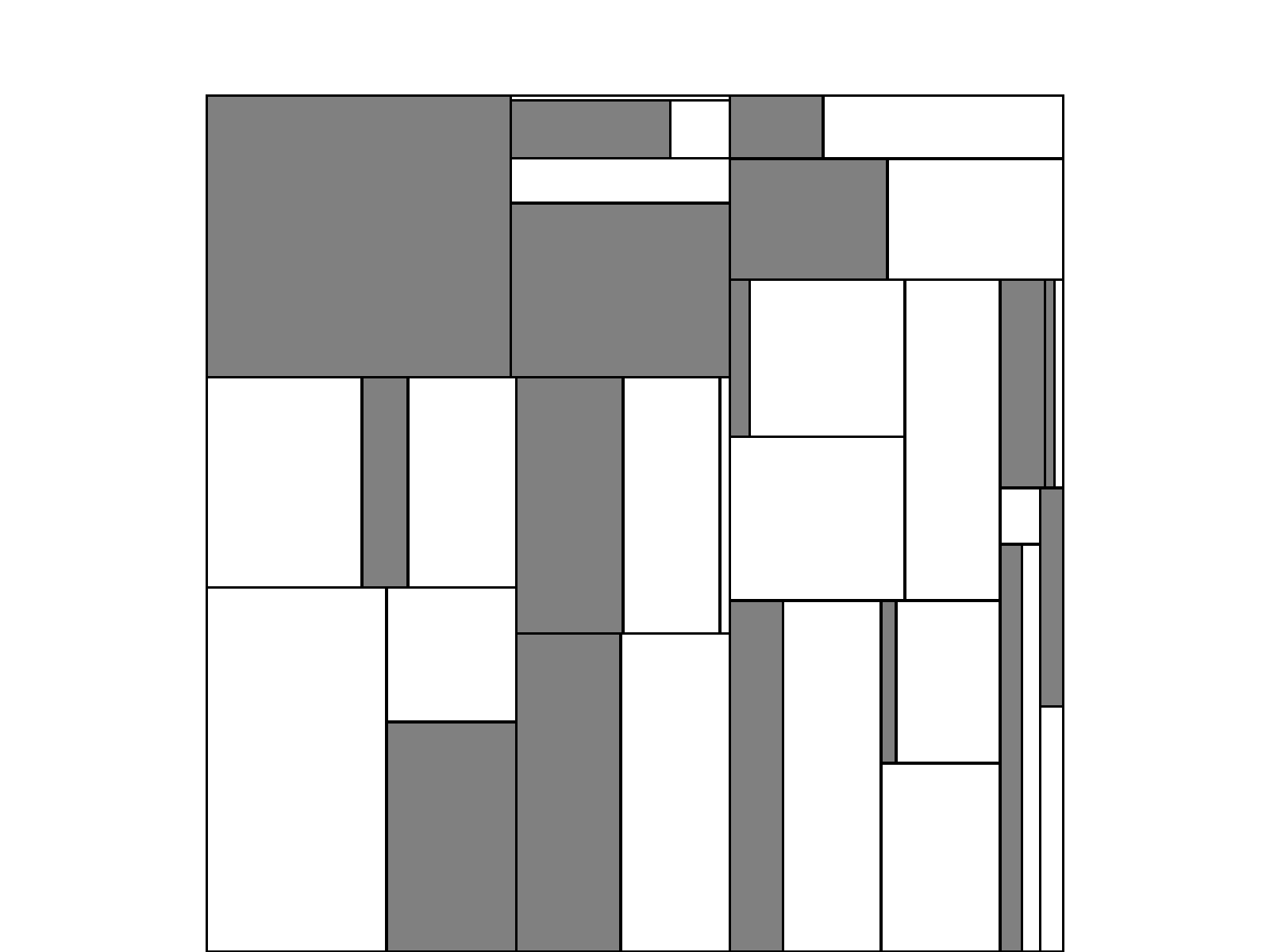}}
	\caption{Treemaps of DMN for CFRP composites. For the UD composite, the network depths $N$ are 5(a), 7(b), 9(c). For the woven composite, the network depths $N$ are 4(d), 6(e), 8(f). The number of active nodes in the bottom layer $N_a$ and the predicted volume fraction of phase 1 $vf_1$ are also shown under each plot.}
	\label{fig:udTree}
\end{figure}

Table \ref{table:cfrp-ud} summarizes the training and test errors for UD and woven RVEs, as well as the volume fractions of phase 1 predicted by the corresponding DMNs with various depths. A linear FE model with 5252 nodes is introduced for the woven composite as a reference, and its accuracy is evaluated on the test dataset. The average test error $\bar{e}^{te}$ and maximum test error of the linear FE model are 6.12\% and 28.3\%, respectively. Measured by $\bar{e}^{te}$, its performance is close to the DMN model with $N=7$ and $N_a=16$. The computational time of the linear FE model for woven composite will be given in Section \ref{sec:time}.
Treemaps are provided in Figure \ref{fig:udTree}. Again, we can conclude that for both UD and woven composites, a DMN with sufficient number of layers is able to extract the volume fraction accurately from the training dataset with pure mechanical information. 

\begin{table}[!t]
	\captionabove{Training results of CFRP RVEs. Average training error $\bar{e}^{tr}$, average test error $\bar{e}^{tr}$, maximum test error and predicted volume fraction $vf_1$  are provided for each DMN. Test errors of the linear FEM model for woven RVE are also shown.} 
	\centering 
	\label{table:cfrp-ud} 
	{\tabulinesep=1.0mm
		\begin{tabu}{c c c c c c c} 
			\hline 
			&&Epochs & Training $\bar{e}^{tr}$ & Test $\bar{e}^{te}$ & Maximum $e_s^{te}$ &  $vf_1$\\
			\hline
			&$N=5$&20000&11.1\%&10.5\%&47.2\%&0.303 (\texttt{+}3.06\%)\\
			UD &$N=7$&20000&0.79\%&0.76\%&3.46\%&0.294 (\texttt{+}0.00\%) \\
			&$N=9$&40000&0.23\%&0.25\%&1.34\%&0.295 (\texttt{+}0.00\%) \\
			\hline
			&$N=4$&20000&34.6\%&35.6\%&83.9\%&0.301 (-27.3\%)\\ 
			&$N=6$ &20000&11.6\%&11.5\%&38.1\%&0.348 (-15.9\%) \\
			Woven&$N=7$ &20000&3.72\%&3.48\%&10.1\%&0.424 (\texttt{+}2.42\%) \\
			&$N=8$ &20000&2.19\%&2.15\%&7.82\%&0.420 (\texttt{+}1.45\%) \\
			&Linear FEM &\textbackslash&\textbackslash&6.12\%&28.3\%&\textbackslash \\
			\hline
	\end{tabu}}
\end{table}

\subsubsection{Online extrapolation}
Ideally, a full three-scale homogenization should be performed by linking every material point in the mesoscale woven RVE with a microscale UD RVE model. However, this is not feasible with the DNS models described above due to the high computational cost. On the other hand, solving the woven RVE implicitly also requires the full information of the overall stiffness tensor of the UD RVE, while the standard FEM solver only provides the values of stress and strain. Therefore, the DNS models of UD and woven RVEs are performed separately, and the obtained two-scale DNS homogenization results will be used to validate our DMN models.

For the UD composite, the fiber phase is considered as an orthotropic elastic material with high strength in the fiber direction. The matrix phase is modeled as an elasto-plastic epoxy material with an exponential hardening law. Its yield stress $\sigma_Y$ is determined by the hardening law as a function of the effective plastic strain $\bar{\varepsilon}^{pl}$, which is a monotonically increasing internal state variable of the plastic material during the deformation. An exponential hardening law is assumed,
\begin{equation}
\sigma^Y(\bar{\varepsilon}^{pl}) = -a_2\exp(-a_1\bar{\varepsilon}^{pl})+a_3,
\end{equation}
where $a_3$ is the ultimate yield stress for large effective plastic strain, $(a_3-a_2)$ represents the yield strength and $a_1$ is a dimensionless hardening constant. An isotropic J2 yield surface is further assumed,
\begin{equation}
f = \bar{\sigma}-\sigma^Y(\bar{\varepsilon}^{pl}) \leq 0,
\end{equation}
where $\bar{\sigma}$ is the von Mises equivalent stress. We choose to formulate the plastic law based on the Green-Lagrange strain and second Piola-Kirchhoff stress. Uniaxial responses of the fiber and epoxy materials are shown in Figure \ref{fig: cfrp} (b). In the two-scale homogenization of woven RVE, the matrix phase shares the same epoxy material as the UD composite. The yarn phase is assumed to be an orthotropic elastic material, and its elastic constants are determined by homogenizing the UD composite. All the material parameters used in the online stage can be found in Table \ref{table:udpara} \footnote{The hardening parameters in the published version were found to be incorrect. The parameters corresponding to the numerical examples are updated.}.
\begin{figure}[t!]
	\centering
	\graphicspath{{Figures/}}
	\subfigure[$\sigma_{11}$ vs. $\varepsilon_{11}$]{\includegraphics[clip=true,trim = 0.0cm 0.0cm 1.0cm 0.5cm,width=0.44\textwidth]{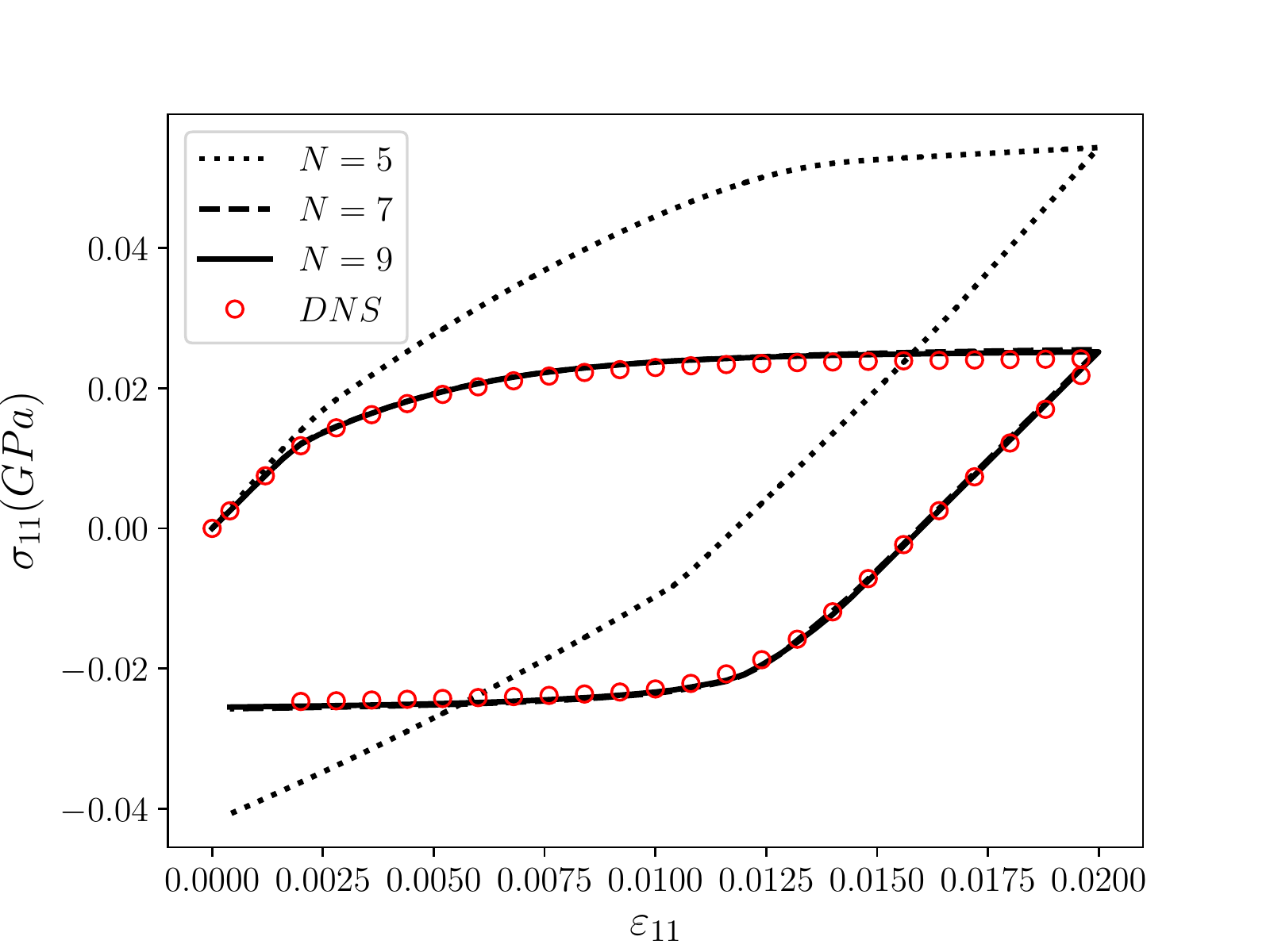}}
	\subfigure[$\sigma_{33}$ vs. $\varepsilon_{33}$]{\includegraphics[clip=true,trim = 0.0cm 0.0cm 1.0cm 0.5cm,width=0.44\textwidth]{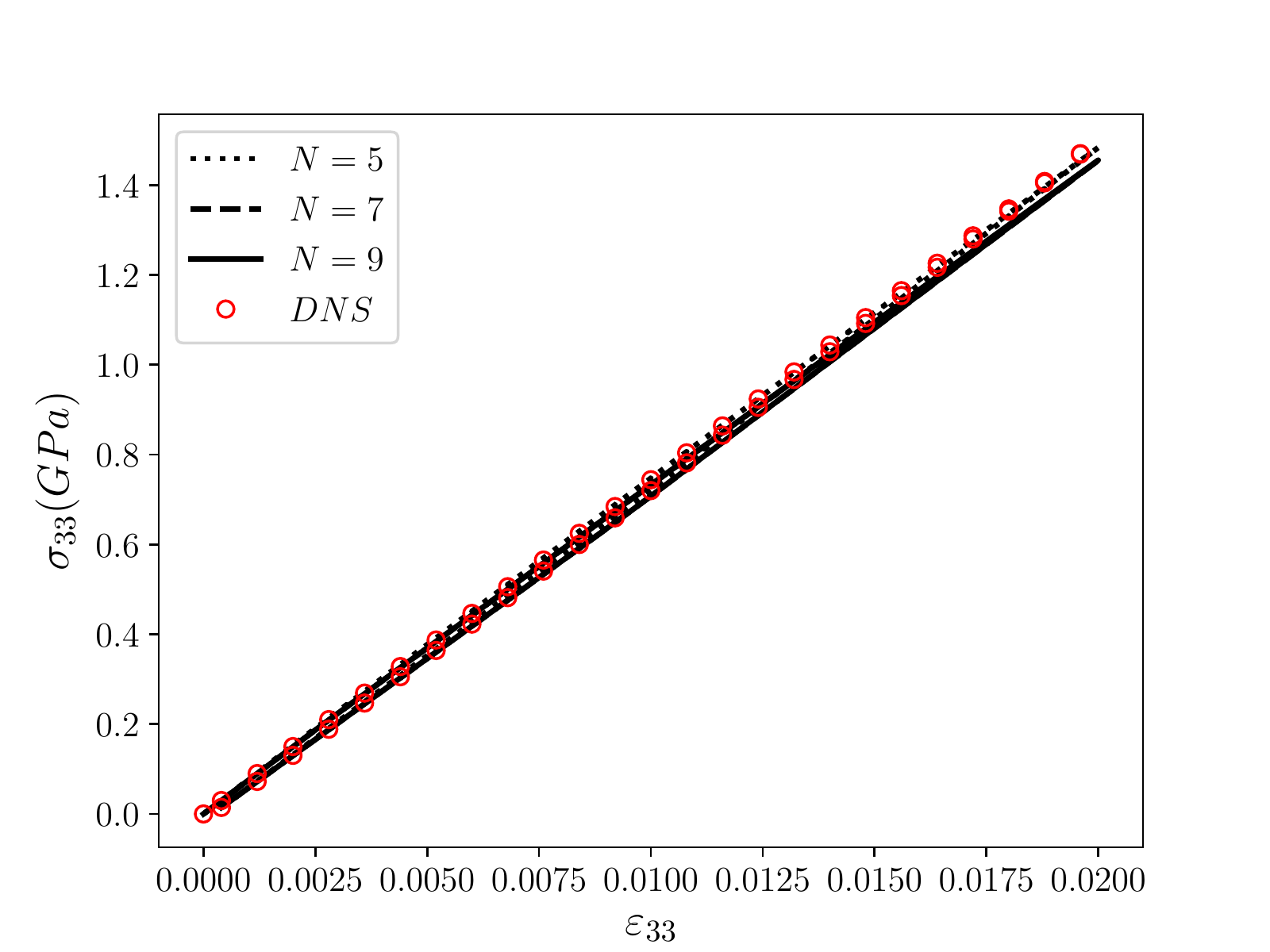}}
	\subfigure[$\sigma_{12}$ vs. $\varepsilon_{12}$]{\includegraphics[clip=true,trim = 0.0cm 0.0cm 1.0cm 0.5cm,width=0.44\textwidth]{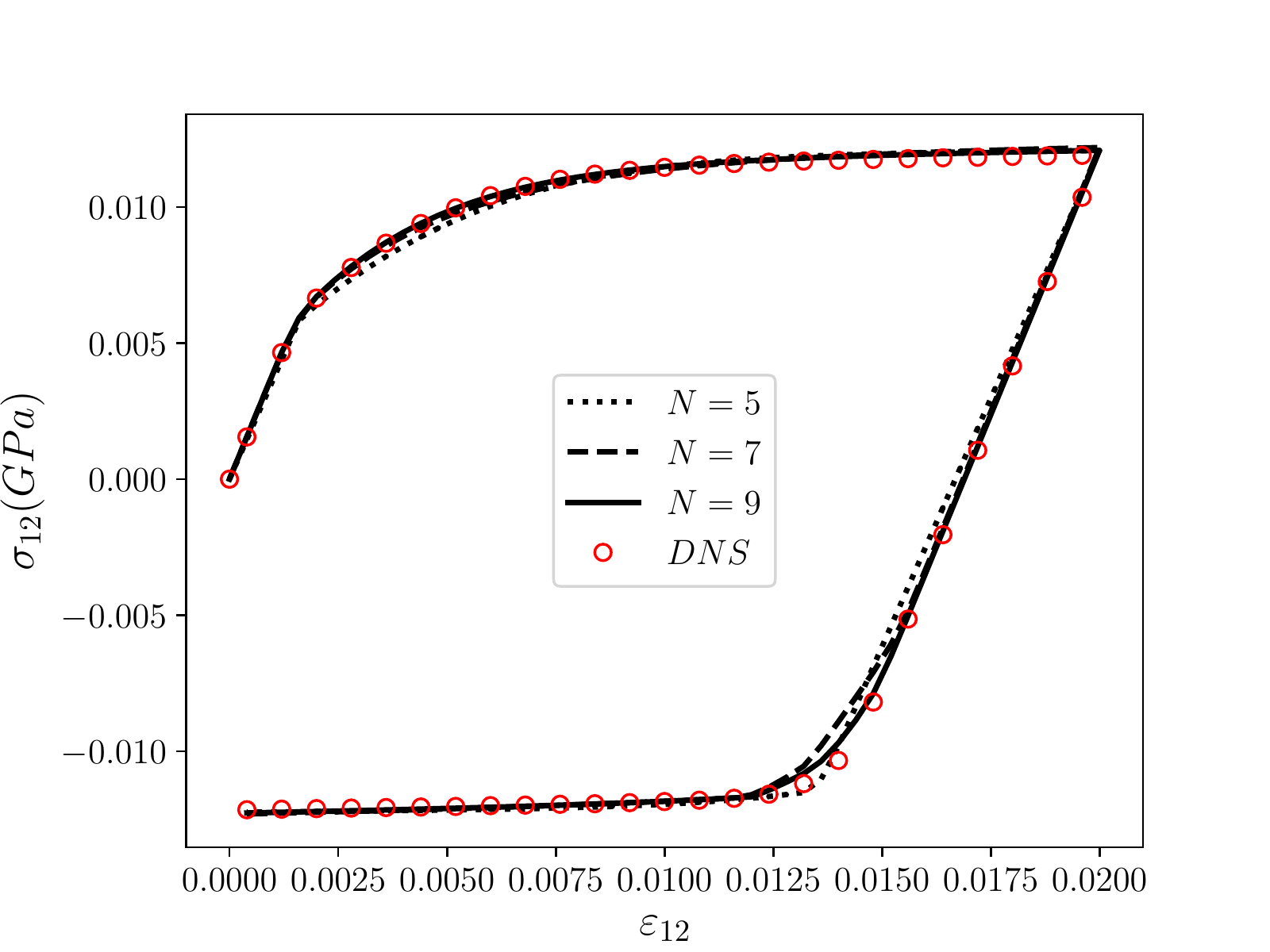}}
	\subfigure[$\sigma_{23}$ vs. $\varepsilon_{23}$]{\includegraphics[clip=true,trim = 0.0cm 0.0cm 1.0cm 0.5cm,width=0.44\textwidth]{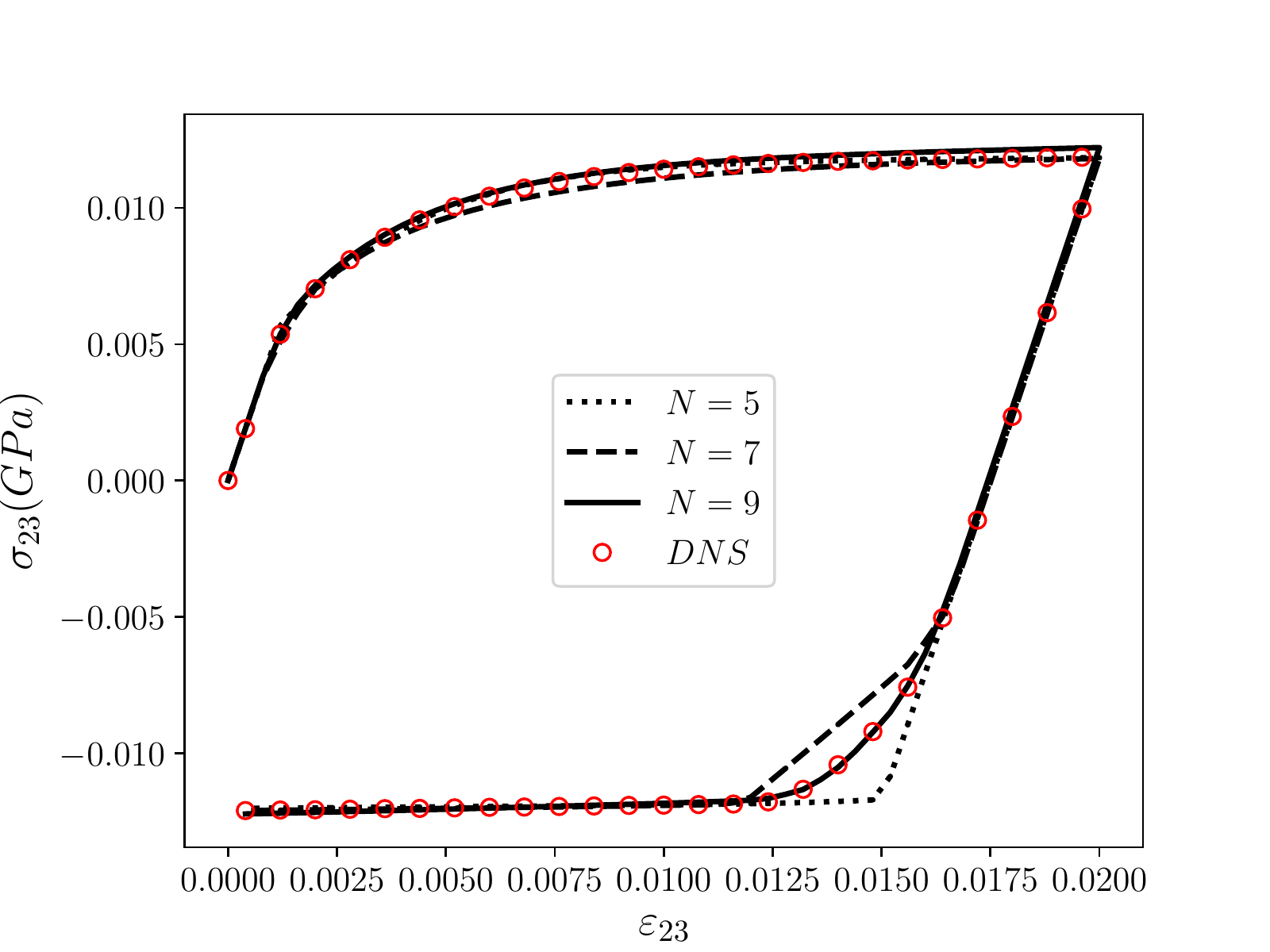}}
	\caption{Loading-unloading stress-strain curves predicted by DNS and DMN for UD composite. Four loading directions were considered, including (a,b) uniaxial tensions and (c,d) shears. The network depths are $N=$5 (dotted), 7 (dashed) and 9 (solid).}
	\label{fig:ud}
\end{figure} 
\begin{table}[htb!]
	\captionabove{Material parameters for the carbon fiber reinforced polymer (CFRP) composite \cite{liu2017reduced}.} 
	\centering 
	\label{table:udpara} 
	{\tabulinesep=1.0mm
		\begin{tabu}{c c c c c c} 
			\hline\hline
			\multirow{4}{*}{Carbon fiber}& $E_1$ (GPa) & $E_2$ (GPa) & $E_3$ (GPa) & $G_{12}$ (GPa) & $G_{13}$ (GPa)\\
			 & 19.8 & 19.8 & 245.0 & 5.9 & 29.2 \\ 
			\cline{2-6}
			&$G_{23}$ (GPa) & $\nu_{12}$ & $\nu_{13}$ & $\nu_{23}$ & \\
			&29.2 & 0.67 & 0.28 & 0.28 &  \\
			\hline\hline
			\multirow{2}{*}{Epoxy} & $E_m$ (GPa) & $\nu_m$ & $a_1$ & $a_2$ (MPa) & $a_3$ (MPa) \\ 
			&3.8 & 0.387 & 200 & 10 & 20 \\
			\hline\hline
			\multirow{4}{*}{Yarn (elastic)} &$E_1$ (GPa) & $E_2$ (GPa) & $E_3$ (GPa) & $G_{12}$ (GPa) & $G_{13}$ (GPa)\\
			&10.2 & 10.2 & 78.8 & 1.95 & 2.39 \\ 
			\cline{2-6}
			&$G_{23}$ (GPa) & $\nu_{12}$ & $\nu_{13}$ & $\nu_{23}$ & \\
			&2.39 & 0.60 & 0.35 & 0.35 &  \\
			\hline
	\end{tabu}}
\end{table}
\begin{figure} [!t]
	\centering
	\graphicspath{{Figures/}}
	\subfigure[$\sigma_{11}$ vs. $\varepsilon_{11}$]{\includegraphics[clip=true,trim = 0.0cm 0.0cm 1.0cm 0.5cm,width=0.44\textwidth]{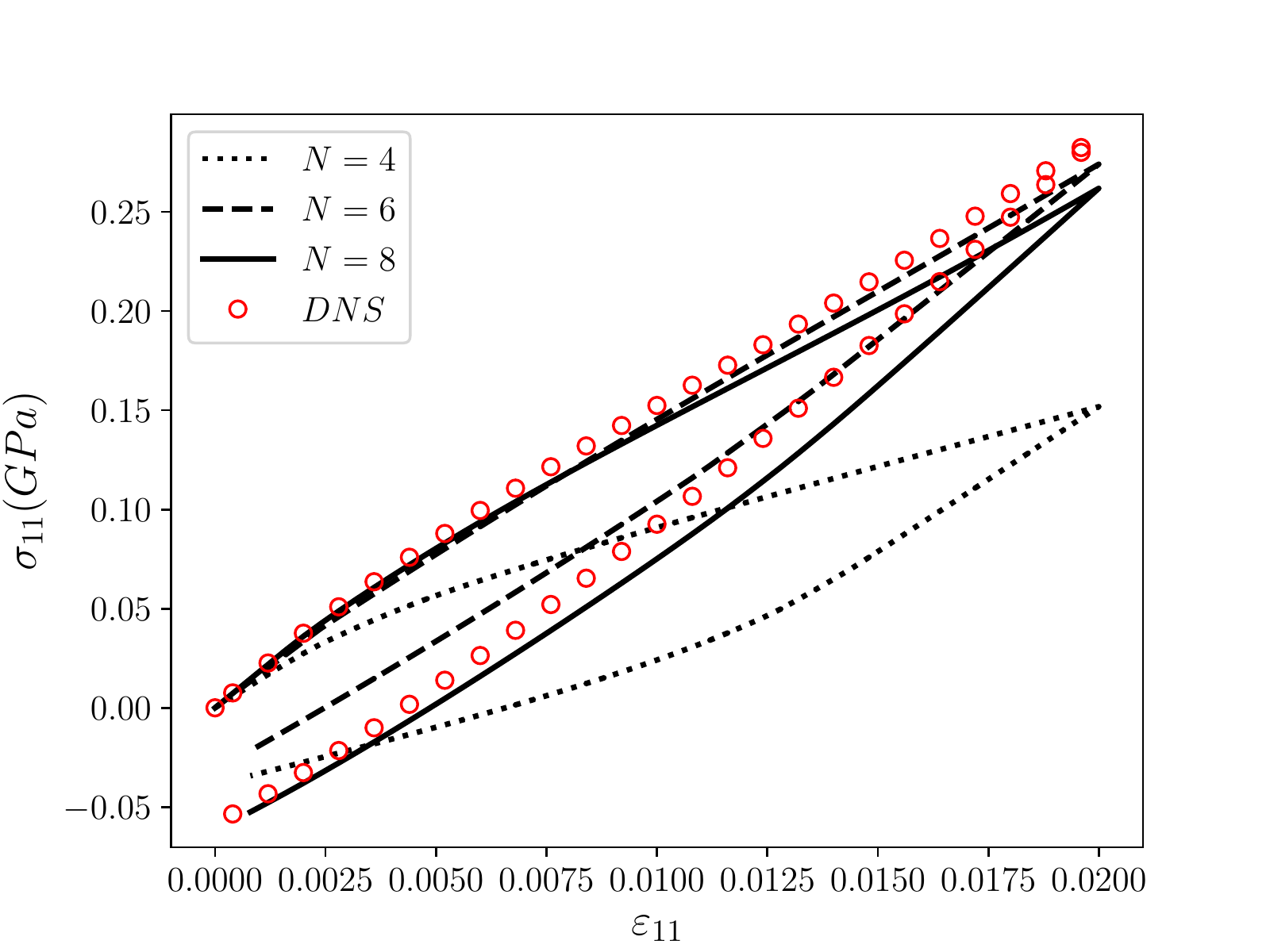}}
	\subfigure[$\sigma_{33}$ vs. $\varepsilon_{33}$]{\includegraphics[clip=true,trim = 0.0cm 0.0cm 1.0cm 0.5cm,width=0.44\textwidth]{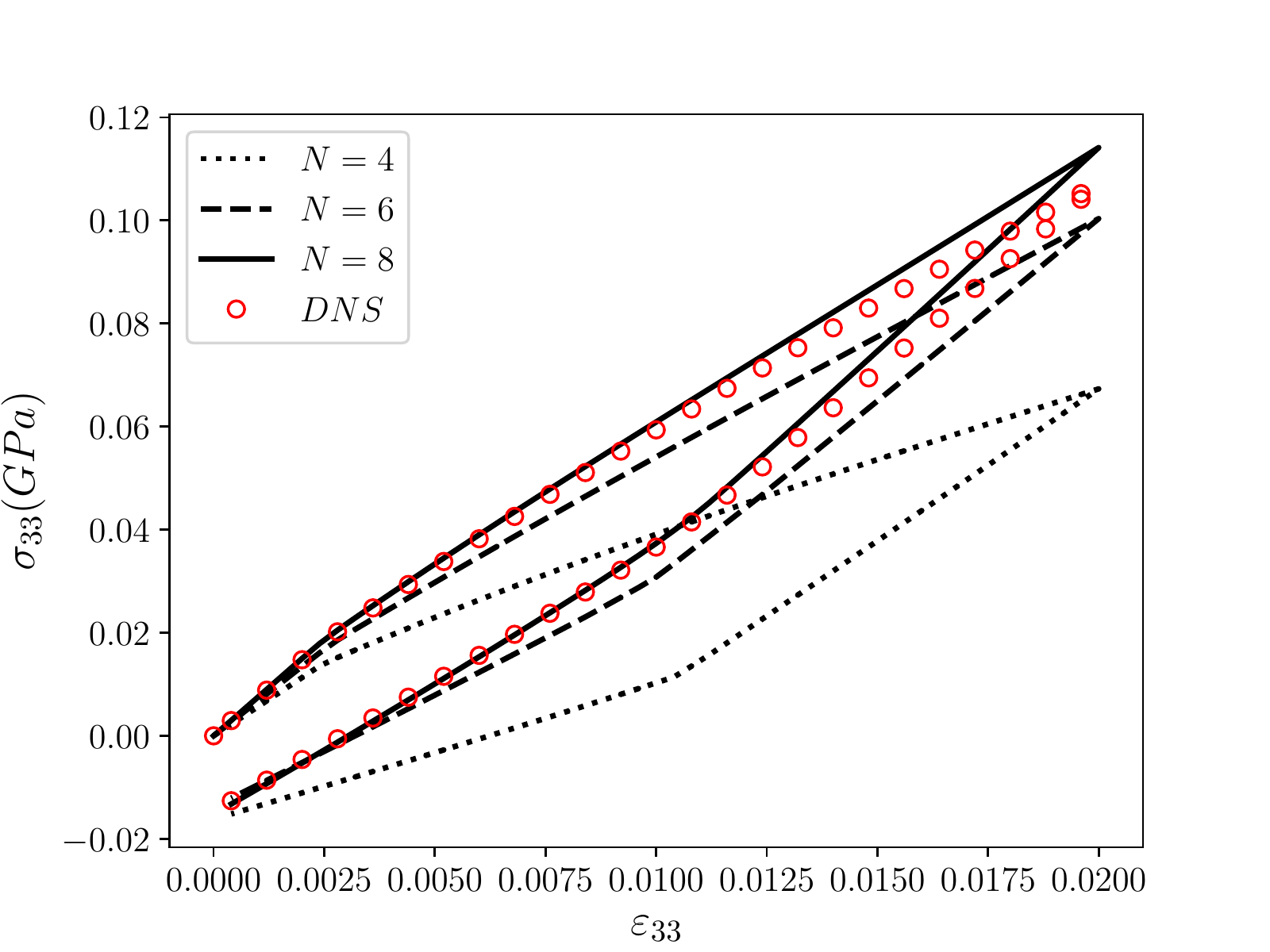}}
	\subfigure[$\sigma_{12}$ vs. $\varepsilon_{12}$]{\includegraphics[clip=true,trim = 0.0cm 0.0cm 1.0cm 0.5cm,width=0.44\textwidth]{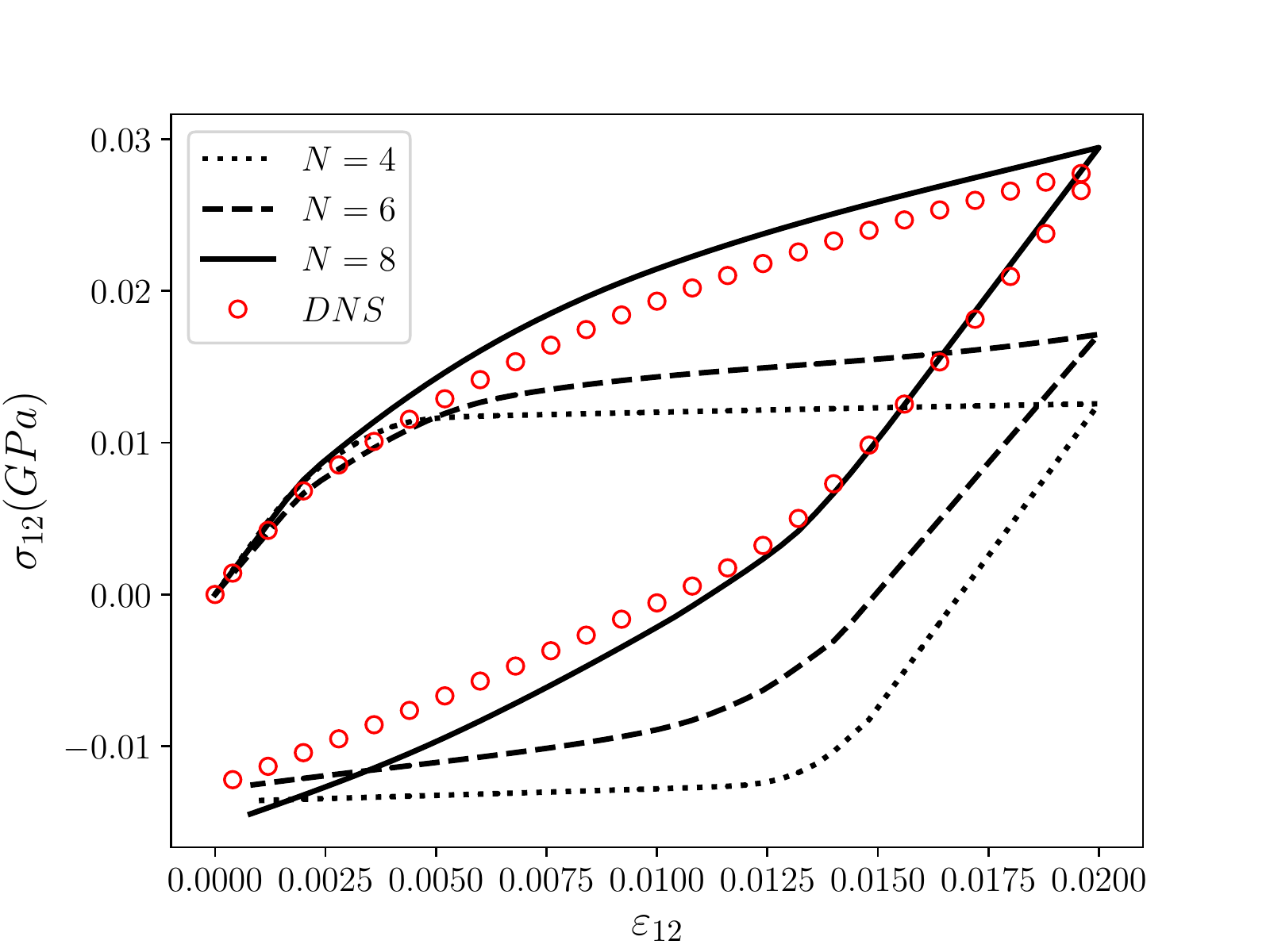}}
	\subfigure[$\sigma_{23}$ vs. $\varepsilon_{23}$]{\includegraphics[clip=true,trim = 0.0cm 0.0cm 1.0cm 0.5cm,width=0.44\textwidth]{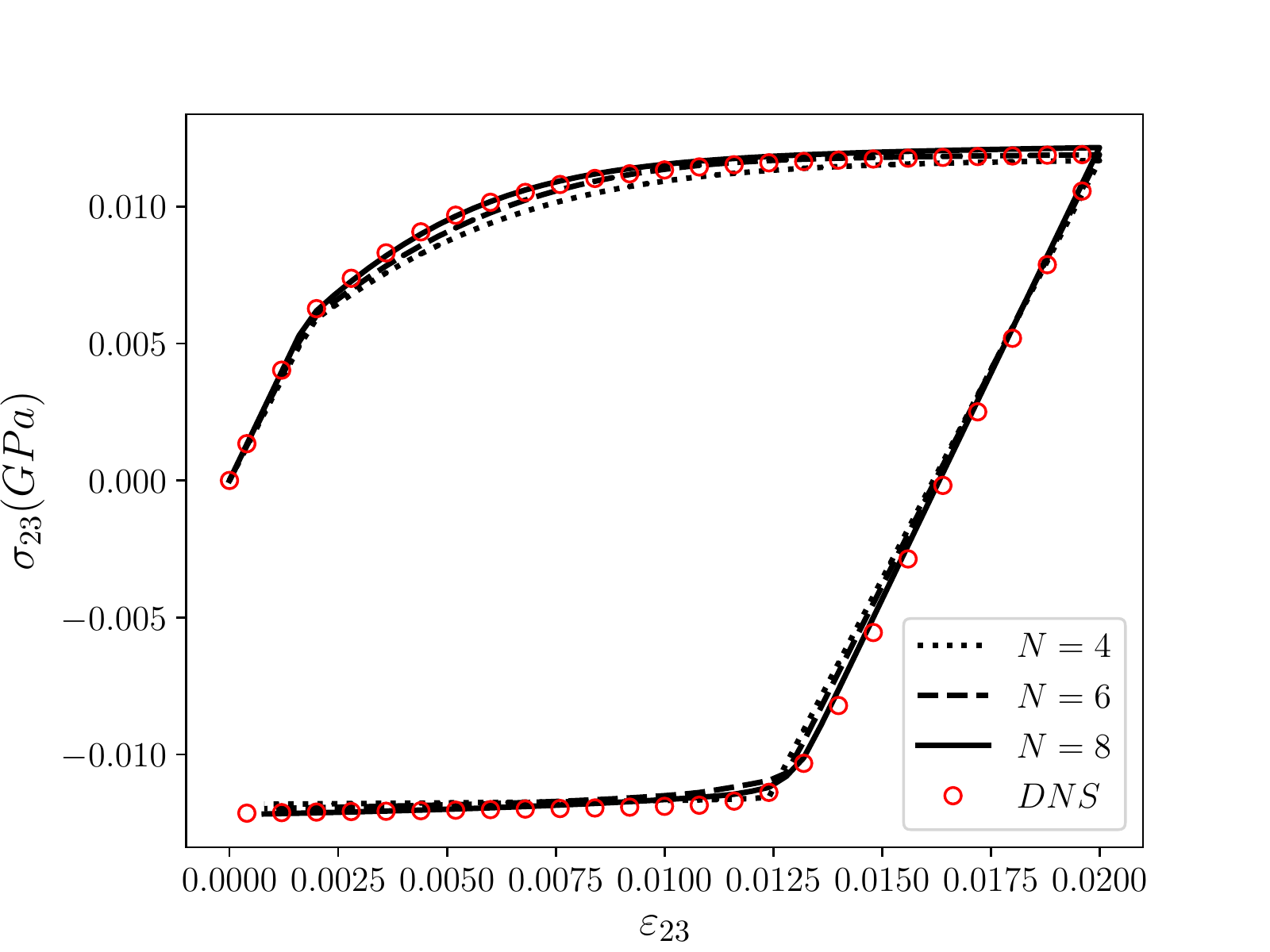}}
	\caption{Loading-unloading stress-strain curves predicted by DNS and DMN for woven composite. Four loading directions were considered: (a,b) uniaxial tensions and (c,d) shears. The network depths are $N=$4 (dotted), 6 (dashed) and 8 (solid).}
	\label{fig:woven}
\end{figure}

The stress-strain curves predicted by DNS and DMN for the UD composite are presented in Figure \ref{fig:ud}. Four loading-unloading conditions were considered: 11, 33 for uniaxial tension and 12, 23 for shear. We can see that the material responses in 33 direction are almost elastic, since the carbon fibers go straightly through the matrix and its Young's modulus ($E_3=245.0$GPa) in this direction is 65 times larger than the one of epoxy matrix ($E_m=3.8$GPa). On the other hand, the overall stress-strain responses in other loading directions are dominated by the hardening of the epoxy matrix. For all these cases, the loading-unloading behaviors can be well captured by DMNs with $N \geq 7$.
\begin{figure} [!t]
	\centering
	\graphicspath{{Figures/}}
	\subfigure[Straightening of a yarn during deformation.]{\includegraphics[clip=true,trim = 4.0cm 0.5cm 6.0cm 0.0cm,width=0.44\textwidth]{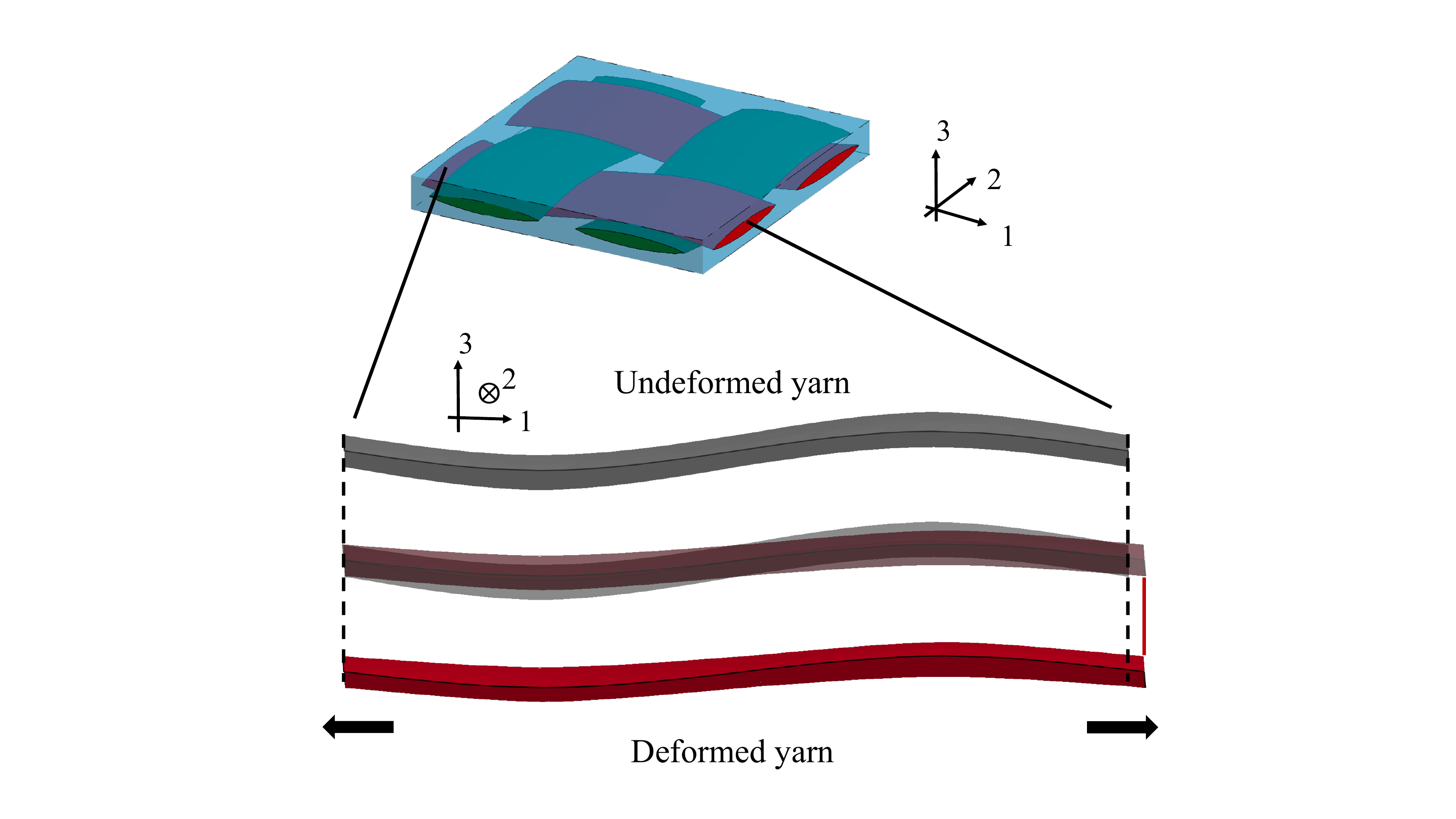}}
	\subfigure[Predictions with or without geometric nonlinearity.]{\includegraphics[clip=true,trim = 0.0cm 0.0cm 1.0cm 0.5cm,width=0.44\textwidth]{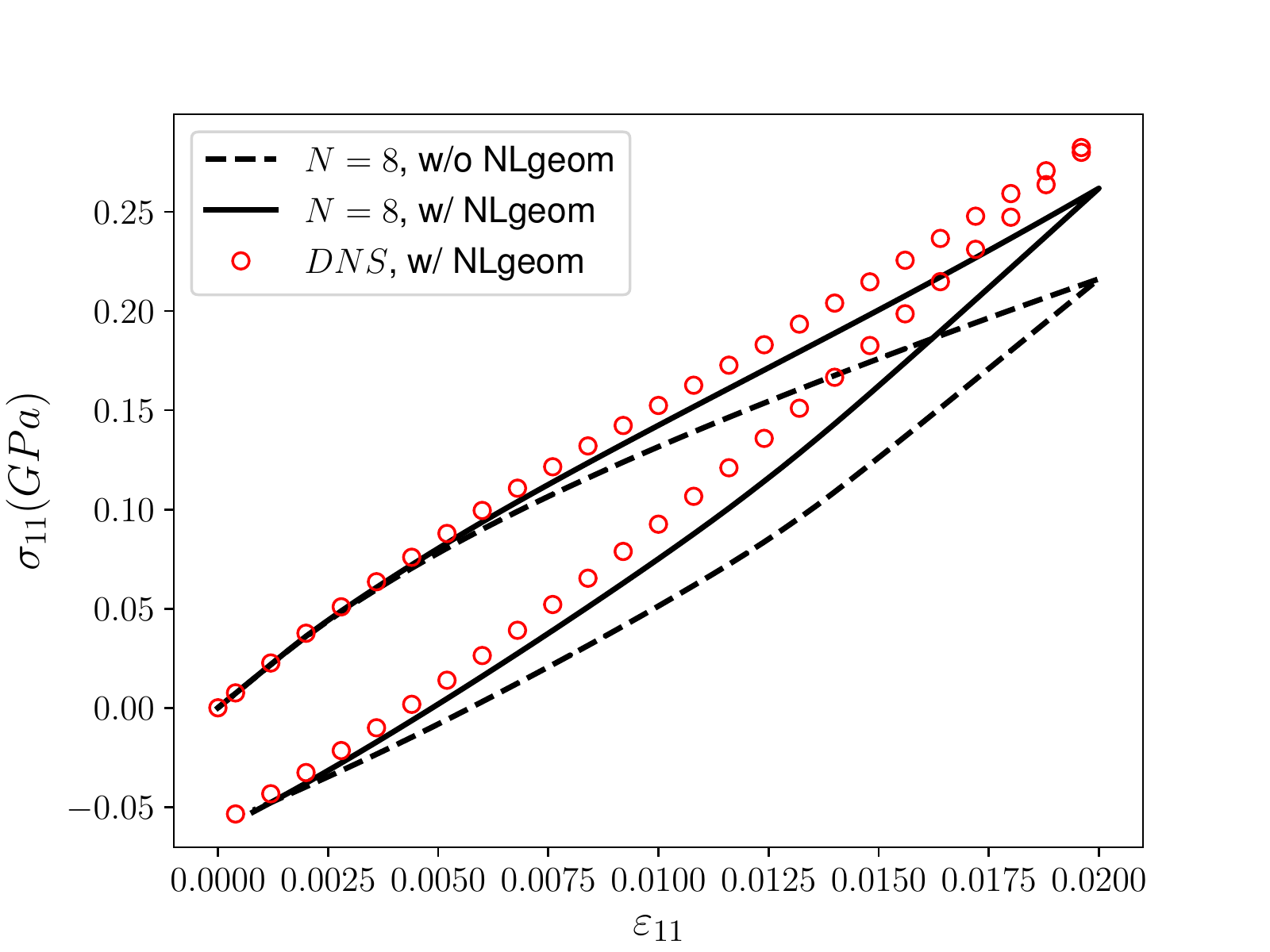}}
	\caption{Stiffening effect of woven composite under uniaxial tension in 11 direction.}
	\label{fig:wovenComp}
\end{figure}

The stress-strain curves predicted by DNS and DMN for the woven composite are presented in Figure \ref{fig:woven}. Similarly, four loading-unloading conditions were considered. Figure \ref{fig:woven} (a) shows the in-plane tension results in 11 direction. The overall stress-strain curve first slightly yields due to matrix hardening, and more interestingly, a stiffening effect is observed after the strain $\varepsilon_{11}$ reaches 1.5\%. As illustrated in Figure \ref{fig:wovenComp} (a), the anisotropic yarn is straightened during the deformation, making it stiffer in the loading direction. Although the overall deformation is small ($<2.0\%$), some segments within the yarn experience relatively large local rotations, which can be captured only if a model takes the geometric nonlinearity into account. To validate this hypothesis, we also loaded the DMN model for $N=8$ based on the small-strain formulation without geometric nonlinearity, and as expected, the stiffening effect was not reproduced as in the DNS or DMN based on finite-strain formulation (see Figure \ref{fig:wovenComp} (b)).

For the out-of-plane tension in 33 direction, the overall responses still show little hardening effect due to the yarn confinements in 11 and 22 directions. By looking at Figure \ref{fig:woven} (d), we can also conclude that the out-of-plane shear in 23 direction is dominated by the epoxy matrix since the overall material is almost perfectly plastic in the end. On the other hand, the in-plane shear behavior in 12 direction is influenced more evenly by the yarn and matrix phases, as the DNS stress-strain curve shown in \ref{fig:woven} (c) has a nontrivial hardening modulus. For all the cases, the predictions from DMN with $N=8$ match the DNS results very well, indicating that our reduced order model preserves the essential microstructural interactions in the original DNS model. One can also see that it will be very challenging to manually develop an empirical material model to accurately capture the complex responses of the woven composite. However, with the help of machine learning techniques, DMN can automatically find the optimum reduced description of the woven composite by learning the hidden physics in the training dataset.
\begin{figure}[!t]
	\centering
	\graphicspath{{Figures/}}
	\includegraphics[clip=true,trim = 1.5cm 4.0cm 1.5cm 4.0cm,width=0.85\textwidth]{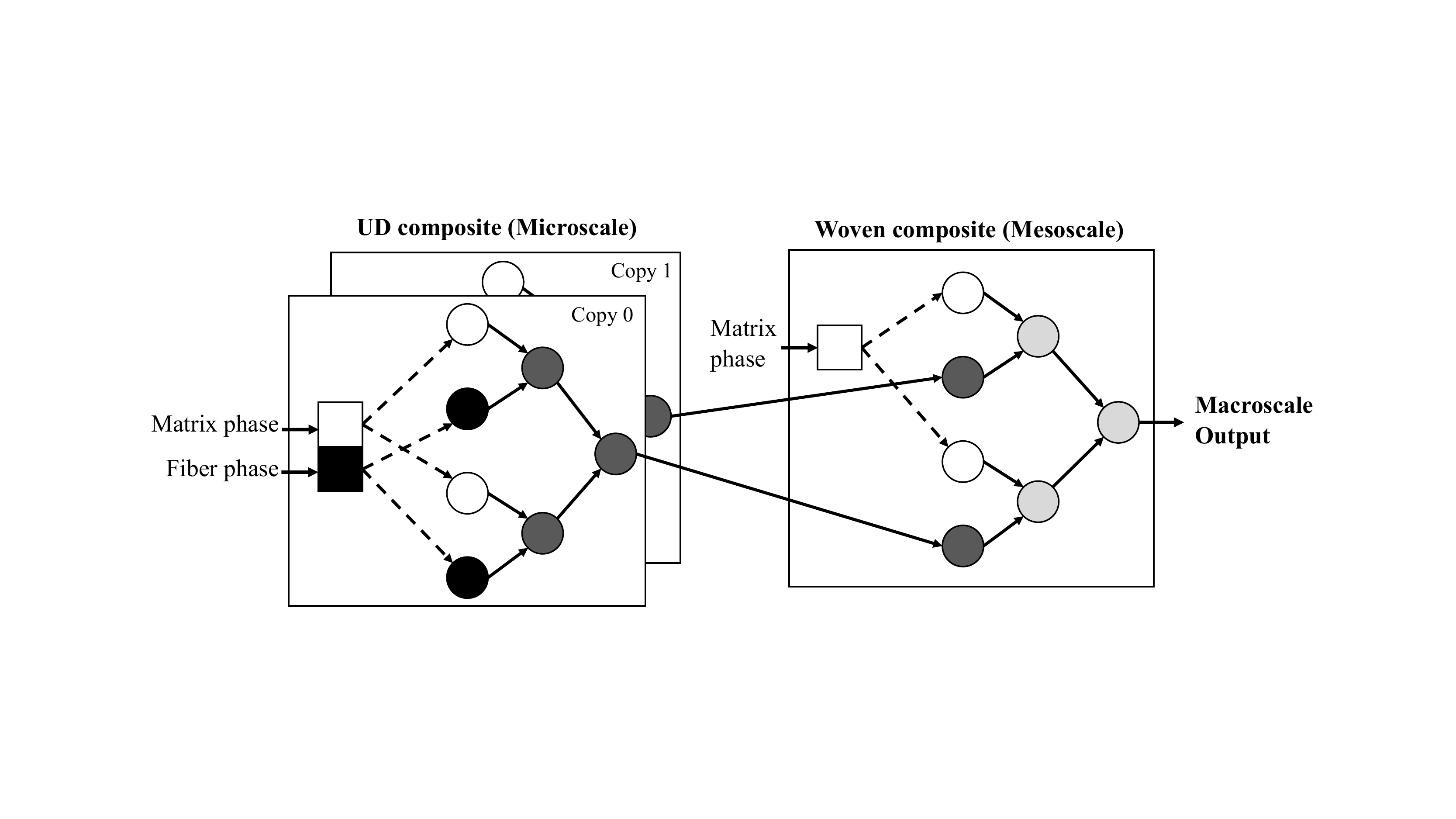}
	\caption{An illustration of three-scale homogenization in CFRP through the concatenation of networks. The material responses of the yarn phase in the mesoscale woven composite are given by the homogenization of the microscale UD composite.}
	\label{fig: threescale}
\end{figure}

Finally, we demonstrate a three-scale homogenization procedure for CFRP composites based on DMN. As shown in Figure \ref{fig: threescale}, it can be realized through the concatenation of networks: Link every active bottom-layer node of the yarn phase in the woven composite with a copy of DMN for the UD composite. As a result, the online calculation of the new integrated network only requires constitutive models for the matrix and fiber materials. Here we take the concatenation of the UD DMN with $N^{ud}=7$ and woven DMN with $N^{woven}=8$ as an example, and we count each active node with an independent material input as a DOF. For the two-scale woven DMN, the number of DOFs is
\begin{equation}\label{eq:dof2}
N_{DOF}^\text{two-scale}=\underbrace{16}_\text{yarn}+\underbrace{22}_\text{matrix} = 38.
\end{equation}
For the three-scale woven DMN, every yarn DOF is replaced by a UD DMN, and the new number of DOFs can be computed as
\begin{equation}\label{eq:dof3}
N_{DOF}^\text{three-scale}=16\times(\underbrace{5}_\text{fiber}+\underbrace{9}_\text{matrix})+\underbrace{22}_\text{matrix} = \underbrace{80}_\text{fiber}+\underbrace{166}_\text{matrix} =246.
\end{equation}
\begin{figure} [!t]
	\centering
	\graphicspath{{Figures/}}
	\subfigure[In-plane tension.]{\includegraphics[clip=true,trim = 0.0cm 0.0cm 1.0cm 0.5cm,width=0.44\textwidth]{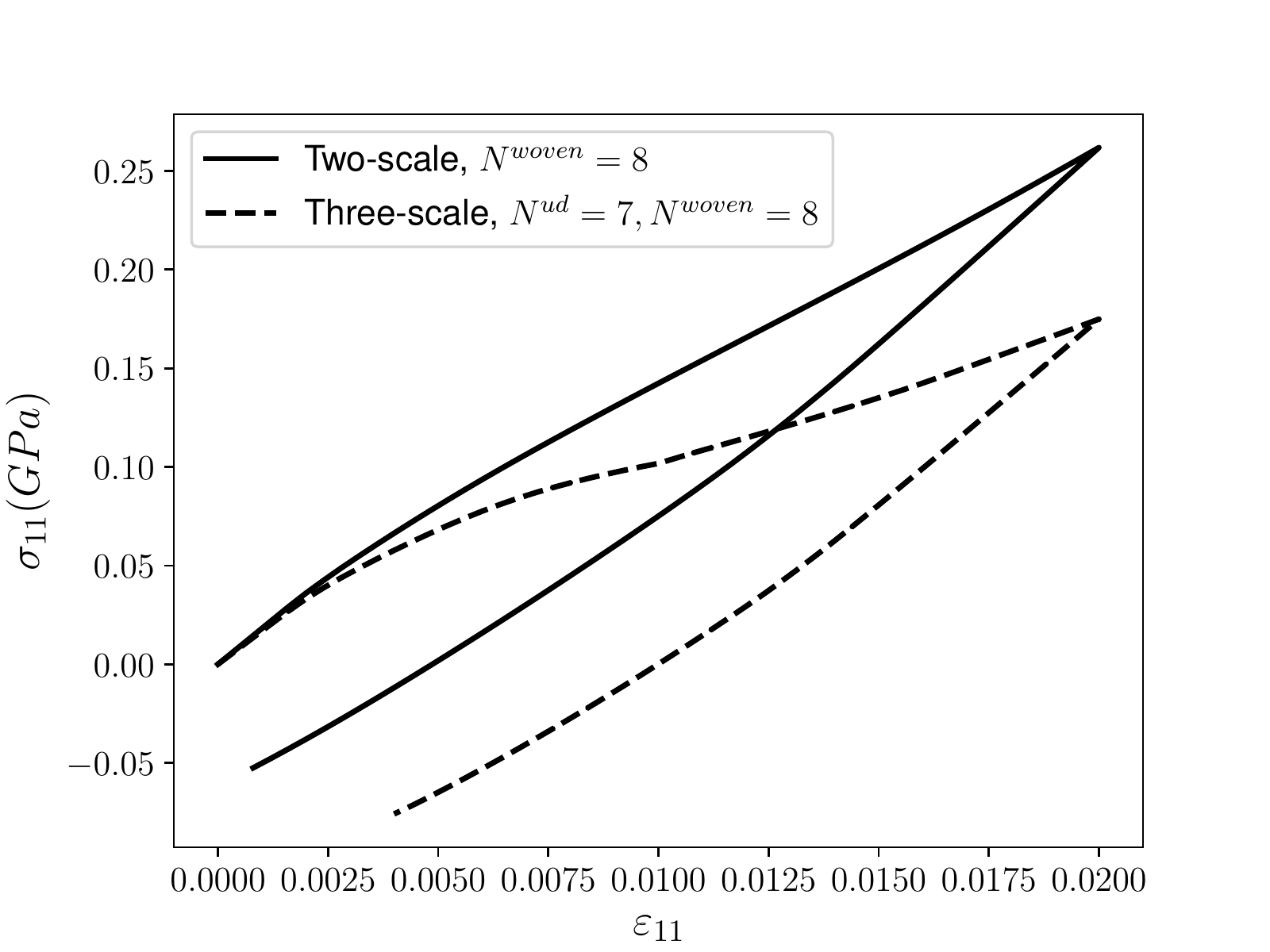}}
	\subfigure[In-plane shear.]{\includegraphics[clip=true,trim = 0.0cm 0.0cm 1.0cm 0.5cm,width=0.44\textwidth]{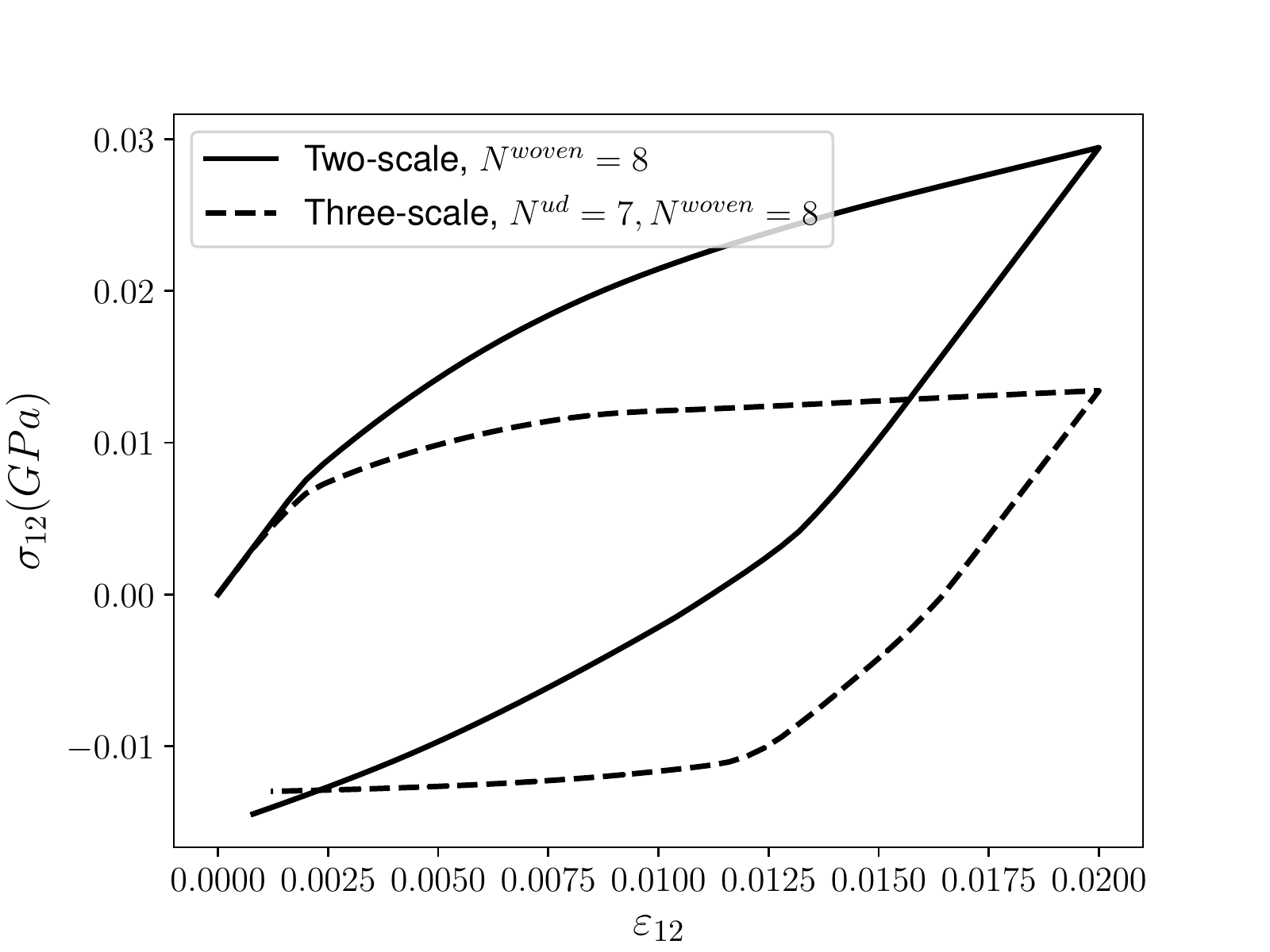}}
	\caption{Three-scale vs. two-scale homogenizations. The network depths of the UD and woven DMNs are 7 and 8, respectively. }
	\label{fig:threescaleResults}
\end{figure}

In the online stage for the three-scale homogenization, we conduct the Newton's method separately for the UD and woven parts in the integrated network. At each newton iteration of the woven DMN, it takes the converged stiffness matrices and residual stresses from the newton iterations of the copies of the UD DMN.  In this way, all the algorithms developed for the two-scale homogenization do not need to be altered. However, for the best efficiency, the three-scale network shown in Figure \ref{fig: threescale} should be solved as a whole in one Newton iteration, so that no extra time is paid to the sub-cyclings in the UD structures.

The in-plane uniaxial tension and shear results predicted by the two-scale and three-scale woven DMNs are compared in Figure \ref{fig:threescaleResults}. Due to the consideration of matrix plasticity in the yarn phase in the three-scale model, more hardening effects are observed in the overall material responses for both cases. After the initial yielding, the three-scale in-plane tension test also shows a stiffening effect due to the straightening of fibers. For the three-scale in-plane shear test, the fully developed hardening behavior is chose to perfectly plasticity, while this phenomena is not captured by the two-scale model due to the lack of plasticity in the homogenized UD material. In summary, more physics within the CFRP system can be considered by the three-scale model based on DMN, while the one based on DNS is computationally infeasible.

\section{A comment on computational cost}\label{sec:time}
Last but not least, we shall discuss the computational cost of DMN in both online and offline stages.  The computational time of generating the dataset and training the network with various depths for the particle-reinforced composite (see Section \ref{sec:particle}) are listed in Table \ref{table:time}. All the processes are parallelized using 10 Intel\textregistered $ $ Xeon\textregistered $ $ CPU E5-2640 v4 2.40 GHz processors. The generation of the training dataset with 400 samples took 39.5 h, or 356 s per sample. Note that to obtain all the components of the overall elastic stiffness matrix for each sample, we need to analyze the same RVE under 6 orthogonal loading conditions. Therefore, one RVE analysis of the particle-reinforced RVE takes around 59 s. Since the generation of each sample is independent, the wall time for generating the training data decreases linearly with the number of processors. 
\begin{table}[!t]
	\captionabove{Offline computational times for the particle-reinforced RVE.} 
	\centering 
	\label{table:time} 
	{\tabulinesep=1.0mm
		\begin{tabu}{c| c |c c c } 
			\hline 
			&Training data generation &\multicolumn{3}{c}{DMN training (20000 epochs)}\\
			&DNS (400 samples) &$\quad N=4\quad$&$\quad N=6\quad$&$\quad N=8\quad$\\
			\hline
			$N_{cpu}$&10&10&10&10\\
		Wall time (h) &39.5&5.4&16.7&43.0 \\
			\hline
	\end{tabu}}
\end{table}

The DMN training times  for $N=4$, 6 and 8 are 5.4, 16.7 and 43.0 h, respectively.  The time cost per epoch decreases as the number of active nodes in the bottom layer is reduced by the model compression algorithm. Here we take the network with $N=8$ as an example. The first 1000 epochs at the beginning of the training took 3.7 h, while the last 1000 epochs only cost 1.9 h. Our current parallelization algorithm is realized by distributing the training samples between various processors, or called ``data level parallelism". It has a good scalability as long as the number of processors is smaller than the mini-batch size in the SGD algorithm. Future improvements will include vectorizing the operations in the building block and parallelizing the computation at the model level, which will allow us to accelerate the training procedure of large-scale DMNs using more processing units (e.g. GPU).
\begin{figure} [!t]
	\centering
	\graphicspath{{Figures/}}
	\subfigure[Hyperelastic particle-reinforced composite.]{\includegraphics[clip=true,trim = 4.2cm 0.1cm 5.9cm 0.5cm,width=0.45\textwidth]{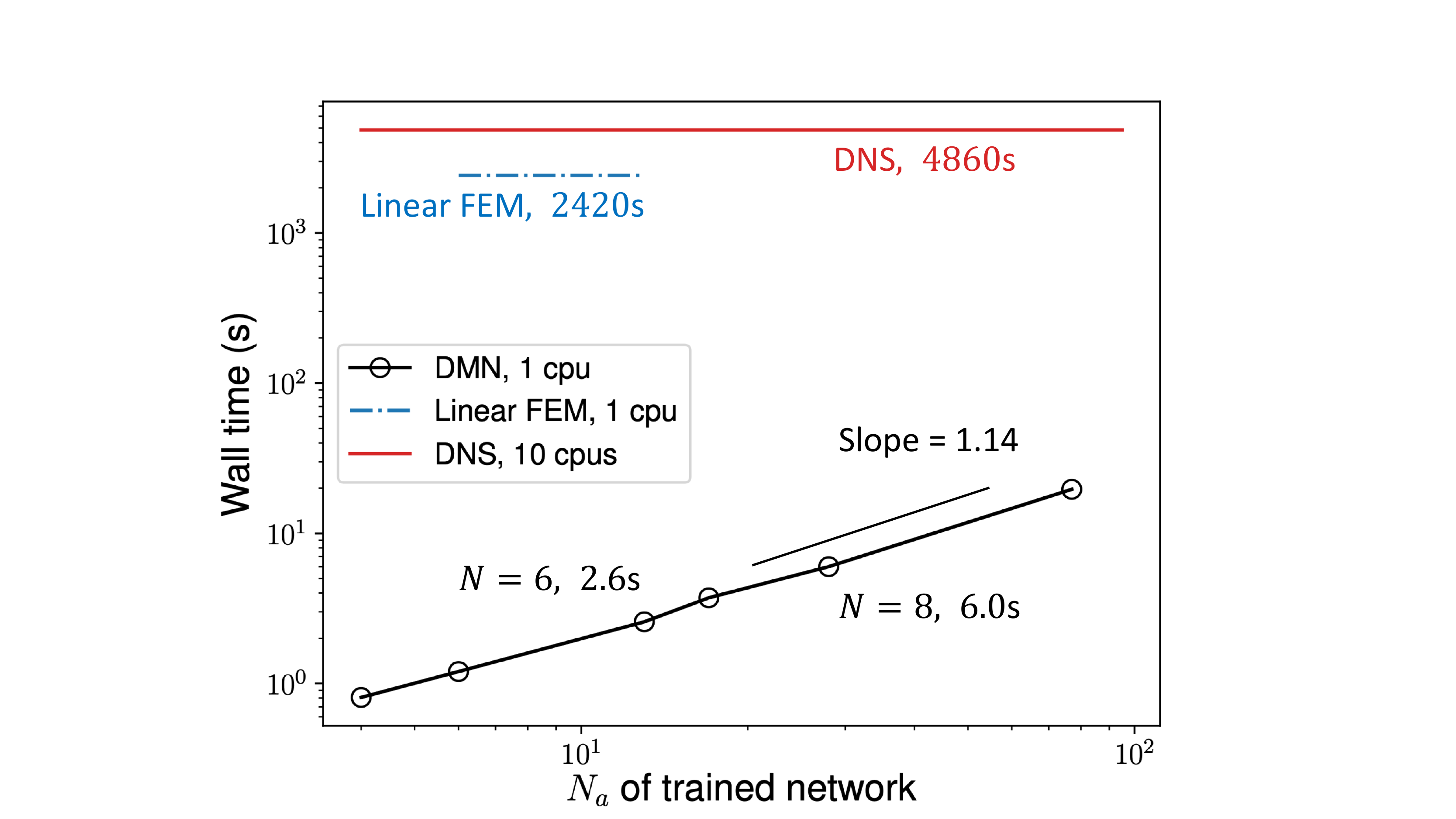}}
	\subfigure[CFRPs with two- and three-scale homogenizations.]{\includegraphics[clip=true,trim = 4.3cm 0.1cm 5.9cm 0.5cm,width=0.45\textwidth]{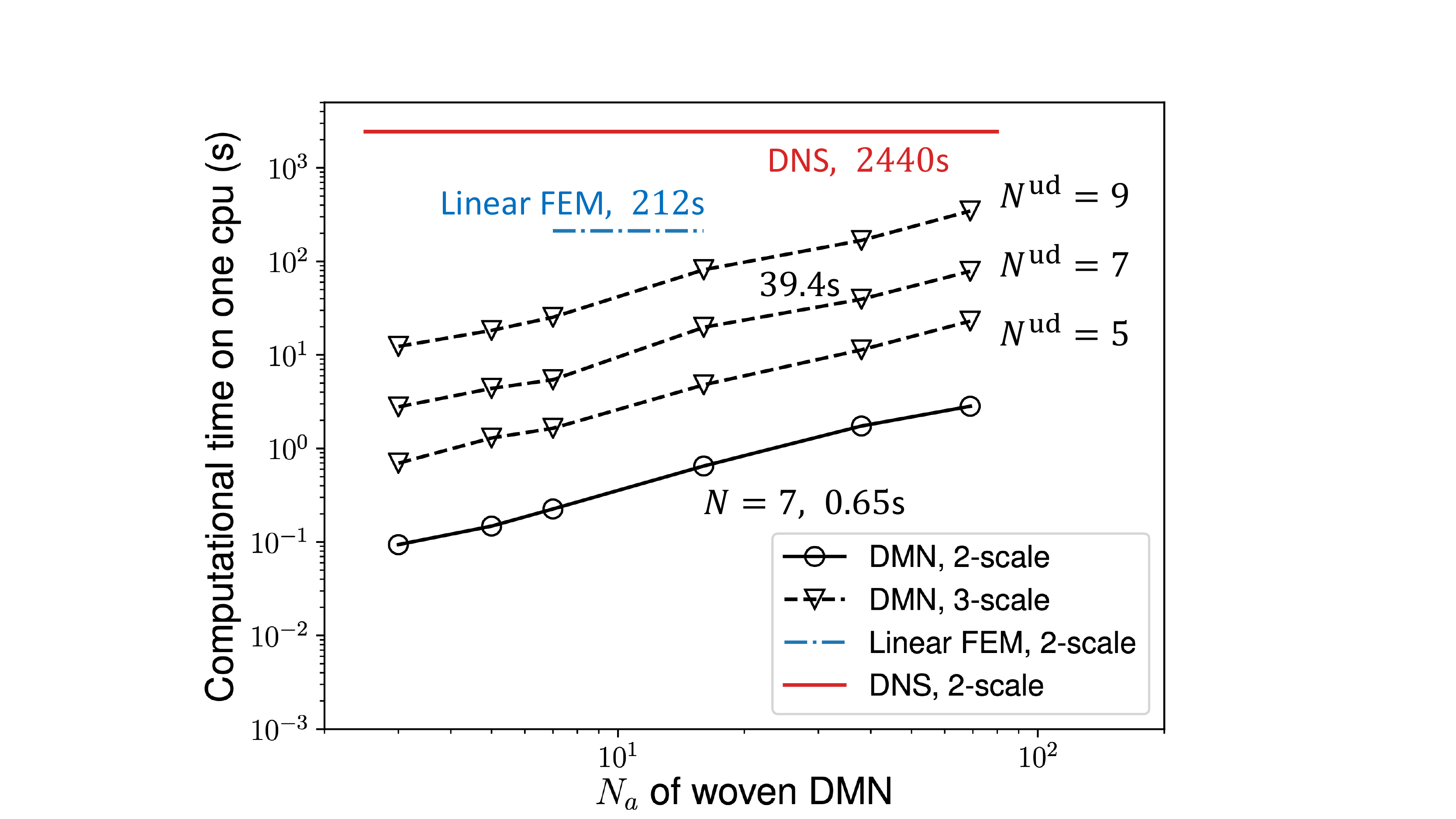}}
	\caption{Computational times of DNS models, linear FE models and DMNs in the online stage.}
	\label{fig:time}
\end{figure}

The computational times of DNS and the online DMNs for the hyperelastic particle-reinforced RVE are given in Figure \ref{fig:time} (a). By using 10 cores, the DNS with 59628 10-node tetrahedron elements required 4860 s. All the tests based on DMN were performed on one core. A linear relationship (slope $\approx1.14$) between the computational time and the number of DOFs $N_a$ is observed for network depth ranging from 4 to 9. When $N=8$ (with $N_a=28$), the online stage of DMN took 6.0 s, which is 810 times faster than the DNS in terms of the wall time, or 8100 times faster if measured by CPU time. Another plot on the computational times for the woven RVE is provided in Figure \ref{fig:time} (b). The DNS was limited to two-scale homogenization where the yarn phase is considered to be elastic, and it cost about 2440 s on one core. Both two-scale and three-scale homogenizations can be realized by DMN.  As we can see from the figure, the two-scale DMN with $N^\text{woven}=8$ took 1.7 s, and the three-scale DMN with $N^\text{woven}=8$ and $N^\text{woven}=7$ took 39.4 s. Ideally, if the concatenated three-scale network is solved simultaneously without sub-cycling as discussed in Section \ref{sec:cfrp}, its computational time is estimated to be 11.0 s based on the numbers of DOFs for the two-scale and three-scale models given in Eq. (\ref{eq:dof2}) and (\ref{eq:dof3}).

It is more objective to evaluate the efficiency of DMN by comparing with a FEM model with similar accuracy. For the simulation of hyperelastic particle-reinforced composite, the linear FE model took 2420 s on one CPU. As shown in Section \ref{sec:particle}, its average test error on 100 random samples (2.30\%) is close to that of the DMN with $N=6$ (1.38\%), which only took 2.6 s on one CPU for the same number of loading steps. At the similar level of accuracy, the DMN is around 930 times faster than the linear FE model.  For the two-scale homogenization of woven composite, the linear FE model is comparable to the DMN with $N=7$ in terms of the average test errors, 6.12\% vs. 3.48\%, while their computational times are 212 s vs. 0.65 s. Due to the complexity of woven microstructure, it is difficult to further coarsen the linear FE mesh with 5252 nodes. However, the material network with $N=7$ for woven composite has only 16 active nodes in the bottom layer, which may correspond to a $4\times2\times2$ FE mesh. Note that the current DMN framework is implemented as Python libraries, and it is possible to improve the efficiency through lower-level programming.

\section{Conclusions}\label{sec:conclusion}
A new framework of deep material network (DMN) for modeling general 3D heterogeneous materials with both material and geometric nonlinearities is developed. Theories of the 3D two-layer building block are derived based on the homogenization equations and interfacial conditions. We show a complete machine learning procedure for training the DMN by first generating offline elastic datasets using DNS, and then optimizing the network fitting parameters with stochastic gradient descent and model compression algorithms. The 3D architectures of DMN are explored for three representative multiscale materials systems, including particle-reinforced hyperelastic rubber composite with Mullins effect, polycrystalline materials with finite-strain rate-dependent crystal plasticity, and elasto-plastic CFRP systems formed by UD and woven composites. For all the examples, we demonstrate the effectiveness of DMN on representing complex RVE microstructures and its accuracy of predicting highly nonlinear behaviors in the online stage. One unique feature of DMN is that it can discover the hidden geometric information, such as phase volume fraction and orientation distributions, from pure mechanical data. While DNS is limited to two-scale homogenization due to its high computational cost, we also show a three-scale DMN of the CFRP system through the concatenation of UD and woven networks, which captures more physics across multiple length scales. The proposed DMN and its machine learning procedure open new possibilities in the design and concurrent simulation of multiscale material systems. The idea of embedding physics into machine-learning models shows promise for the future of data-driven multiscale mechanics and physics.

\section*{Acknowledgment}
Z. Liu and C.T. Wu warmly thank Dr. John O. Hallquist of LSTC for his support to this research. Z. Liu would like to thank Cheng Yu for his part in helpful discussions. This is a preprint version of an article published in \text{Journal of the Mechanics and Physics of Solids}. The final authenticated version is available online at: \url{https://doi.org/10.1016/j.jmps.2019.03.004}.

\appendix
\section{Elementary rotation matrices}\label{ap:ap1}
In small-strain formulation, the elementary rotation matrices shown in Eq. (\ref{eq:smallR}) are given by
\begin{equation}
\text{X}_{(1,1)}=1,\quad\textbf{X}_{([2,3,4],[2,3,4])}(\alpha)=\textbf{r}^p(\alpha),\quad \textbf{X}_{([5,6],[5,6])}(\alpha)=\textbf{r}^v(\alpha);
\end{equation}
\begin{equation*}
\text{Y}_{(2,2)}=1,\quad\textbf{Y}_{([1,3,5],[1,3,5])}(\beta)=\textbf{r}^p(-\beta),\quad \textbf{Y}_{([4,6],[4,6])}(\beta)=\textbf{r}^v(-\beta);
\end{equation*}
\begin{equation*}
\text{Z}_{(3,3)}=1,\quad\textbf{Z}_{([1,2,6],[1,2,6])}(\gamma)=\textbf{r}^p(\gamma),\quad \textbf{Z}_{([4,5],[4,5])}(\gamma)=\textbf{r}^v(\gamma).
\end{equation*}
The in-plane and output-plane rotation matrices $\textbf{r}^p$ and $\textbf{r}^v$ for an arbitrary angle $\theta$ are defined in Mandel notation as
\begin{equation}
\textbf{r}^p(\theta)=\begin{Bmatrix}
\cos^2\theta&\sin^2 \theta&\sqrt{2}\sin\theta\cos\theta\\
\sin^2 \theta&\cos^2\theta&-\sqrt{2}\sin\theta\cos\theta\\
-\sqrt{2}\sin\theta\cos\theta&\sqrt{2}\sin\theta\cos\theta&\cos^2\theta-\sin^2\theta\\
\end{Bmatrix},
\quad\textbf{r}^v(\theta)=\begin{Bmatrix}
\cos \theta&-\sin\theta\\
\sin\theta&\cos \theta\\
\end{Bmatrix}.
\end{equation}
In finite-strain formulation, the elementary rotation matrices shown in Eq. (\ref{eq:finiteR} are given by
\begin{equation}
\text{X}^f_{(1,1)}=1, \quad \textbf{X}^f_{([2,3,4,5],[2,3,4,5])}(\alpha)=\textbf{r}^{pf}(\alpha),\quad \textbf{X}^f_{([6,8],[6,8])}(\alpha)=\textbf{X}^f_{([7,9],[7,9])}(\alpha) = \textbf{r}^{vf}(\alpha);
\end{equation}
\begin{equation*}
\text{Y}^f_{(2,2)}=1, \quad \textbf{X}^f_{([1,3,6,7],[1,3,6,7])}(\beta)=\textbf{r}^{pf}(-\beta),\quad \textbf{Y}^f_{([4,9],[4,9])}(\beta)=\textbf{X}^f_{([5,8],[5,8])}(\beta) = \textbf{r}^{vf}(-\beta);
\end{equation*}
\begin{equation*}
\text{Z}^f_{(3,3)}=1, \quad \textbf{Z}^f_{([1,2,8,9],[1,2,8,9])}(\gamma)=\textbf{r}^{pf}(\gamma),\quad \textbf{Z}^f_{([4,6],[4,6])}(\gamma)=\textbf{Z}^f_{([5,7],[5,7])}(\gamma) = \textbf{r}^{vf}(\gamma),
\end{equation*}
where the in-plane and output-plane rotation matrices are
\begin{equation}
\textbf{r}^{pf}(\theta)=\begin{Bmatrix}
\cos^2\theta&\sin^2 \theta&\sin\theta\cos\theta&\sin\theta\cos\theta\\
\sin^2 \theta&\cos^2\theta&-\sin\theta\cos\theta&-\sin\theta\cos\theta\\
-\sin\theta\cos\theta&\sin\theta\cos\theta&\cos^2\theta&-\sin^2\theta\\
-\sin\theta\cos\theta&\sin\theta\cos\theta&-\sin^2\theta&\cos^2\theta\\
\end{Bmatrix},\quad
\textbf{r}^{vf}(\theta)=\begin{Bmatrix}
\cos\theta&-\sin\theta\\
\sin\theta&\cos \theta\\
\end{Bmatrix}
\end{equation}
for an arbitrary angle $\theta$. 

\section{Gradients for training}\label{ap:ap2}
Derivatives need to be derived for computing the gradient vector of the cost function in the training process. Since the procedure of deriving the derivatives is standard based on the chain rule, we will only provide some essential steps here.  All the derivatives of the building block that will appear during the training process are listed as below,
\begin{equation}
\dfrac{\partial \boldsymbol{\mathfrak{h}}_C}{\partial \bar{\textbf{C}}^1},\dfrac{\partial \boldsymbol{\mathfrak{h}}_C}{\partial \bar{\textbf{C}}^2},\dfrac{\partial \boldsymbol{\mathfrak{h}}_C}{\partial w^1},\dfrac{\partial \boldsymbol{\mathfrak{h}}_C}{\partial w^2},\dfrac{\partial \boldsymbol{\mathfrak{r}}_C}{\partial \textbf{C}},\dfrac{\partial \boldsymbol{\mathfrak{r}}_C}{\partial \alpha},\dfrac{\partial \boldsymbol{\mathfrak{r}}_C}{\partial \beta},\dfrac{\partial \boldsymbol{\mathfrak{r}}_C}{\partial \gamma}.
\end{equation}

The derivative of $(\hat{\textbf{C}}_{345})^{-1}$ with respect to an arbitrary variable $x$ (e.g. volume fraction or components in the stiffness matrix) will be necessary for computing the gradient of the strain concentration tensor $\textbf{s}^1$ in the homogenization function $ \boldsymbol{\mathfrak{h}}_C$:
\begin{equation}
\dfrac{\partial \left(\hat{\textbf{C}}_{345}\right)^{-1}}{\partial x}=-\left(\hat{\textbf{C}}_{345}\right)^{-1}\dfrac{\hat{\textbf{C}}_{345}}{\partial x}\left(\hat{\textbf{C}}_{345}\right)^{-1}.
\end{equation}
Moreover, the derivatives of volume fraction $f_1$ with respect to the weights $w^1$ and $w^2$ are
\begin{equation}
\dfrac{\partial f_1}{w^1} = \dfrac{f_2}{w^1+w^2},\quad \dfrac{\partial f_1}{w^2} = -\dfrac{f_1}{w^1+w^2}.
\end{equation}

Direct computation of derivatives of the rotation function $ \boldsymbol{\mathfrak{r}}_C$ could be complicated. Instead, it is decomposed into three steps by utilizing Eq. (\ref{eq:r3}), so that the derivatives of stiffness matrices with respect to the corresponding rotation angles are
\begin{equation}
\dfrac{\partial\bar{\textbf{C}}}{\partial \gamma}=-\textbf{Z}'(-\gamma)\textbf{C}^{\alpha\beta}\textbf{Z}(\gamma)+\textbf{Z}(-\gamma)\textbf{C}^{\alpha\beta}\textbf{Z}'(\gamma),
\end{equation} 
\begin{equation*}
\dfrac{\partial\textbf{C}^{\alpha\beta}}{\partial \beta}=-\textbf{Y}'(-\beta)\textbf{C}^\alpha\textbf{Y}(\beta)+\textbf{Y}(-\beta)\textbf{C}^\alpha\textbf{Y}'(\beta),
\end{equation*}
\begin{equation*}
\dfrac{\partial\textbf{C}^\alpha}{\partial \alpha}=-\textbf{X}'(-\alpha)\textbf{C}\textbf{X}(\alpha)+\textbf{X}(-\alpha)\textbf{C}\textbf{X}'(\alpha).
\end{equation*}
Specifically, $\textbf{X}'(\alpha)$ denotes the derivative of the rotation matrix $\textbf{X}$ to its rotation angle $\alpha$. Furthermore, the derivatives of $\bar{\textbf{C}}$ with respect to the components in ${\textbf{C}}^{\alpha\beta}$ can be written as
\begin{equation}\label{eq:a}
\dfrac{\partial \bar{C}_{ij}}{\partial{C}^{\alpha\beta}_{kl}}=Z_{ik}(-\gamma)Z_{lj}(\gamma).
\end{equation}
Similarly, we have
\begin{equation}
\dfrac{\partial {C}_{ij}^{\alpha\beta}}{\partial{C}^\alpha_{kl}}=Y_{ik}(-\beta)Y_{lj}(\beta) \quad \text{and}\quad \dfrac{\partial {C}_{ij}^\alpha}{\partial{C}_{kl}}=X_{ik}(-\alpha)X_{lj}(\alpha).
\end{equation}

\bibliography{references_ROM}
\end{document}